\documentclass[ALICE,manyauthors]{cernphprep}
\usepackage[comma,square,numbers,sort&compress]{natbib}
\usepackage{hyperref}
\usepackage{lineno}
\usepackage{inputenc}
\usepackage{fontenc}
\usepackage{hyperref}
\usepackage{graphicx}  
\usepackage{dcolumn}   
\usepackage{bm}        
\usepackage{amssymb}   
\usepackage{amsfonts}
\usepackage{graphics}
\usepackage{grffile}   
\usepackage{epsfig}
\usepackage{units}
\usepackage[usenames]{color}
\usepackage[normalem]{ulem} 
\usepackage{lineno}
\usepackage{color}
\usepackage{subfigure}
\newcommand{ \be }{\begin{linenomath*}\begin{eqnarray}}
	\newcommand{ \ee }{\end{eqnarray}\end{linenomath*}}
\newcommand{ \la }{\langle}

\newcommand{ \ra }{\rangle}

\def \mean#1 {{\la #1 \ra}}

\newcommand{ \pp}{pp}
\newcommand{ \pPb}{p--Pb}
\newcommand{ \PbPb}{Pb--Pb}
\newcommand{ \AuAu}{Au--Au}
\newcommand{ \dAu}{d--Au}

\newcommand{ \pt }{p_{\rm T}}
\newcommand{ \gevc }{GeV/\textit{c}}

\newcommand{ \pti}{\textit{p}_{\rm T,i}}

\newcommand{ \dptOne}{\rm{d}\textit{p}_{\rm T,1}}
\newcommand{ \dptTwo}{\rm{d}\textit{p}_{\rm T,2}}

\newcommand{ \phiOne}{\varphi_1}
\newcommand{ \phiTwo}{\varphi_2}

\newcommand{ \etaOne }{\eta_1}
\newcommand{ \etaTwo }{\eta_2}

\newcommand{ \Deta }{\Delta \eta}

\newcommand{ \DetaDphi }{\left(  \Delta \eta, \Delta \varphi \right)}

\newcommand{ \rhoTwo}{\rho_2}

\newcommand{ \RTWO}{$R_2$\ }

\newcommand{ \PTWO}{$P_2$\ }

\newcommand{\rhoDptDpt}{\la \Delta \textit{p}_{\rm T}  \Delta \textit{p}_{\rm T} \ra }

\newcommand{\dedx}{d$E$/d$x$}
\definecolor{dgreen}{cmyk}{1.,0.,1.,0.2}        
\definecolor{orange}{cmyk}{0.,0.353,1.,0.}    

\def \snn {\sqrt{\textit{s}_{_{\rm NN}}}}
\def \bp  {{\bf p}}

\begin{document}

\begin{titlepage}
\PHyear{2018}
\PHnumber{118}      
\PHdate{9 May}
%

\title{Two particle  differential transverse momentum and number
density correlations in \pPb\ and \PbPb\ at the LHC}\ShortTitle{Two particle  differential transverse momentum and number
density correlations}

\Collaboration{ALICE Collaboration\thanks{See
Appendix~\ref{app:collab} for the list of collaboration members}}
\ShortAuthor{ALICE Collaboration} 

\begin{abstract}
We present measurements of two-particle differential number correlation functions R2 and transverse momentum correlation functions $P_2$, obtained from \pPb\ collisions at 5.02 TeV and   \PbPb\  collisions at 2.76 TeV. The results are obtained using charged particles in the pseudorapidity range $|\eta| <$ 1.0, and transverse momentum range $0.2 < \pt < 2.0$   \gevc\  as a function of pair separation in pseudorapidity, $|\Delta\eta|$, azimuthal angle, $\Delta\varphi$, and for several charged-particle multiplicity classes.  Measurements are carried out for like-sign and unlike-sign charged-particle pairs separately and combined to   obtain charge-independent and charge-dependent correlation functions.  We study the evolution of the width of the near-side peak  of these correlation functions  with collision centrality. Additionally,  we study Fourier decompositions of the correlators in $\Delta\varphi$ as a function of the pair separation $|\Delta\eta|$. Significant differences in the dependence of their harmonic  coefficients on multiplicity classes are found.   These differences   can   be exploited, in theoretical models, to obtain further insight into charged-particle production and transport  in
heavy-ion collisions. Moreover, an upper limit of non-flow contributions to flow coefficients $v_n$ measured in \PbPb\ collisions based on the relative strength of Fourier coefficients measured in \pPb\ interactions is estimated. 
\end{abstract}
\end{titlepage}

\section{INTRODUCTION}
Measurements carried out at the Relativistic Heavy Ion Collider (RHIC)
and the Large Hadron Collider (LHC) during the last decade indicate that a  strongly interacting Quark--Gluon Plasma (sQGP) is produced in heavy nuclei
collisions at  high beam energies~\cite{Adams:2005dq,Adcox:2004mh,Arsene20051,Back:2004je}.
In particular, observations of strong elliptic
flow and theoretical studies based on relativistic hydrodynamics indicate that this
matter behaves as a very low specific shear viscosity (shear viscosity over entropy density ratio)
fluid~\cite{Adams:2004bi,Aamodt:2011by,Heinz:2013th,PhysRevLett.105.252302}. Additionally, the observed suppression of high transverse momentum ($\pt$) single-hadron
production  as well as dihadron correlations, in heavy-ion
collisions, compared to elementary \pp\
interactions, showed that the produced matter is rather opaque~\cite{Adams:2003im,Abelev:2009ah,Adler:2002tq,PhysRevLett.91.072301,PhysRevLett.88.022301,PhysRevLett.91.172302,PhysRevC.69.034910,PhysRevLett.96.202301,PhysRevC.75.024909,PhysRevLett.97.152301,Sickles:2004jz}.  Furthermore, studies of two- and multi-particle correlation functions  unravelled  several unanticipated
correlation features~\cite{Adler:2002tq, Adare:2008ae,PhysRevLett.97.162301,PhysRevC.77.011901,PhysRevLett.95.152301,PhysRevLett.97.052301,PhysRevLett.98.232302,PhysRevC.75.034901,Adler:2005ee}, including a near-side correlation peak (i.e., the prominent and relatively narrow peak centered at $\Delta\varphi=0$, $|\Delta\eta|=0$ observed in two-particle correlation functions) broadening, the appearance
of a near-side elongated ridge in relative pseudorapidity, as well as a
strong suppression or modification of the away-side correlation peak
relative to the one observed in pp collisions
~\cite{Adams:2004pa,Abelev:2009ag,Abelev:2009ah}. Extensive
studies were carried out, both at RHIC and LHC energies, to fully
characterize and understand the underlying causes of these features.
Significant progress was achieved with the realization that fluctuations in
the initial spatial configuration of colliding nuclei can greatly
influence  the measured correlations, most
particularly the development of odd and higher harmonics in the azimuthal particle distributions (anisotropic flow)~\cite{AlversRoland}. However, a quantitative assessment of  the
magnitude  and impact of non-flow effects on measured correlations  requires further investigations.
Non-flow effects may   arise from
resonance decays or low-multiplicity hadronization processes
associated with mini-jets, string fragmentation, or color tube
break-up~\cite{Dumitru:2008wn,Gavin:2008ev,Dusling:2009ar,Romatschke:2006bb,Shuryak:2007fu}. However, it remains unclear how these different particle production
mechanisms influence the shape and strength of correlation functions
and what their relative contributions might be. It is also unclear how
the surrounding environment associated with these processes
can alter two- and multi-particle correlation functions.  In an effort
to shed light on some of these questions, we consider
additional observables and types of correlation functions.

In this work, we present  measurements of $R_{2}$, a differential
two-particle number correlation function and  a differential transverse momentum correlation function, defined below, and
identified as $P_{2}$~\cite{Sharma:2008qr}.  The two
correlation functions are studied in \pPb\ collisions at
$\snn = $ 5.02 TeV and \PbPb\ collisions at   
$\snn = $ 2.76 TeV as a function of charged-particle
pair relative pseudorapidity, $\Delta\eta$, and relative azimuthal
angle, $\Delta\varphi$, as well as produced charged-particle multiplicity (corresponding to collision centrality in \PbPb). The observable $P_{2}$
features an explicit dependence on the transverse
momentum of the produced particles that provides sensitivity to
the correlation ``hardness,'' i.e., how low and high momentum particles contribute to the correlation dynamics.
Combined measurements of number and transverse-momentum correlations
  provide further insight into mechanisms of particle production and transport in
nucleus-nucleus collisions. The measurements presented in this work
 thus provide additional quantitative constraints on existing
models of collision dynamics used towards the characterization of the
matter produced in high-energy nucleus-nucleus collisions.

The $R_2$ and $P_2$ correlation functions are first reported independently for like-sign (LS) and unlike-sign (US) particles given they feature distinct dependences on particle production mechanisms. In particular, US pair correlations are expected to be rather sensitive to neutral resonances decays. The US and LS correlations are then combined to obtain charge-independent (CI) and charge-dependent (CD) correlation functions, defined in Sec.~\ref{Sec:Definition}. At high collisional energy, one expects energy-momentum conservation to play a similar role in US and LS correlations. The CD correlations obtained by subtracting LS from US correlations are then largely driven by charge conservation. Comparison of LS, US, CI, and CD correlations thus enables a detailed characterization of the particle production and transport  processes involved in heavy-ion collisions.  The study of CD correlations, in particular, shall then provide strong constraints on particle production models.

In order to obtain a  detailed characterization of the  $R_2$ and $P_2$ correlation functions,
 their shape is studied as a function of collision centrality and pair separation in pseudorapidity.
The width of the correlation functions, most particularly  their charge-dependent components $R_2^{\rm (CD)}$ and $P_2^{\rm (CD)}$,
are  sensitive to
charged-particle creation mechanisms and time of origin~\cite{Bass:2000az,Jeon:2001ue,Pratt:2015jsa,PhysRevC.90.064911}, momentum conservation~\cite{Pratt:2010zn,Borghini:2007ku,Borghini:2006yk}, as well as transport phenomena such as
radial flow~\cite{Voloshin:1999yf,Voloshin:2005qj,Pruneau:2007ua} and  diffusion processes~\cite{AbdelAziz:2006jv,Gavin:2006xd,Gavin:2011gr,Gavin:2008ta}.  
We report the longitudinal (pseudorapidity) and azimuthal widths of the near-side peaks of  the $R_2$ and $P_2$  correlators as a function of charged-particle multiplicity and longitudinal (pseudorapidity) pair separation. 
 Fourier decompositions are studied  as a function of pseudorapidity pair separation in order to obtain a detailed characterization of
flow and non-flow contributions to these correlation functions.

This paper is organized as follows. Section~\ref{Sec:Definition}
presents the definition of the observables $R_2$ and $P_{2}$ and
briefly discusses their properties. In Sec.~\ref{Sec:Experiment},
the experimental  setup and experimental methods used to acquire and
analyze the data are discussed, while the methodology used to measure the
$R_2$ and $P_2$ observables is described in
Sec.~\ref{Sec:methodology}. Systematic effects are considered in
Sec.~\ref{Sec:systematics}. Measurements of the  $R_2$ and $P_{2}$
correlation functions are reported in
Sec.~\ref{Sec:results}. 
Results are discussed in Sec.~\ref{Sec:discussion} and  summarized in Sec.~\ref{Sec:summary}.

\section{OBSERVABLES DEFINITION}
\label{Sec:Definition}

Single-and two-particle invariant cross sections  integrated over the $\pt $ range of interest are represented as
\be
\rho_1(\eta,\varphi) = \frac{1}{\sigma_1} \frac{d^2\sigma_1}{d\eta d\varphi};  \hspace{0.2in}
\rho_2(\eta_1,\varphi_1,\eta_2,\varphi_2) = \frac{1}{\sigma_2} \frac{d^4\sigma_2}{d\eta_1 d\varphi_1d\eta_2 d\varphi_2},
\ee
where $\rho_1$ and $\rho_2$ represent single- and two-particle densities, $\sigma_1$ and $\sigma_2$ represent single- and 
two-particle cross sections, while $\eta$ and $\varphi$ represent the pseudorapidity and azimuthal angle of produced particles. 
 
Two-particle correlations are determined based on normalized cumulants
defined according to
\be\label{Eq:r2-4D} R_2(\varphi_1,\eta_1,\varphi_2,\eta_2) =
\frac{\rho_2(\varphi_1,\eta_1,\varphi_2,\eta_2)}{\rho_1(\varphi_1,\eta_1)\rho_1(\varphi_2,\eta_2)}
- 1.
\ee
Given that the primary interest lies in the correlation strength as a function of pair separation,
one integrates over all coordinates  taking into account  
experimental acceptance  to  obtain the correlation
functions $R_2(\Delta\varphi,\Delta\eta)$ according to
\be
\label{Eq:r2-2D} \nonumber
R_2(\Delta\varphi,\Delta\eta) &=& \frac{1}{\Omega(\Delta\eta)}\int \rm{d}\varphi_1 \rm{d}\varphi_2 \rm{d}{\bar\varphi}
\delta(\Delta\varphi -\varphi_1 + \varphi_2)\delta(\bar\varphi - 0.5(\varphi_1 + \varphi_2)) \\
& &\times \int \rm{d}\eta_1 \rm{d}\eta_2 \rm{d}{\bar\eta}
\delta(\Delta\eta -\eta_1 + \eta_2) \delta(\bar\eta -0.5(\eta_1 + \eta_2)) R_2(\varphi_1,\eta_1,\varphi_2,\eta_2),
\ee
where  the azimuthal angles  $\varphi_1$ and $\varphi_2$ are measured in the range $[0,2\pi]$ whereas the pseudorapidities $\eta_1, \eta_2$ are measured in the range $[-1,1]$. The factor $\Omega(\Delta\eta)$ represents the width of the acceptance in $\bar\eta=(\eta_1+\eta_2)/2$ at a given $\Delta\eta=\eta_1-\eta_2$. The azimuthal angle difference, $\Delta\varphi=\varphi_1-\varphi_2$, is shifted to fall within  the range $[-\pi/2,3\pi/2]$. The integration is carried out across all values  of $\bar\varphi=(\varphi_1+\varphi_2)/2$.

Different observables can be defined which  are sensitive to
the correlation between the transverse momentum of produced
particles. Integral correlations  expressed in
terms of  inclusive and  event-wise averages of the product
$\Delta \textit{p}_{\rm T,i} \Delta \textit{p}_{\rm T,j}$ (where $\Delta\pti=\pti-\la p_{\rm
T} \ra$) of particle pairs $i \ne
j$ have been reported~\cite{Sharma:2008qr,Ravan:2013lwa,Adams:2005aw,Adamova:2003pz,Adamova:2008sx,Abelev:2014ckr}.
A generalization to differential correlation functions with dependences
on the relative azimuthal angles and pseudorapidities of particles is
straightforward when expressed in terms of inclusive
averages denoted $\la \Delta \pt  \Delta \pt \ra$~\cite{Sharma:2008qr}. In
this study, measurements of transverse momentum correlations are reported 
in terms of a dimensionless correlation function $P_{2}$ defined as a
ratio of the differential correlator $\la \Delta \pt  \Delta \pt \ra$ to 
the square of the average transverse momentum
\be \label{Eq:p2-2D}
P_{2} (\Delta\eta,\Delta\varphi) = \frac{\rhoDptDpt \DetaDphi}{\la \pt\ra^2}
= \frac{1}{\la \pt\ra^2} \frac{\int\limits_{p_{\rm
T,min}}^{\textit{p}_{\rm T,max}} ~ {\rhoTwo(\bp_1,\bp_2)\, ~ \Delta
\textit{p}_{\rm T,1} \Delta \textit{p}_{\rm T,2} ~~ \dptOne \dptTwo}}
{\int\limits_{\textit{p}_{\rm T,min}}^{\textit{p}_{\rm T,max}} {\rhoTwo(\bp_1,\bp_2) ~~
\dptOne \dptTwo}},
\label{Eq:Rpt}
\ee
where  $\la \pt
\ra = \int \rho_1 ~ \pt ~ \rm{d}\pt~ / \int \rho_1 ~ \rm{d}\pt$
is the inclusive average momentum of produced particles in an event
ensemble. Technically, in this analysis, integrals of the numerator
and denominator of the above expression are first evaluated in four
dimensional space as functions of $\eta_1$, $\varphi_1$, $\eta_2$, and
$\varphi_2$. The ratio is calculated and  subsequently
averaged over all coordinates, similarly as for $R_2$, as discussed
above. For the sake of simplicity, the inclusive momentum $\la \pt \ra$
is considered independent of the particle's pseudorapidity. This
approximation is justified by the limited pseudorapidity range
of this analysis and by prior observations of the approximate invariance
of $\la \pt \ra$ in the central rapidity ($\eta \approx 0$) region~\cite{Adams:2003xp}.

By construction, $P_{2}$ is a measure of two-particle transverse momentum
correlations: it is positive whenever particle pairs emitted at
specific  azimuthal angle and pseudorapidity differences are more likely
to both have transverse momenta higher (or  lower) than the $\pt$ average, and negative
when a high $\pt$ particle ($\pt>\la \pt\ra$) is more likely to be accompanied by a low
$\pt$ particle ($\pt<\la \pt\ra$). For instance, particles emitted within a jet typically have higher
$\pt$ than the inclusive average. Jet particles therefore contribute a
large positive value to $P_{2}$. Hanbury-Brown--Twiss (HBT)   correlations, determined by pairs of identical particles with $\textit{p}_{\rm T,1} \approx \textit{p}_{\rm T,2}$  likewise
contribute positively to this correlator. However, bulk correlations involving a mix of low and high momentum correlated particles can contribute both positively and negatively.

The $R_2$ and $P_2$ correlation functions reported in this work are determined for unidentified charged-particle pairs in the 
range $0.2 < \pt < 2.0 $  \gevc\ and are considered untriggered correlation functions. Differential correlation functions offer multiple
advantages over integral correlations as they provide more detailed
information on the particle correlation structure and
kinematical dependences.
They can also be corrected for instrumental effects more reliably than measurements of  integral  correlations. Such corrections for instrumental effects on $R_2$ and $P_2$ correlation functions are discussed in  Sec.~\ref{Sec:methodology}.

The LS and US correlation functions are additionally combined to
obtain charge-independent (CI) and charge-dependent (CD)
correlation functions defined according to
\be\label{Eq:CI}
O^{\rm (CI)}&=&  \frac{1}{2}\left( O^{\rm (US)} +  O^{\rm (LS)} \right) =  \frac{1}{4}\left( O^{(+, -)} + O^{(-, +)} +  O^{(+, +)} +  O^{(-, -)} \right),
\\ \label{Eq:CD}
O^{\rm (CD)} & =& \frac{1}{2}\left( O^{\rm (US)} - O^{\rm (LS)} \right)=  \frac{1}{4}\left( O^{(+, -)} + O^{(-, +)} -  O^{(+, +)} -  O^{(-, -)} \right),
\ee
where $O$ represents either of the observables $R_2$ and $P_{2}$.  

Charge-independent correlators $O^{\rm (CI)}$ measure the average correlation strength
between all charged particles, whereas charge-dependent correlators
$O^{\rm (CD)}$ are sensitive to the difference between correlations of
US particles and those of LS particles.
At high collision energies, such as those achieved at the LHC,  negatively and positively charged particles are produced in approximately equal quantities and are found to have very similar $\pt$ spectra~\cite{Abelev:2012wca}.  The impact of  energy-momentum conservation on particle correlations is thus  expected to be essentially the same for US and LS  pairs.  The $O^{\rm (CD)}$ correlators consequently suppress the influence of energy-momentum conservation and provide particular
sensitivity to unlike-sign charge pair creation and transport processes.
The charge-dependent correlation function $R_{2}^{\rm (CD)}$, in particular,
should in fact feature similar sensitivity to charge pair $(+,-)$ creation
as the charge balance function $B$ defined according to
\be\label{Eq:balanceFunction}
B(\Delta\eta) =
\frac{1}{2} \left( \frac{\rho_2^{(+,-)} - \rho_2^{(+,+)} }{\rho_1^{(+)}} + \frac{\rho_2^{(-,+)} -
\rho_2^{(-,-)} }{\rho_1^{(-)}}  \right)
\ee
and proposed by Pratt et al. to investigate the evolution of quark production
in heavy-ion collisions~\cite{Bass:2000az, Jeon:2001ue, Pratt:2003gh}.
Several measurements and theoretical studies of the balance function have
already been reported. The STAR experiment has measured balance functions
in  \AuAu, \dAu, and \pp\ collisions at $\snn = $130  and $200$ GeV~\cite{Adams:2003kg, Aggarwal:2010ya,Abelev2010239,Adamczyk:2015yga}.
More recently, the  ALICE collaboration reported  observations of a narrowing of the
balance function with increasing produced charged-particle multiplicity ($N_{\rm ch}$) in  \PbPb\   collisions
at $\snn = $ 2.76 TeV, as well as in \pPb\ collisions at $\snn =$  5.02 TeV, and \pp\
collisions at $\snn = $ 7 TeV~\cite{Abelev:2013csa,Adam:2015gda}. Measurements in \AuAu\ and \PbPb\ are
in qualitative agreement with the  scenario, proposed by Pratt et al.~\cite{Bass:2000az, Jeon:2001ue, Pratt:2003gh},
of two-stage quark production in high-energy central heavy-ion collisions but observations
of a narrowing of the balance function with increasing $N_{\rm ch}$ in \pPb\  and \pp\ put this simple
interpretation into question.
At RHIC, and even more  at LHC energies, the number of positively and  negatively charged particles produced
in the range $|\eta|<1.0$ are nearly equal. Hence, the observable $R_2$ and the
balance function are thus related according to 
\be\label{Eq:balanceFunctionConnec}
R_{2}^{\rm (CD)}(\Delta\eta) = \frac{B(\Delta\eta)}{\rho_1^{(+)} + \rho_1^{(-)}}.
\ee
This implies that the narrowing of the balance function observed in most
central collisions, relative to peripheral collisions, is matched by
a reduction of the width of
the charge-dependent correlation function, $R_{2}^{\rm (CD)}$. Additionally,
given the observables $R_2$ and $P_2$ are both dependent on integrals of the
two-particle density $\rho_2(\vec p_1, \vec p_2)$, one
might expect a similar longitudinal narrowing of  $P_2$ with collision centrality.
However,  the explicit dependence of $P_2$'s   on the product $\Delta \pt \Delta \pt$ implies it might have a different sensitivity to 
the collision system's radial expansion (radial flow)  relative to that of $R_2$. A comparison of the centrality dependence of the longitudinal
widths of the $R_2$ and $P_2$ correlations may then provide additional insight
into the system's evolution and particle production dynamics, as well as put new
constraints on models designed to
interpret the observed narrowing of the balance function and the near-side
ridge~\cite{Cheng:2004zy}.


\section{ALICE DETECTOR AND DATA ANALYSIS}
\label{Sec:Experiment}

The analysis and results reported in this paper are based on data acquired with the ALICE detector~\cite{Aamodt:2008zz} 
during the $\snn = $ 2.76 TeV \PbPb\ run in 2010 and the $\snn = $ 5.02 TeV \pPb\ run in 2013. The reported correlation functions 
are measured for charged particles detected within the Inner Tracking System (ITS)~\cite{Nouais:2001cd} and the Time 
Projection Chamber (TPC)~\cite{2010NIMPA.622..316A}. The ITS and TPC are housed within a large solenoidal 
magnet producing  a uniform longitudinal magnetic field of 0.5 T. Together they provide charged-particle track reconstruction and 
momentum determination with full coverage in azimuth and in the pseudorapidity range $|\eta|<1.0$. Data were acquired with 
a minimum bias (MB) trigger primarily based on the V$0$ detector, which also served for \PbPb\ collision centrality and \pPb\ 
multiplicity class selection. 
This detector consists of sub-systems V$0$A  and V$0$C which  cover  the pseudorapidity ranges $2.8 < \eta < 5.1$ and $-3.7 < \eta < -1.7$, respectively.
Detailed descriptions of the ALICE detector, its subsystems, and triggers, as well as their 
respective performance, were reported 
elsewhere~\cite{Aamodt:2008zz,Nouais:2001cd,Baechler:2004ge,Beole:2012asa,Carminati:2004fp,Alessandro:2006yt,Abelev:2014ffa}.

The primary vertex of a collision is reconstructed based on charged-particle tracks measured with  the ITS and 
TPC detectors. Events were included in this analysis if  at least one accepted charged-particle track contributed 
to the primary vertex reconstruction and if they featured only one primary vertex. The primary vertex was furthermore 
required to be  within $\pm$10 cm from the nominal interaction point along the beam direction to ensure a uniform 
$\eta$ acceptance within the TPC. The fraction of pile-up events in the analysis sample is found to be negligible 
after applying dedicated pile-up removal criteria~\cite{Abelev:2014ffa}. Event filtering based on primary vertex 
selection criteria yielded samples of approximately $14 \times 10^{6}$ \PbPb\ events and $81 \times 10^{6}$ \pPb\ events.

The centrality of \PbPb\ collisions is estimated from the total signal amplitude measured by the V$0$ detectors using a 
standard ALICE procedure~\cite{Aamodt:2010cz, Abelev:2013qoq}. Nine collision centrality classes corresponding to 
0--5\% (most central collisions), 5--10\%, 10--20\%, 20--30\%, up to 70--80\% fractions of the total cross section 
were used in the analysis. The most peripheral collisions, with a fractional cross section $>80$\%, are not included in 
this analysis to avoid issues encountered with limited collision vertex reconstruction and trigger efficiencies. The 
\pPb\ data are similarly analyzed in terms of multiplicity classes. An ALICE analysis reported in~\cite{Adam:2016jfp} 
showed that in \pPb\ collisions, the produced charged-particle multiplicity is only loosely related to the collision impact 
parameter. So while it is appropriate to analyze the data in terms of multiplicity classes based on their fractional cross 
sections, these classes cannot be considered a direct indicator of the impact parameter in those collisions. They are 
representative, nonetheless, of qualitative changes in the particle production. Our analysis goal is thus to identify and 
document changes and trends in the shape and strength of the $R_2$ and $P_2$ correlators as a function of these 
multiplicity classes.

The analysis was restricted to primary particles, i.e., particles produced by strong interactions. Contamination from 
secondary charged particles (i.e., particles originating from weak decays such as neutral kaons ($K_\textit{S}^0$) 
and lambdas ($\Lambda^0$), conversions and secondary hadronic interactions in the detector material) is suppressed 
with track selection criteria based on charged-tracks' distance of closest approach (DCA) to the primary interaction 
vertex of the collision. Only ``bulk'' charged-particle tracks measured in the transverse-momentum range 
$0.2 < \textit{p}_{\rm T} < 2.0$~\gevc\ were selected. Particles in this momentum range constitute the dominant fraction 
of the produced particles and are believed to be primarily the product of non-perturbative interactions. They thus constitute 
the main focus of this work towards the characterization of the systems produced in \pPb\ and \PbPb\ collisions.

In order to suppress contamination from spurious and incorrectly reconstructed tracks, charged-particle tracks were 
included in the analysis only if they consisted of  at least 70 out of a maximum of 159 reconstructed TPC space points, 
and featured a momentum fit with a  $\chi^{2}$-value per degrees of freedom smaller than 4. Additionally, tracks identified 
as candidate daughter tracks of reconstructed secondary weak-decay topologies were also rejected. The DCA of 
extrapolated trajectories to the primary vertex position was restricted to less than 3.2 cm along the beam direction 
and less than 2.4 cm in the transverse plane. These selection criteria are broad and chosen to provide a high reconstruction 
efficiency. As such they are susceptible to some contamination of the primary track sample from secondary particles, such 
as charged hadrons produced by weak decays of $K_\textit{S}^0$ mesons and $\Lambda^0$ baryons. One verified, 
however, with the applications of more stringent DCA requirements, that such secondary decays have a relatively small 
impact on the measured correlation functions. These and other systematic effects are discussed in 
Sec.~\ref{Sec:systematics}. In addition, contamination of the primary track sample by electrons originating from 
$\gamma$-conversions and $\pi^{0}$-Dalitz decays is suppressed based on measurements of the tracks specific 
ionization energy loss (d$E$/d$x$) carried out with the TPC. Average energy losses are evaluated based on a 
truncated average method described in~\cite{Yu:2013dca}. The pion, kaon, proton, and electron specific energy loss 
dependence on momentum is used to reject tracks compatible with an electron hypothesis. Tracks with a d$E$/d$x$ 
within 3$\sigma$ of the expectation value for electrons and outside of 3$\sigma$ away of the expectation values for 
pions, kaons and protons, were excluded from the analysis. Further rejection of electrons produced by 
$\gamma$-conversions was accomplished by imposing a minimum invariant mass value of 0.05 \gevc\ $^2$ on 
all charged-particle pairs considered for inclusion in the analysis. Variations of these selection criteria, discussed in 
Sec.~\ref{Sec:systematics}, were studied to quantify systematic effects resulting from hadron losses and contamination 
by secondaries.

The above criteria lead to a reconstruction efficiency of about 80\% for primary particles and contamination from 
secondaries of about 5\% at $p_{\rm{T}}$ = 1 \gevc\ ~\cite{Abelev:2013bla}. No filters were used to suppress 
like-sign (LS) particle correlations resulting from HBT effects, which produce a strong and narrow peak 
centered at $\Delta\eta,\Delta\varphi=0$ in LS correlation functions.

\section{ANALYSIS METHODOLOGY}
\label{Sec:methodology}

\subsection{Two-particle correlations}

The correlation
functions $R_{2}$ and $P_{2}$    are nominally independent of detection efficiencies,
bin-by-bin  in $\Delta \eta$ and $\Delta \varphi$,
provided they are invariant during the data
accumulation period and independent of event characteristics and conditions~\cite{Pruneau:2002yf,Sharma:2008qr}.
However,  particle detection efficiencies are found to exhibit a small
dependence on the position of the  primary vertex,  $v_{z}$. This  creates  extraneous
structures in  the correlation observables $R_2$ and $P_2$  at $\Delta \eta \approx 0$ and  near $|\Delta \eta| \approx 2$.
Studies of these effects~\cite{Ravan:2013lwa,Tarini:2011xxa} showed they can be
properly suppressed by measuring the single- and two-particle
yields in narrow bins of $v_{z}$ and calculating $R_2$ and $P_2$ as  averages of
correlations measured in each $v_{z}$ bin.  In this work, it is found that distortions can be
reasonably well suppressed by using 0.5 cm wide $v_{z}$ bins. Given the fiducial
$v_{z}$ range of  $|v_{z}|<$ 10 cm, this  suggests the analysis would have to be carried out in  40 $v_{z}$ bins and thus 40 sets of histograms. Instead, one
uses a weight technique in which
single- and two-particle histograms are incremented with $v_{z}$ dependent weights
pre-calculated to equalize the detection response across the entire
fiducial acceptance~\cite{Ravan:2013lwa}.
Weights,  $w_{\pm}(\eta,\varphi,\pt,v_{z})$,  are calculated independently
for positively and negatively charged particles, positive and
negative magnetic field polarities,  as the inverse of
raw (i.e., uncorrected) particle yields, $N^{\pm}(\eta,\varphi,\pt,v_{z})$,
determined as a function of  pseudorapidity, $\eta$, azimuthal angle, $\varphi$,
transverse momentum, $\pt$, and the
vertex position $v_{z}$ of the events.
The analysis reported in this work was carried out with weights calculated in 
40 bins in $v_{z}$ in the range $|v_{z}|<10$ cm, 72 bins in $\varphi$
(full azimuthal coverage), 20 bins in $\eta$ in the range $|\eta|<1.0$,
and 18 bins in $\pt $  in the range $0.2 < \pt  < 2.0$ \gevc.
The analysis proceeded in  two stages. In the first stage,
all events were processed to determine weights according to
\be
w_{\pm}(\eta,\varphi,\pt,v_{z}) = \frac{N^{\pm}_{avg}(\pt)}{N^{\pm}(\eta,\varphi,\pt,v_{z})},
\ee
where $N^{\pm}_{avg}$ represents a $\pt$-dependent average of particle yields measured at all $\varphi$, $\eta$, and $z$. Calculated weights were used in the second stage to analyze all events and obtain raw number densities $\rho_1(\eta,\varphi)$  and $\rho_2(\etaOne,\phiOne,\etaTwo,\phiTwo)$, as well as $\pt$-dependent quantities.  Single particle histograms, pair histograms, and $\pt $ histograms  were incremented with  weights $w_{\pm}(\eta,\varphi,\pt,v_{z})$,   $w_{\pm}(\eta_1,\varphi_1,p_{\rm T,1},v_{z})w_{\pm}(\eta_2,\varphi_2,p_{\rm T,2},v_{z})$,  and $p_{T,1}p_{T,2}w_{\pm}(\eta_1,\varphi_1,p_{\rm T,1},v_{z})w_{\pm}(\eta_2,\varphi_2,p_{\rm T,2},v_{z})$, respectively. These histograms were  used   to calculate the correlators according to Eqs. (\ref{Eq:r2-4D}--\ref{Eq:p2-2D}).

\begin{figure}[h!]
\includegraphics[width=0.39\linewidth]{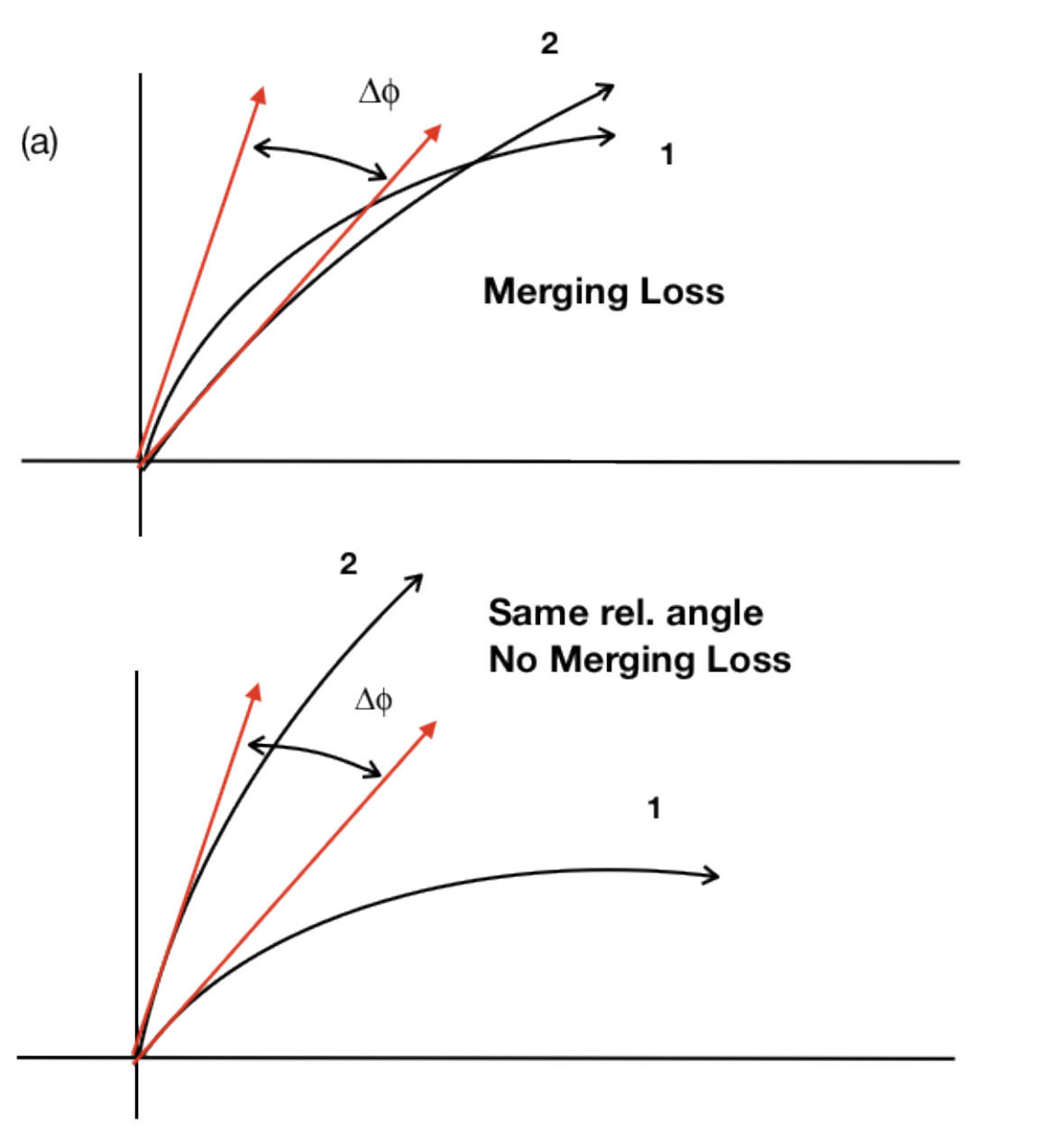}
\includegraphics[width=0.49\linewidth]{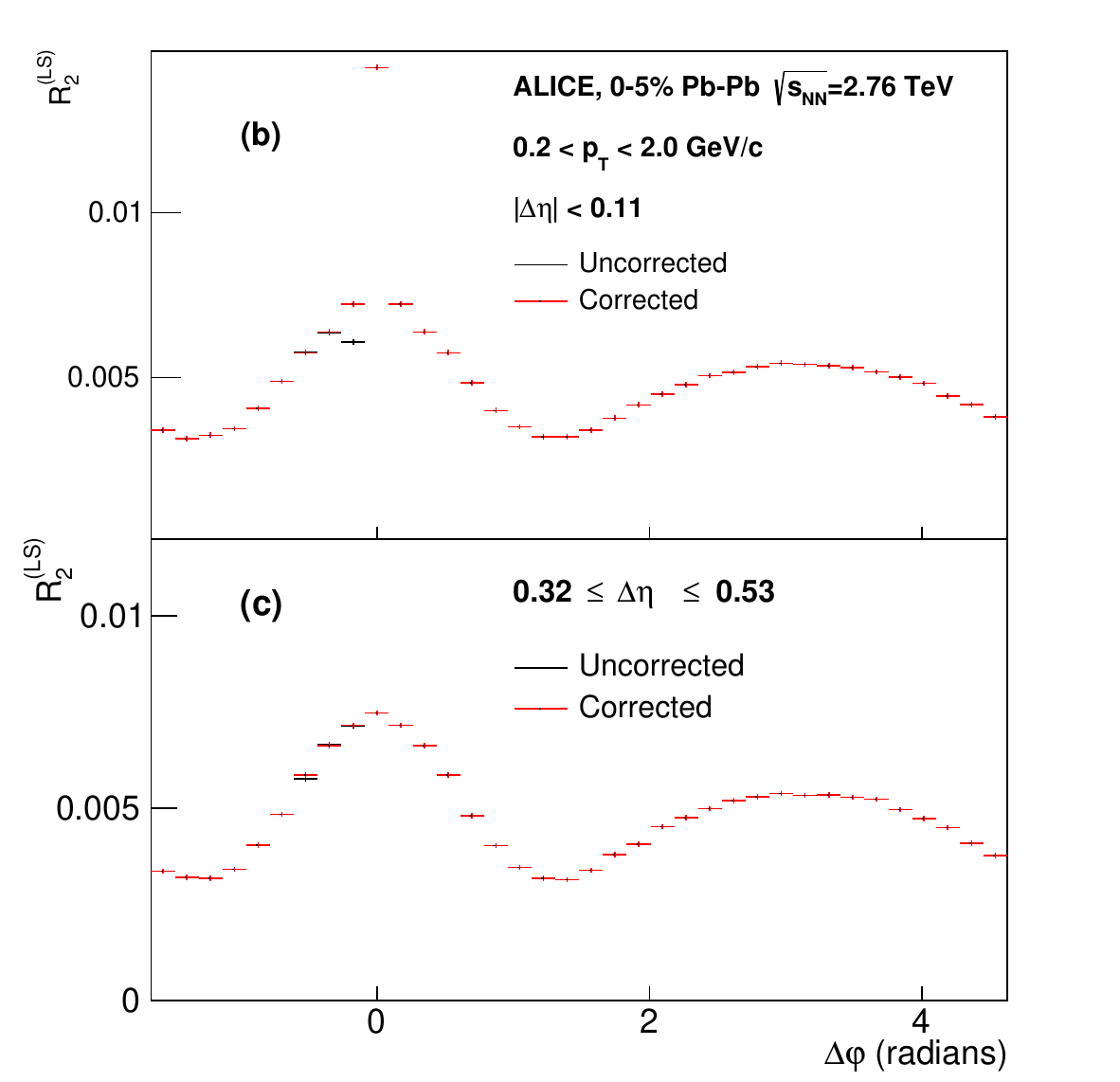}
\caption{(a) Schematic illustration of  cowboy (top) and sailor (bottom) track topologies for like-sign pairs; (b) Projection of the correlator   $R_{2}^{\rm (--)}$
onto $\Delta\varphi$ for LS pairs in the range $|\Delta\eta|<0.11$; and (c)  in the range $0.32 \le \Delta\eta < 0.53$.}
\label{fig:comboy}
\end{figure}

The correlators $R_{2}$ and $P_{2}$ were measured for the particle pair charge combinations
$(+, -)$, $(-, +)$, $(+, +)$, and  $(-, -)$ separately. For a symmetric collision
system such as \PbPb, correlations between particles are symmetric under
independent reflections   $\Delta\eta \rightarrow -\Delta\eta$ and
$\Delta\varphi \rightarrow -\Delta\varphi$. The measured pair yields were first checked for detector effects.  They are  indeed symmetric under
reflections $\Delta\eta \rightarrow -\Delta\eta$ and
$\Delta\varphi \rightarrow -\Delta\varphi$. The correlation functions
$R_2$ and $P_2$ measured in \PbPb\  collisions are thus fully symmetrized
in $\Delta\eta$ and  $\Delta\varphi$. In the case of the  \pPb\   collision system,
the lack of reflection symmetry $z \rightarrow -z$  implies that only $\Delta \varphi$
symmetry is expected. In principle, the pair correlations, much like the
single particle yields, could then feature  a non-symmetric and arbitrarily complex
dependence on $\Delta \eta$. In practice, one finds that the forward
($\Delta \eta>0$) and backward ($\Delta \eta<0$) correlation yields are equal
within the statistical and systematic uncertainties of the measurement, owing, most
likely, to the  narrow $\eta$  range of the detector acceptance relative to the very
wide rapidity span of particles produced at LHC energies.
The correlation functions $R_2$ and $P_2$ reported for  \pPb\   collisions are thus
also fully symmetrized in $\Delta\eta$ and  $\Delta\varphi$.  Additionally, one
observes that the correlation functions
of $(+, +)$ and $(-, -)$ pairs are equal within statistical uncertainties. One thus does not report them
independently. Overall, given the symmetry of $(+,-)$ and $(-,+)$
correlations and the observed equality of $(+,+)$ and $(-,-)$ correlations, one
averages the former to obtain unlike-sign (US)  and the latter to obtain like-sign (LS)
$R_2$ and $P_2$ correlation functions that  are fully symmetrized for both
collision systems.
The  weight correction procedure works  very well for single particle losses  but does not
address pair losses, most particularly those associated with track crossing and merging topologies for pairs with $\Delta \eta \approx 0$. We exploit the
expected $\Delta\varphi$ symmetry of the correlation functions by using lossless ``sailor" pair topologies to correct for losses observed with  ``cowboy" topologies~\cite{Agakishiev:2011st}. For like-sign pairs, the two topologies are distinguished, for a given magnetic field polarity, as schematically illustrated in Fig.~\ref{fig:comboy} (a), by counting pairs based on a momentum ordering technique: pairs featuring $p_{{\rm T},2} >p_{{\rm T},1}$ and $\Delta\varphi_{21} = \varphi_2-\varphi_1>0$ are counted at   
$\Delta\varphi >0$ as pair incurring no losses, whereas pairs at $p_{{\rm T},2} >p_{{\rm T},1}$ and $\Delta\varphi_{21} = \varphi_2-\varphi_1<0$
are counting at  $\Delta\varphi < 0$ as pair incurring losses. In the  $\Delta \eta < 0.2$ range where such losses occur, it is thus sufficient to use pairs with 
$\Delta\varphi >0$ to correct the yield of pairs with $\Delta\varphi  < 0$. Projections of $R_2^{\rm (--)}$, displayed in Fig.~\ref{fig:comboy}, show that losses associated with cowboy topologies are strongest at  $|\Delta \eta| < 0.11$ and negligible at
$|\Delta \eta| > 0.32$. A similar technique based on charge ordering is used for unlike-sign tracks. Unfortunately, this technique does not enable full efficiency correction  for track pairs with $|\Delta\eta| < 0.3$  and
$|\Delta\varphi| \approx 0$ radians. The $3\times 3$ bin region centered
at $\Delta\eta = \Delta\varphi = 0$ is thus under corrected. The two-dimensional correlators reported in this work
are then plotted without those bins. Note, however, that the calculation of the near-side peak widths, discussed in this work, do include the central $3\times 3$ bins and the potentially incomplete efficiency loss correction is treated as source of systematic error.

The  azimuthal dependence, $\Delta\varphi$, of the correlation function was studied by performing   a Fourier decomposition in several narrow ranges of $\Delta \eta$.
The Fourier decompositions were carried out using projections of the $R_{2}^{\rm (CI)}$ and $P_{2}^{\rm (CI)}$
distributions onto $\Delta \varphi$ from a number of $\Delta\eta$ ranges.  Given the $R_{2}^{\rm (CI)}$ and $P_{2}^{\rm (CI)}$ distributions reported in this work are symmetric by construction, the decompositions are limited
to cosine terms exclusively and are further limited to include terms of orders $n$=1 to $n$=6 
\be
f(\Delta\varphi) = b_o(\Delta\eta) + 2 \times \sum_{n=1}^6{b_n(\Delta\eta) \cos(n \Delta\varphi)},
\label{Eq:fourierDecomp}
\ee
in which  $b_0$ and $b_n$ are $\Delta\eta$ dependent fit coefficients. One finds that the inclusion of $n>6$ terms does not significantly improve the fits of the $\Delta\varphi$ projections and that these higher order coefficients  are not significant. Although the inclusion of $n=5, 6$ terms does improve the fits, these coefficients typically have sizable uncertainties and  are thus not explicitly reported in this work.

In the case of \RTWO and \PTWO measured in  \PbPb\ distributions,  one anticipates  that, at large $|\Delta\eta|$, the Fourier coefficients $b_n$ to be predominantly driven by flow effects determined by the collision system geometry.
It is then useful to compare the Fourier coefficients $v_n$ obtained with
Eq. (\ref{Eq:fourierDecomp}) to flow coefficients obtained
with the  scalar--product method~\cite{Voloshin:2008dg,Borghini:2001vi} briefly described in Sec.~\ref{Sec:scalarProduct}.
One thus defines and reports, in the following, the harmonic coefficients $v_n[R_2]$
and $v_n[P_{2}]$ calculated from the coefficients $b_n$ obtained
from fits  of projections of $R_{2}(\Delta\varphi)$ and $P_{2} (\Delta\varphi)$,
respectively, according to 
\be
v_n [O] = \rm sign(b_n) \times \sqrt{\frac{|b_n|}{1+b_0} }
\label{Eq:vnDefinition1}
\ee
where $O$ represents either of \RTWO or $P_{2}$. The $\rm sign(b_n)$ and the absolute value are used to account for the fact that the Fourier decomposition fits yield negative  coefficients in some cases, particularly in \pPb\ collisions  and for high orders $n>4$. Flow-like behavior, with sizable $v_2$ and $v_3$ coefficients, has been observed in  \pPb\  collisions~\cite{Abelev:2014mda}.  However,  as discussed in Sec.~\ref{Sec:fd}, Fourier decompositions  carried out in this work  produce negative values for  coefficients  $b_1$, $b_3$, and $b_4$ at large $|\Delta\eta|$ pair separations.  Results of decompositions of $R_{2}$ or $P_{2}$ measured in  \pPb\  collisions are thus reported in terms of the coefficients $b_n$ exclusively.

\subsection{Measurements of $v_n$ coefficients with the scalar-product method}
\label{Sec:scalarProduct}

The scalar-product (SP) method~\cite{Voloshin:2008dg,Borghini:2001vi,ABELEV:2013wsa,Abelev:2014pua,Yin:2012sk,Adam:2016nfo}, a two-particle correlation method, is used to extract the $v_n$ coefficients according to
\begin{equation}
v_{n}\{{\rm SP}\} = \frac{\langle {\bf u}_{n, k} \frac{{\bf Q}_{n}^{*}}{M} \rangle} {\sqrt{\langle \frac{{\bf Q}_{n}^{a}}{M^{a}}  \frac{{{\bf Q}_{n}^b}^{*}} {M^{b}} \rangle}},
\label{Eq:mth_sp}
\end{equation}
where ${\bf u}_{n, k}=\exp(in\varphi_k)$ is the unit vector of the particle of interest (POI) $k$, ${\bf Q}_{n}$ is the
event flow vector, $M$ is the event multiplicity and $n$ is the harmonic number. The full event is divided into
two independent sub-events $a$ and $b$ composed of tracks from different pseudorapidity intervals with flow
vectors ${\bf Q}_{n}^{a}$ and ${\bf Q}_{n}^{b}$ and multiplicities $M^{a}$ and $M^{b}$. The angle brackets denote
averages over all selected particles and events. The notation $Q^*$ represents the complex conjugate of   $Q$.

The $x$ and $y$ components of the flow vector ${\bf Q}_{n}$ are
\begin{align}
Q_{n,x} = \sum_l \cos(n \varphi_l) \\
Q_{n,y} = \sum_l \sin(n \varphi_l),
\label{Eq:mth_Q_vector}
\end{align}
where the sum is carried over all reference particles (RPs) $l$ in the relevant (sub-)event.

Unidentified charged particles from a certain $\pt$ interval are taken as POIs and correlated with RPs from the full
$\pt$ range. The sub-events $a$ and $b$ are defined within the pseudorapidity range $-1.0<\eta<-0.1$ and $0.1<\eta<1.0$, which 
results in a pseudorapidity gap of $|\Delta\eta|>0.2$ that reduces non-flow contributions. To further suppress non-flow effects, a 
pseudorapidity gap of $|\Delta\eta|>0.9$ is also employed by selecting $a$ and $b$ within $-1.0<\eta<-0.45$ and $0.45<\eta<1.0$. 
The POIs are taken from $a$ and the RPs from $b$ and vice-versa. Non-uniformities in the detector azimuthal acceptance influence 
the $v_n$ coefficients at a level of less than 0.1\%.

\section{SYSTEMATIC UNCERTAINTIES}
\label{Sec:systematics}

Sources of systematic effects were investigated to assess their impact on the two-dimensional correlation functions, their projections onto the $\Delta\eta$ and $\Delta\varphi$ axes, the width of the near-side peak of the CI and CD correlation functions, and the coefficients extracted from the $\Delta\eta$ dependent Fourier decompositions  of $\Delta\varphi$ projections of the CD correlations, as well as  on the flow coefficients extracted with  the scalar-product method.
Systematic effects are considered significant if the maximum span of variations obtained by varying a given parameter (or condition) exceeded the statistical uncertainties of the observable considered or if variations were observed for the same data sample.
Contributions of sources yielding   significant deviations were  found to be uncorrelated and thus added in quadrature to obtain the total systematic uncertainties reported in  Tabs.~1--3  and all plots presented in this paper.

One first considers  systematic effects on the overall amplitude of the correlation functions.
The $R_{2}$ and $P_{2}$ correlators were determined with \PbPb\  data samples collected with positive and negative magnetic   field configurations. Peak correlator amplitude differences  obtained with the two field configurations were typically small for US and LS correlators and had maximum values  of 1.4\%    and  1.9\% for $R_2$ and $P_2$ correlators, respectively. These values    were adopted as   systematic uncertainties associated with distortions of the solenoidal magnetic  field, the TPC electric field, and corrections for space charge effects. Given the amplitude and shape  of the correlators is dependent on the produced particle multiplicity, systematic effects associated with the collision and multiplicity selection were assessed by repeating the \PbPb\ and \pPb\ analyses  with alternative multiplicity estimators. In the case of \PbPb\ collisions, the SPD track multiplicity was used as an alternative
centrality estimator, and it was found that  the amplitude of the $R_2$ and $P_2$ correlation functions changed from the default analysis by at most 1.6\% and 1.9\%, respectively.
In the case of \pPb\ collisions, correlation amplitudes observed when using the V0-A and V0-C detectors for the definition of multiplicity classes were compared and one did not find statistically significant differences~\cite{PhysRevC.91.064905}. No systematic uncertainty is thus assigned to this contribution in  \pPb\   collision measurements.

Minor contributions to the systematic
uncertainties arise from the selection of the $v_z$-vertex fiducial range. Globally, correlation functions obtained with
the nominal range of $|v_z|<10$ cm, used in this analysis,  exhibit amplitude differences  smaller than 1\% relative to those
obtained with a more restrictive vertex position range of $|v_z|<6$ cm. Additionally, it is found that increasing the vertex bin width  (used in the correction weight calculation) by a factor of two yielded correlation amplitude changes by at most 4\% relative to the nominal bin size reported in this work.

Systematic  uncertainties also arise from the charged-particle track definition and track quality selection criteria. These uncertainties were examined  by repeating the
correlation analyses using track selection criteria distinct from the nominal criteria described in Sec.~\ref{Sec:Experiment}.
The varied track quality criteria included the minimal number of TPC space points per track, the maximum $\chi^{2}$ per
degree of freedom obtained in  the momentum fit, as well as  the maximum track distance of closest approach (DCA) to the primary vertex (both along the beam direction and in the transverse plane). Variations of these track quality selection criteria typically have a rather small impact on the amplitude of the correlation functions (up to 0.8\% for $R_2$ and 1.2\% for $P_2$), but nonetheless have measurable effects on the width of the near side peak of the CI and CD correlation functions listed in Tab. 1.

The differences between correlation functions obtained with charged-particle tracks reconstructed with only TPC hits (known as TPC tracks), TPC tracks refitted to include the primary vertex, and so called hybrid tracks, which include a mixture of TPC tracks with vertex refit and tracks that also include one or several hits in the ITS, were considered. Amplitude  differences between correlation functions obtained with TPC tracks only and TPC tracks with a primary vertex refit are typically small, i.e., less than 5\%,  but the  $R_{2}$ and $P_{2}$ CI correlation functions
exhibit differences as large as  8\% and 15\%, respectively, in  the  range $|\Deta| < 0.6$, $|\Delta\varphi|<0.6$,  in the most central collisions.  The impact of these amplitude changes on the width and shape of the correlation functions is summarized in Tabs.~1--2. Correlation functions, most particularly $P_2^{(CD)}$ correlations,  obtained with hybrid tracks featured significant distortions associated with TPC sector boundary. Correlation functions obtained with these tracks were thus not included
in our assessment of  systematic effects associated with the track quality and the track reconstruction algorithm.

Uncertainties associated with the criteria used for rejection of electron contamination were studied by
varying the selection criteria on deviations from the expected Bethe-Bloch parameterization  of the specific ionization energy loss, \dedx, for electrons from 3$\sigma$ to 5$\sigma$. Changes in the correlation function amplitude were smaller than 1.3\%
for both collision systems and all multiplicity classes.

Systematic uncertainties associated with the track-by-track efficiency and contamination corrections were studied using simulated \pPb\  and \PbPb\ collisions produced with the HIJING event generator~\cite{Wang:1991hta,Gyulassy:1994ew} and propagated through a GEANT3~\cite{Brun:1119728} model of the ALICE detector.  Correlation functions obtained  at
the event generator
level were compared  with those obtained
after taking full account of detector effects. Deviations are typically negligible in non-central collisions. Maximum discrepancies  of about 1.6\%  were found in the most central \PbPb\  collisions. No measurable effects were observed in the most peripheral \PbPb\ collisions and \pPb\ collisions.

Systematic uncertainties on the width of the near-side of the CI and CD correlation functions were studied by repeating the analysis  with the  variations discussed earlier in this section. Additionally,  the effect of the incomplete efficiency correction in the $(\Delta\eta,\Delta\varphi)=(0,0)$ bin was studied by arbitrarily doubling the correlation yield in that bin. Such a change produces width reductions smaller than 3\%. All systematic uncertainty contributions to the near-side peak widths are listed in Tab.~1.

Systematic effect studies pertaining specifically to the determination of the  azimuthal dependence  of the correlations, and most particularly the Fourier decomposition coefficients extracted from $R_2$ and $P_2$ LS, US, and CI correlation functions were also carried out. These correlation functions were initially determined with 72 bins in $\Delta\varphi$ but rebinned to 36 bins to suppress some residual  effects on the Fourier decomposition fits, particularly in the case of the $P_{2}$ correlation functions. Studies showed, however, that  the coefficients extracted from $R_{2}$
are less sensitive to rebinning, within statistical uncertainties, while  coefficients obtained
in  fits of $P_{2}$ for $n \geq 2$ did exhibit greater sensitivity to the rebinning. One finds the fit coefficients are
stable, with rebinning, for  0--50\% collision centralities (\PbPb), but measurable variations were observed for more peripheral bins.  For central \PbPb\  collisions,  systematic shifts for $n \geq 1$ coefficients were found to be  smaller than 5\%  while shifts as large as  13\% were obtained in \PbPb\  peripheral collisions. Distortions were
far smaller for $R_2$ and $P_2$ correlation functions measured in \pPb\ collisions. The systematic uncertainties associated with distortions are estimated to be less than one percent for this system.

The $v_n$ coefficients extracted using the scalar-product method were studied under variations of the number of the TPC space points (varied from 70 to 100), the collision centrality determination, the $v_z$ binning, charged-particle track definition, different magnetic field polarities, criteria for electron rejection, and various other aspects of the detector response. Systematic uncertainties inferred from these studies are presented in Tab.~3.  We also studied the impact of the detector response based on GEANT simulations of  HIJING~\cite{Wang:1991hta, Gyulassy:1994ew} and AMPT~\cite{Lin:2004en} events. We compared $v_n$ coefficients evaluated directly from the models with those obtained  from reconstructed tracks (i.e., tracks obtained from a simulation of the detector performance) and assessed maximum systematic uncertainties of 3\%, 4\% and 5\% for $v_2$, $v_3$ and $v_4$, respectively.

Systematic uncertainties associated with the extraction of the average correlation function widths $\langle\Delta\eta\rangle$,
discussed in Sec. \ref{Sec:nearsidepeakwidth}, are summarized in Tab.~\ref{Tab:Sys_width}, whereas typical values of
systematic uncertainties of  the flow harmonic $v_{n}$ coefficients measured in \PbPb\  collisions, reported in Sec. \ref{Sec:fd},
are summarized in Tab.~\ref{Tab:Sys_pbpb_vn}. Similarly,
systematic uncertainties associated with the Fourier decomposition coefficients $b_n$ obtained for  \pPb\   collisions  are summarized
in  Tab.~3. Systematic uncertainty values listed in these
tables correspond to maximum differences encountered for each system and across all
multiplicity classes, and all pseudorapidity ranges considered in this analysis.

\begin{table}[h]
\centering
\begin{tabular}{l c c c }
\hline\hline
Category & Correlation function & \PbPb\  & p--Pb
\\
\hline
&$R_{2}$ & $1.6\%$ & $-$ \\[-1ex]
\raisebox{1.5ex}{Magnetic field} &$P_{2}$
& $1.9\%$ & $-$\\[1ex]
\hline
&$R_{2}$ & $0.3\%$ & $-$\\[-1ex]
\raisebox{1.5ex}{Centrality determination} &$P_{2}$
& $0.7\%$ & $-$\\[1ex]
\hline
&$R_{2}$ & $1.9\%$ & $2.8\%$\\[-1ex]
\raisebox{1.5ex}{$z$-vertex binning} &$P_{2}$
& $2.8\%$ & $3.6\%$\\[1ex]
\hline
&$R_{2}$ & $2.4\%$ & $2.9\%$\\[-1ex]
\raisebox{1.5ex}{Track selection} &$P_{2}$
& $3.4\%$ & $3.9\%$\\[1ex]
\hline
&$R_{2}$ & $0.4\%$ & $0.6\%$\\[-1ex]
\raisebox{1.5ex}{Electron rejection} &$P_{2}$
& $0.9\%$ & $0.8\%$\\[1ex]
\hline
&$R_{2}$ & $0.14\%$ & $-$\\[-1ex]
\raisebox{1.5ex}{Tracking efficiency} &$P_{2}$
& $0.26\%$ & $-$\\[1ex]
\hline
&$R_{2}$ & $3\%$ & $3\%$\\[-1ex]
\raisebox{1.5ex}{$\Delta\eta =0$, $\Delta\varphi =0$ bin} &$P_{2}$
& $3\%$ & $3\%$\\[1ex]
\hline
&$R_{2}$ & $4.6\%$ & $5.8\%$\\[-1ex]
\raisebox{1.5ex}{Total} &$P_{2}$
& $5.1\%$ & $6.1\%$\\[1ex]

\hline
\hline
\end{tabular}
\caption{Maximum systematic uncertainties of the correlation widths, $\langle\Delta\eta\rangle$. Values marked with a dash are too small to be measurable. Total uncertainties are obtained as sums in quadrature of individual contributions.}
\label{Tab:Sys_width}
\end{table}

\begin{table}[th]
\centering
\begin{tabular}{l c r r r}
\hline\hline
Category & Method &$v_{2}$ &$v_{3}$ &$v_{4}$\\
\hline
&$R_{2}$ & $1.1\%$ & $0.6\%$ & $1.4\%$\\[-1ex]
\raisebox{1.5ex}{Magnetic field} &$P_{2}$ & $1.4\%$ & $0.9\%$ & $1.6\%$\\[1ex]
& $SP$ & - & - & -\\[1.5ex]
\hline
&$R_{2}$ & $0.7\%$ & $0.7\%$ & $1.1\%$\\[-1ex]
\raisebox{1.5ex}{Centrality determination} &$P_{2}$ & $0.5\%$ & $0.8\%$ & $1.6\%$\\[1ex]
& $SP$ & 1.0\% & 1.0\% & 1.0\% \\[1.5ex]
\hline
&$R_{2}$ & $1.6\%$ & $2.0\%$ & $3.2\%$\\[-1ex]
\raisebox{1.5ex}{Vertex-Z binning} &$P_{2}$ & $1.9\%$ & $2.8\%$ & $3.7\%$\\[1ex]
& $SP$ & - & - & - \\[1.5ex]
\hline
&$R_{2}$ & $3.5\%$ & $3.2\%$ & $5.3\%$\\[-1ex]
\raisebox{1.5ex}{Track selection} &$P_{2}$ & $4.9\%$ & $4.9\%$ & $6.2\%$\\[1ex]
& $SP$ & 2.2\% & 2.2\% & 2.2\% \\[1.5ex]
\hline
&$R_{2}$ & $0.6\%$ & $0.3\%$ & $0.8\%$\\[-1ex]
\raisebox{1.5ex}{Electron rejection} &$P_{2}$ & $1.0\%$ & $0.8\%$ & $1.3\%$\\[1ex]
& $SP$ & - & - & - \\[1.5ex]
\hline
&$R_{2}$ & $0.4\%$ & $0.2\%$ & $0.7\%$\\[-1ex]
\raisebox{1.5ex}{Efficiency effect} &$P_{2}$ & $1.2\%$ & $0.9\%$ & $1.6\%$\\[1ex]
& $SP$ & 3.0\% & 4.0\% & 4.0\% \\[1.5ex]
\hline
&$R_{2}$ & $3.0\%$ & $6.0\%$ & $8.0\%$\\[-1ex]
\raisebox{1.5ex}{$\Delta\varphi$ binning} &$P_{2}$ & $7.0\%$ & $11.0\%$ & $13.0\%$\\[1ex]
& $SP$ & - & - & - \\[1.5ex]
\hline
&$R_{2}$ & - & - & - \\[-1ex]
\raisebox{1.5ex}{No. of TPC clusters} &$P_{2}$ & - & - & -\\[1ex]
& $SP$ & 2.0\% & 2.0\% & 5.0\% \\[1.5ex]
\hline
&$R_{2}$ & - & - & - \\[-1ex]
\raisebox{1.5ex}{Comparison to Monte Carlo} &$P_{2}$ & - & - & - \\[1ex]
& $SP$ & 3.0\% & 4.0\% & 5.0\% \\[1.5ex]
\hline
&$R_{2}$ & 5.1\% & 7.2 \%& 10.3\% \\[-1ex]
\raisebox{1.5ex}{Total} &$P_{2}$ & 9\% & 12.5\% & 15.0\% \\[1ex]
& $SP$ & 5.3\% & 6.5\% & 8.5\% \\[1.5ex]
\hline
\end{tabular}
\caption{Systematic uncertainties on $v_{n}$ from $R_{2}$, $P_{2}$ and SP in \PbPb\  collisions. Values marked with a dash are too small to be measurable or not applicable. Total uncertainties are obtained as sums in quadrature of individual contributions.}
\label{Tab:Sys_pbpb_vn}
\end{table}

\begin{table}[h]
\centering
\begin{tabular}{l c c c c c}
\hline\hline
Category &Correlation function &$b_{1}$ &$b_{2}$ &$b_{3}$ &$b_{4}$
\\
\hline
&$R_{2}$ & $1.4\%$ & $1.2\%$ & $1.9\%$ & $2.7\%$\\[-1ex]
\raisebox{1.5ex}{$z$-vertex binning} &$P_{2}$
& $2.0\%$ & $1.7\%$ & $2.2\%$ & $3.2\%$\\[1ex]
\hline
&$R_{2}$ & $8.3\%$ & $6.4\%$ & $8.1\%$ & $8.9\%$\\[-1ex]
\raisebox{1.5ex}{Track selection} &$P_{2}$
& $10.8\%$ & $9.3\%$ & $10.9\%$ & $11.0\%$\\[1ex]
\hline
&$R_{2}$ & $0.9\%$ & $0.2\%$ & $0.7\%$ & $0.9\%$\\[-1ex]
\raisebox{1.5ex}{Electron rejection} &$P_{2}$
& $0.7\%$ & $0.9\%$ & $0.8\%$ & $1.0\%$\\[1ex]
\hline
&$R_{2}$ & $0.1\%$ & $0.2\%$ & $0.7\%$ & $1.3\%$\\[-1ex]
\raisebox{1.5ex}{$\Delta\varphi$ binning} &$P_{2}$
& $0.2\%$ & $0.6\%$ & $1.1\%$ & $2.0\%$\\[1ex]
\hline
&$R_{2}$ & $8.5\%$ & $6.5\%$ & $8.4\%$ & $9.4\%$\\[-1ex]
\raisebox{1.5ex}{Total} &$P_{2}$
& $11.0\%$ & $9.5\%$ & $11.2\%$ & $11.7\%$\\[1ex]
\hline
\end{tabular}
\label{Tab:Sys_ppb_bn}
\caption{Maximum  systematic uncertainties on $b_{n}$ coefficients obtained  from $R_{2}$ and $P_{2}$ in  \pPb\   collisions. Total errors are obtained as sums in quadrature of individual contributions.}

\end{table}
\section{RESULTS}
\label{Sec:results}

Measurements of the correlation functions $R_2$
and $P_{2}$ for LS  and US particle pairs are presented in Sec.~\ref{Sec:ChargeExclusiveResults}  while 
charge-independent (CI), and charge-dependent (CD) correlation
functions constructed from these are presented in
Sec.~\ref{Sec:ChargeIndependent} and Sec.~\ref{Sec:ChargeDependent}, respectively.
The amplitude, shape  and width of $R_2$ and $P_2$ CI  and
CD correlations are  sensitive to the particle production
dynamics as well as the system evolution. Several phenomena may in fact
contribute in shaping the azimuthal and longitudinal dependence of these correlation
functions, including anisotropic and radial flow,  thermal diffusion~\cite{AbdelAziz:2006jv},
as well as  two-stage quark production~\cite{Pratt:2015jsa}.
A detailed characterization of the longitudinal and azimuthal  profiles of both
CI and CD correlation functions is thus of  interest in order to further
improve the understanding of these competing mechanisms and effects.
Section~\ref{Sec:nearsidepeakwidth} presents analyses of the correlation
function longitudinal and azimuthal widths and their evolution with increasing produced particle multiplicity. Section~\ref{Sec:fd} reports  studies of  Fourier decompositions of azimuthal
projections of  $R_2$ and $P_2$   as a function of the
longitudinal separation of particle pairs. Altogether, these different studies enable the characterization of flow and non-flow components in \PbPb\ and \pPb\ collisions.

\subsection{Like-sign and unlike-sign correlation functions}
\label{Sec:ChargeExclusiveResults}

\begin{figure}[h!]
\includegraphics[width=0.46\linewidth]{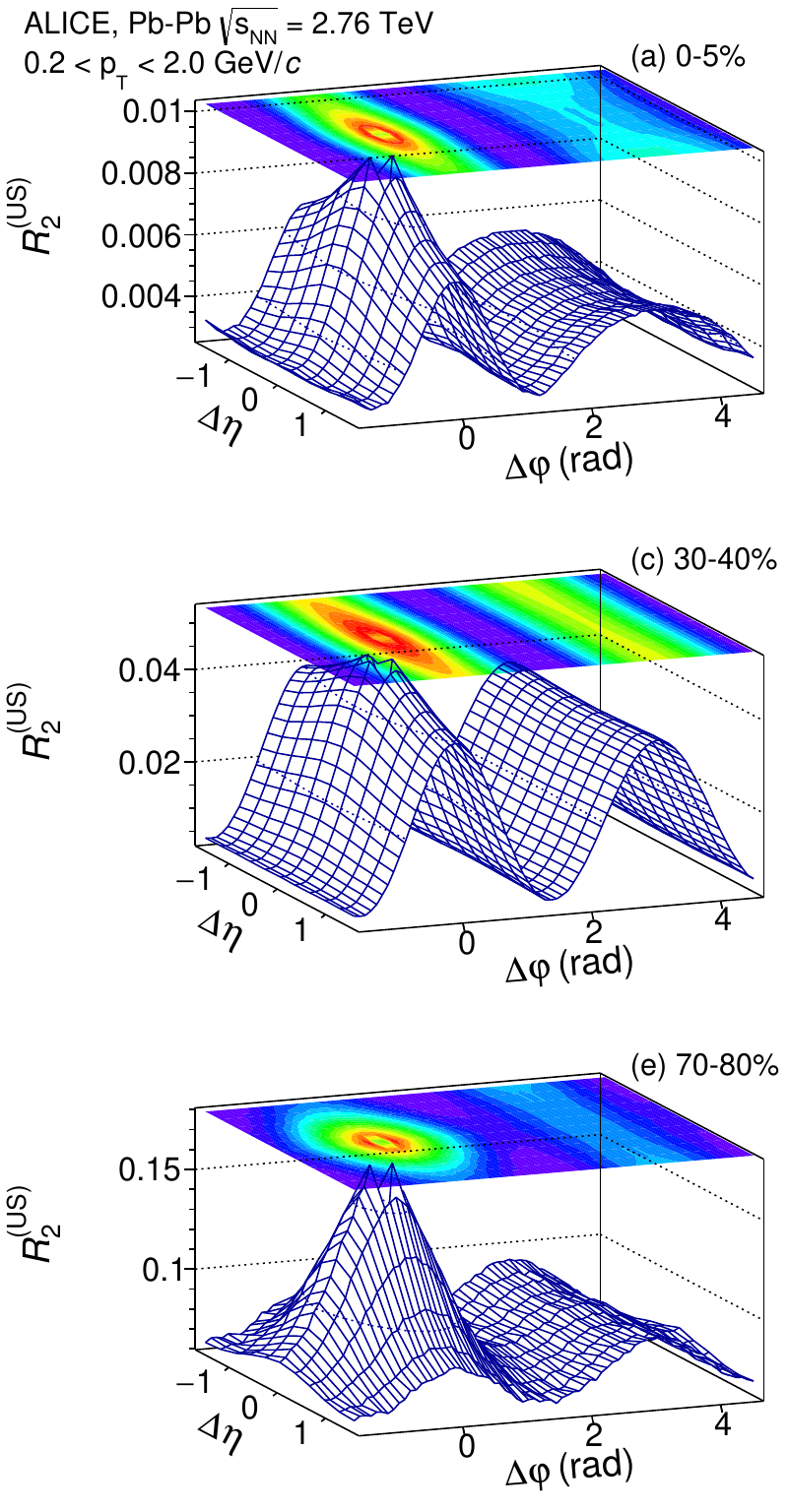}
\includegraphics[width=0.46\linewidth]{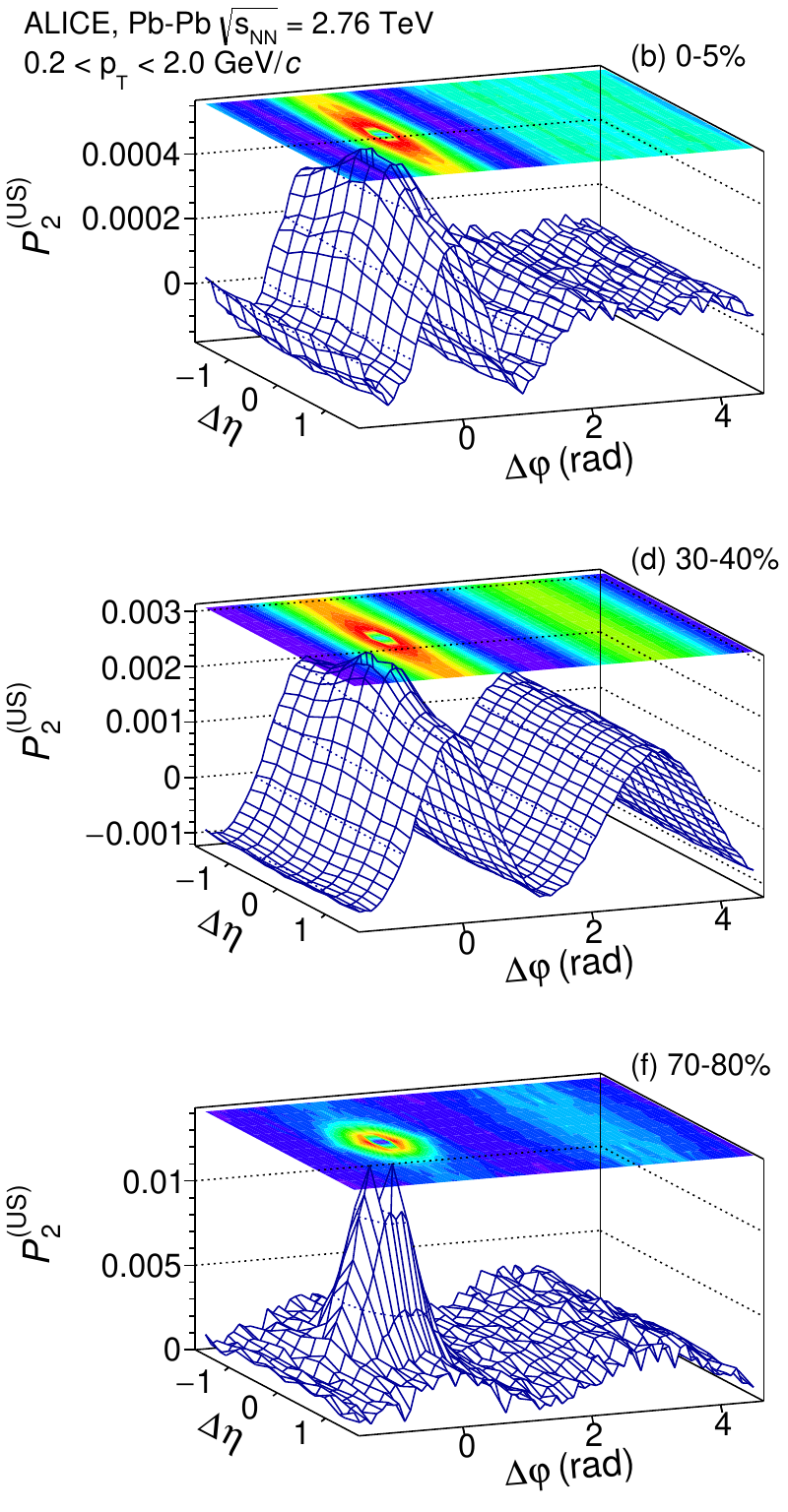}
\caption{Correlation functions $R_{2}^{(\rm US)}$ (left column) and $P_{2}^{(\rm US)}$ (right column) of charged hadrons in the range $0.2 < \pt < 2.0$   \gevc\  measured in  \PbPb\  collisions at $\snn = $ 2.76 TeV for selected centrality classes.}
\label{Fig:Corr_2D_ChPM_PbPb}
\end{figure}
\begin{figure}[h!]
\includegraphics[width=0.46\linewidth]{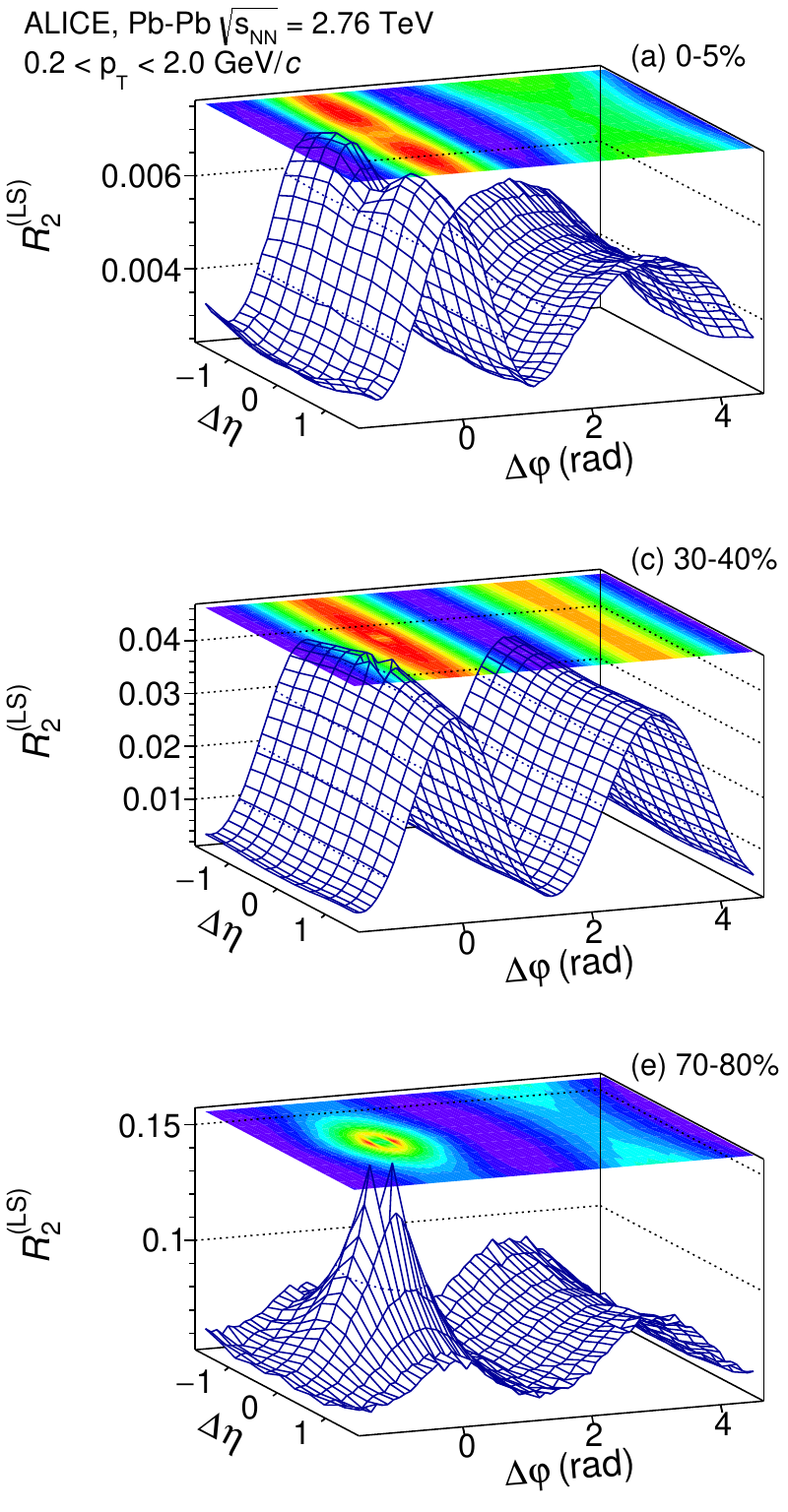}
\includegraphics[width=0.46\linewidth]{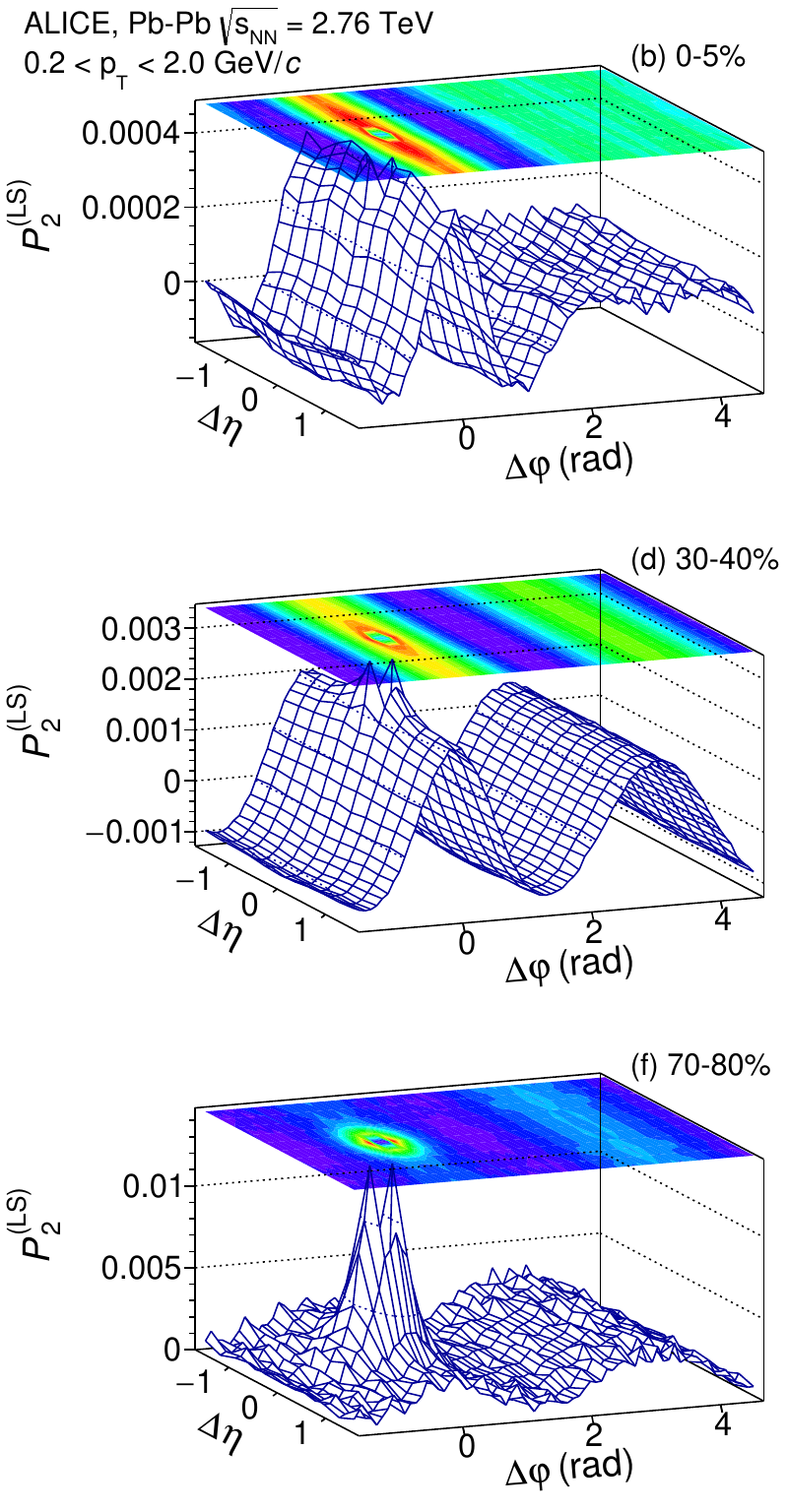}
\caption{Correlation functions $R_{2}^{(\rm LS)}$ (left column) and $P_{2}^{(\rm LS)}$ (right column) of charged hadrons in the range $0.2 < \pt < 2.0$   \gevc\  measured in  \PbPb\  collisions at $\snn = $ 2.76 TeV for selected centrality classes.}
\label{Fig:Corr_2D_ChPP_PbPb}
\end{figure}
\begin{figure}[h!]
\includegraphics[width=0.46\linewidth]{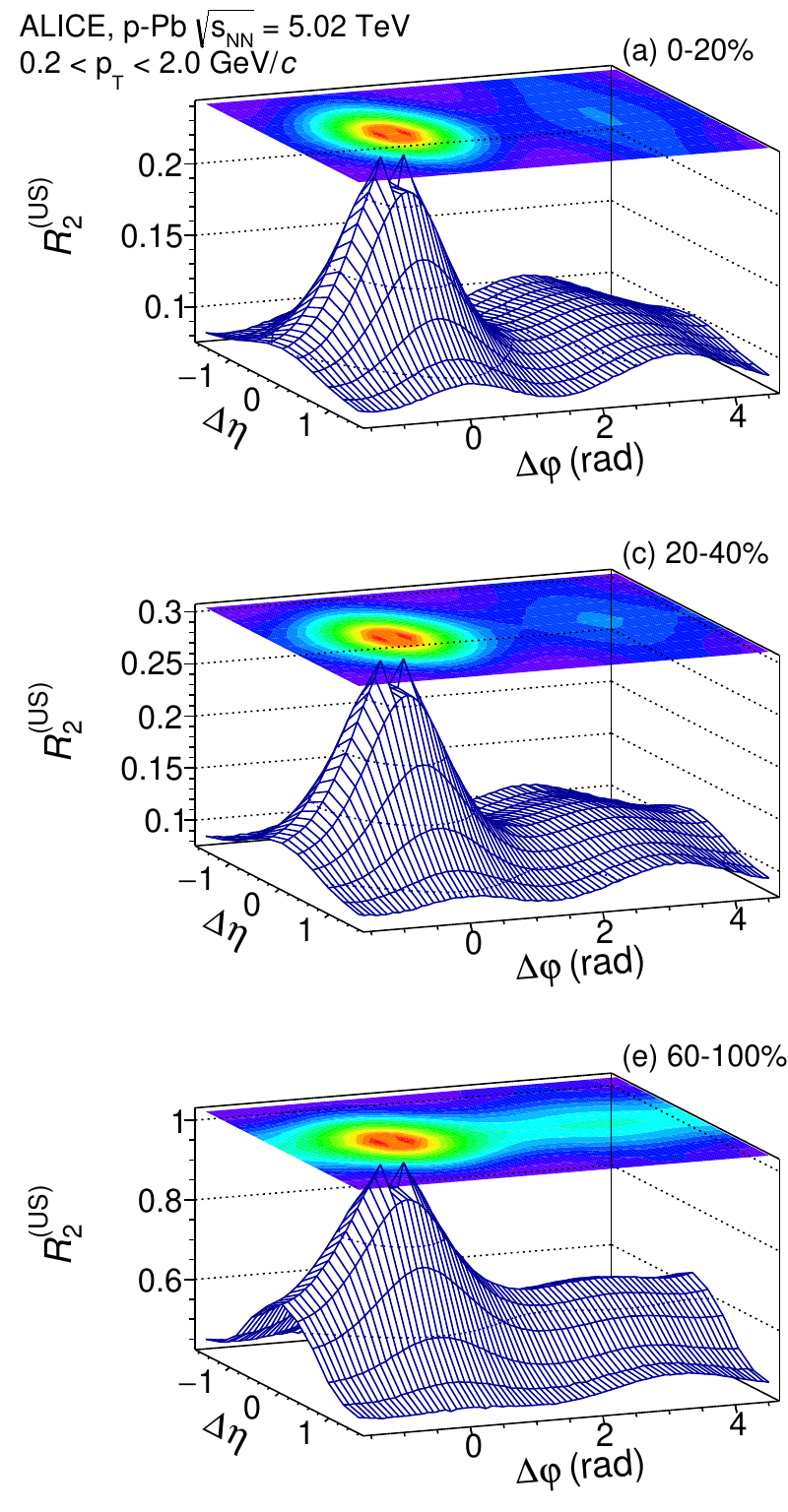}
\includegraphics[width=0.46\linewidth]{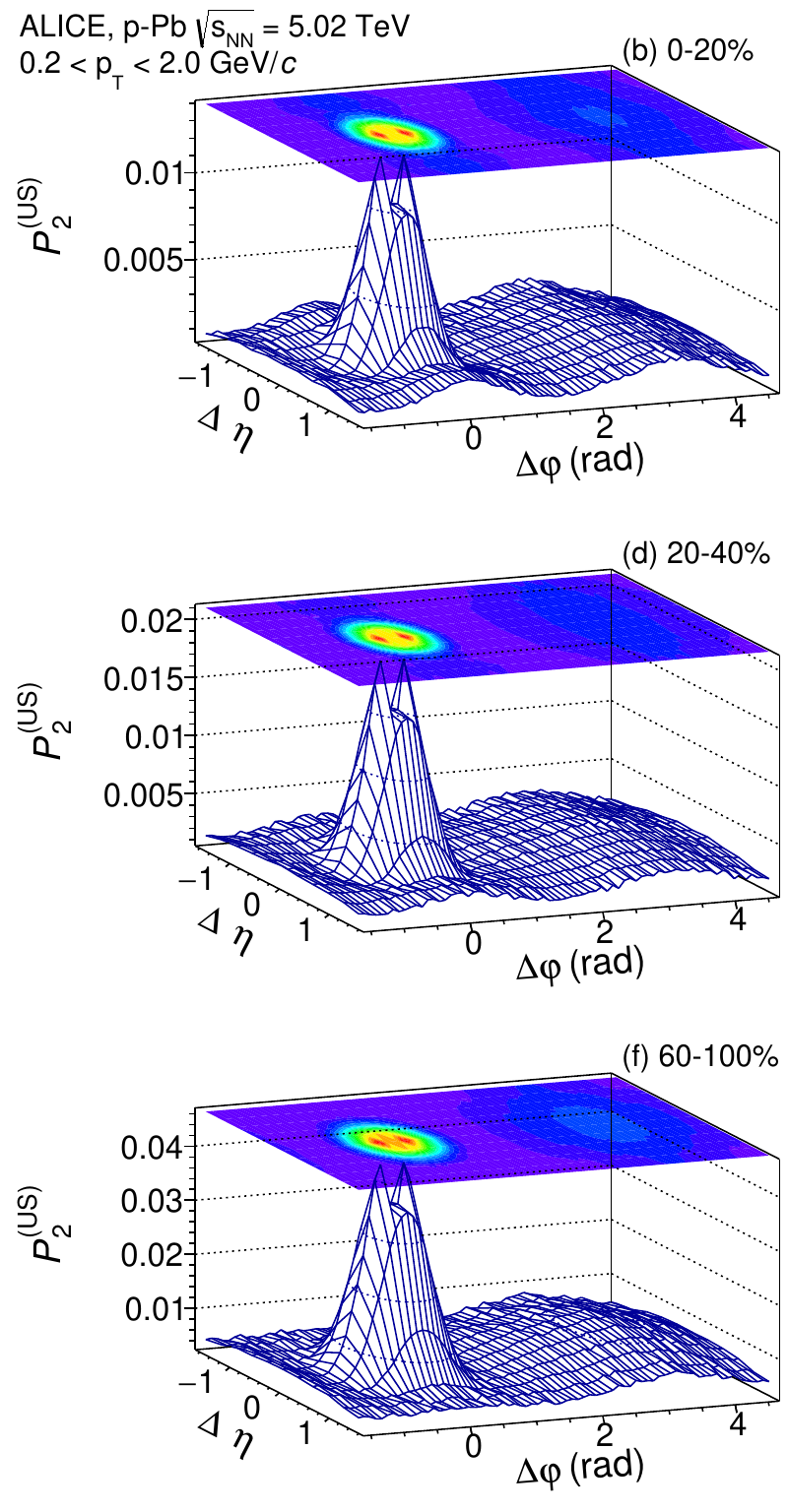}
\caption{Correlation functions $R_{2}^{(\rm US)}$ (left column) and $P_{2}^{(\rm US)}$ (right column) of charged hadrons in the range $0.2 < \pt < 2.0$   \gevc\  measured in  \pPb\  collisions at $\snn = $ 5.02 TeV for selected multiplicity classes.}
\label{Fig:Corr_2D_ChPM_pPb}
\end{figure}
\begin{figure}[ht!]
\includegraphics[width=0.46\linewidth]{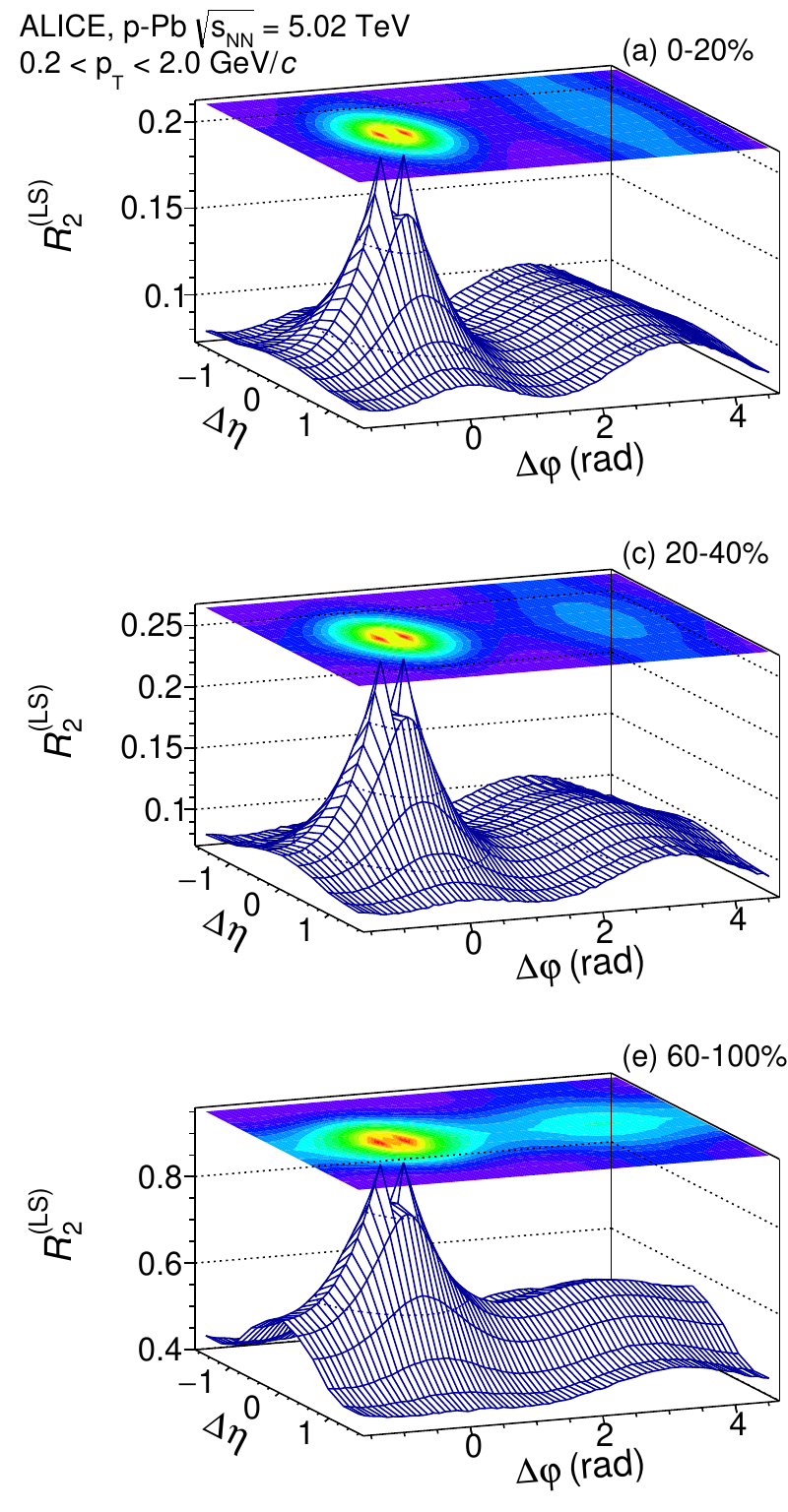}
\includegraphics[width=0.46\linewidth]{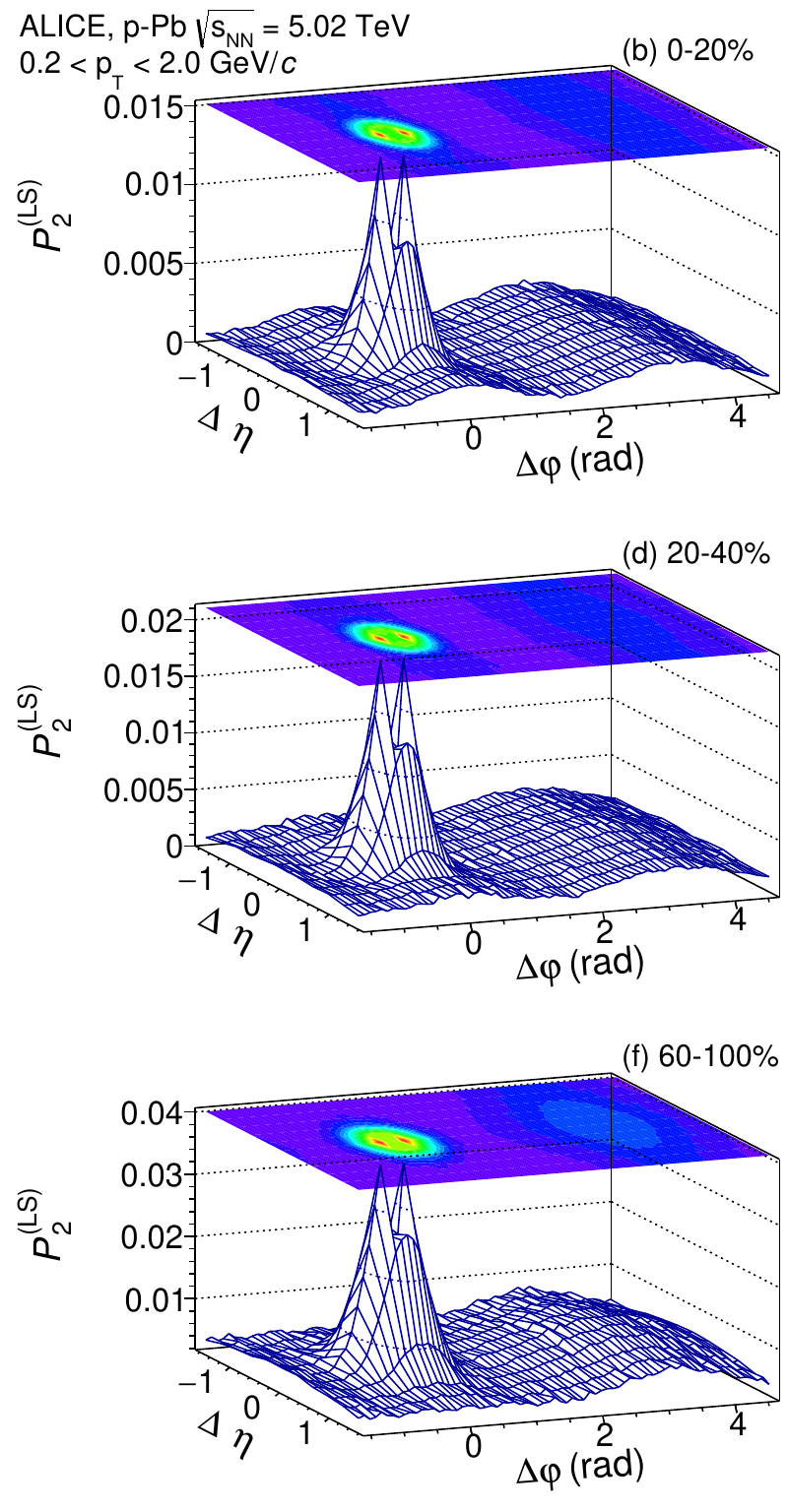}
\caption{Correlation functions $R_{2}^{(\rm LS)}$ (left column) and $P_{2}^{(\rm LS)}$ (right column) of charged hadrons in the range $0.2 < \pt < 2.0$   \gevc\  measured in  \pPb\  collisions at $\snn = $ 5.02 TeV for selected multiplicity classes.}
\label{Fig:Corr_2D_ChPP_pPb}
\end{figure}

The $R_2$ and $P_2$ correlation functions measured in \PbPb\  collisions are displayed in Figs.~\ref{Fig:Corr_2D_ChPM_PbPb}--\ref{Fig:Corr_2D_ChPP_PbPb} for unlike- and like-sign pairs for three representative  multiplicity classes corresponding to 70--80\% (peripheral collisions), 30--40\% (mid-central collisions) and
0--5\% (most central collisions) fractions of the cross section. The corresponding correlation functions measured in  \pPb\   collisions are  shown  in Figs.~\ref{Fig:Corr_2D_ChPM_pPb}--\ref{Fig:Corr_2D_ChPP_pPb}  for  event multiplicity classes  corresponding to fractions of cross sections of 60--100\%, 20--40\% and 0--20\%.  These do not unambiguously map to distinct \pPb\ collision impact parameter or centrality.

One observes that the $R_{2}(\Delta\eta,\Delta\varphi)$ and $P_{2}(\Delta\eta,\Delta\varphi)$ correlation functions
measured in \PbPb\ and \pPb\  exhibit
similar  trends with increasing multiplicity.  Although they have quite different amplitudes, owing to the $\Delta \pt  \Delta \pt $
dependence of $P_2$, one  finds correlation amplitudes to be largest
in peripheral \PbPb\  collisions and low multiplicity classes in \pPb. Furthermore, the amplitudes   of  the $R_2$ and $P_2$   correlation functions qualitatively exhibit similar   decreasing trends with increasing particle multiplicity, reaching the smallest values in the 5\% and 20\% highest multiplicity classes in  \PbPb\  and  \pPb\   collisions, respectively. A similar dependence on  produced particle multiplicity has been observed for both triggered and untriggered number correlation functions~\cite{Voloshin:2004th,PhysRevC.75.034901,Alver:2009id,Adare:2008ae,Abelev:2013csa,Abelev2013267,PhysRevC.77.011901,Aamodt:2011by},    but
is reported  for the first time, in this work,  for the $P_2$ observable. It results in a large part from
the  increasing number of  elementary interactions
(e.g., parton--parton interactions) associated  with the growing geometrical overlap of the colliding nuclei.

In addition, the $R_2$ and $P_2$ correlation functions exhibit a strong near-side peak in 70--80\% \PbPb\ collisions. This
peak  is noticeably narrower, along both the $\Delta\eta$ and $\Delta\varphi$ axes,
in the $P_2$ correlations,  a feature we
study quantitatively  in Sec.~\ref{Sec:nearsidepeakwidth}. Both $R_2$ and $P_2$
correlations are strongly modified in higher multiplicity collisions  with the emergence of strong $\Delta \varphi$ modulations, known to arise from anisotropic flow in \PbPb\ collisions. Although the near-side peak
remains  an important feature of US correlations, in all multiplicity classes, it
appears significantly overshadowed by flow-like modulations  in the 5\% highest multiplicity LS correlations.
One additionally  finds that the $R_2$ correlations are positive,
although, as cumulants, they are not required to be, while the $P_2$
correlations feature $\Delta\varphi$ ranges where the correlation strength is
negative. Such negative values reflect   $\Delta\varphi$ intervals in which, on average, the  $\pt $ of one particle 
might be found above  $\la \pt \ra$, while the other
is below $\la \pt \ra$, effectively yielding a negative $\Delta \pt  \Delta \pt $ value.
One also observes that the $P_2$ and $R_2$ away-side
(i.e., for $\Delta\varphi \sim \pi$)
dependence on the relative pseudorapidity, $\Delta \eta$, are qualitatively
different. While $R_2$ features a bowed shape, i.e., a concave  dependence
on $\Delta\eta$  with a minimum at $\Delta \eta = 0$,   the
away-side strength of the $P_2$ correlation is essentially flat, i.e., independent
of $\Delta\eta$ within uncertainties.  Similar concave dependences were also reported by the CMS collaboration in
high-multiplicity pp collisions~\cite{Khachatryan:2010gv}  and by the STAR collaboration in 5\% central \AuAu\ collisions~\cite{Agakishiev:2011pe}.

Another interesting difference between $R_2$ and $P_2$, visible in US (Fig.~\ref{Fig:Corr_2D_ChPM_PbPb}) and  LS (Fig.~\ref{Fig:Corr_2D_ChPP_PbPb})  correlations,  involves their away-side dependence
on $\Delta \varphi$  in  the 5\% highest multiplicity
collisions. One finds that the away-side of $P_2$ exhibits a broad structure extending over the full range
of the measured $\Delta \eta$ acceptance and features a weak double hump
structure with a minimum at $\Delta\varphi =\pi$ and side peaks located
approximately at $\Delta \varphi = \pi \pm \pi/3$, while the $R_2$ correlation
function, in the same multiplicity class,  exhibits a convex dependence on $\Delta \varphi$. It is worth
noting, however, that double hump structures similar to that observed in $P_2$
have already been reported with triggered and untriggered number correlations,
albeit only for A--A  collision centralities in the range 0--2\%~\cite{Aamodt:2011by,CMS:2013bza} or after subtraction
of a $v_2$ flow background  in less central collisions~\cite{Adams:2005ph,Adler:2005ee,Adare:2008ae}. These  features were
initially associated with conical particle emission~\cite{Wiedemann:2000za,CasalderreySolana:2004qm,Vitev:2005yg,Chaudhuri:2005vc,Stoecker:2004qu,Ruppert:2005uz,Ruppert:2005dw,Renk200719,Majumder:2005sw,Qin:2007ej,Adler:2005ee}  but are now understood to be caused by  strong triangular
flow ($v_3$) originating from initial state fluctuations in A--A
collisions~\cite{AlversRoland}. The  $P_2$ correlation function
features  a double hump structure already in the 5\% \PbPb\ collisions, by contrast to the
more central collisions required to identify a similar structure in $R_2$. This
suggests that $P_2$ correlations are more sensitive to the presence of the
triangular flow component~\cite{Adam:2017ucq}. We thus carry out a
comparative analysis of the Fourier decompositions of the $R_2$ and $P_2$
correlation functions both as a function of collision centrality and
pseudorapidity difference  in Sec.~\ref{Sec:fd}.


We contrast the near-side peaks of LS and US correlation functions and their evolution
with produced particle multiplicity.  The  $R_2^{\rm (US)}$ and
$P_2^{\rm (US)}$  correlation functions feature    stronger near-side peaks than the $R_2^{\rm (LS)}$ and
$P_2^{\rm (LS)}$ correlation functions  in equivalent multiplicity classes. Additionally, the amplitudes of the near-side peaks of the US correlation functions  decrease in   higher multiplicity classes but remain  an essential feature of both $R_2$ and $P_2$.  By contrast, the $R_2^{\rm (LS)}$ and
$P_2^{\rm (LS)}$ near-side peaks not only weaken
in amplitude, but essentially disappear  in higher multiplicity classes in \PbPb,
leaving behind  near-side structures with a complicated dependence
on $\Delta \eta$. The LS correlation functions  measured at the highest
multiplicities (Fig.~2) hint that the $R_2$ and $P_2$ are sensitive
to different aspects of the correlation dynamics, which one discusses in
greater details in Sec.~\ref{Sec:discussion}.

We next focus  on the US and LS correlation functions measured
in  \pPb\   collisions, displayed in Figs.~\ref{Fig:Corr_2D_ChPM_pPb}--\ref{Fig:Corr_2D_ChPP_pPb}.
We find that both the $R_2$ and $P_2$ correlation functions
feature   prominent near-side peaks similar to those observed in most  peripheral \PbPb\ collisions. Unlike \PbPb\ collisions, however,  the near-side peaks of both $R_2$ and $P_2$ dominate the correlation functions
irrespective of their multiplicity class, although the peak amplitude
decreases, as expected, with increasing  particle multiplicity.
Flow-like $\Delta\varphi$ modulations are observed in the 20--40\% and 0--20\% multiplicity
classes that are qualitatively similar to those reported by the CMS
collaboration~\cite{Khachatryan:2015waa} in very-high multiplicity
triggered events  and those observed by the ALICE collaboration
for charged particles in the range $0.2<\pt < 3.0$  \gevc\    in
the same multiplicity classes~\cite{Abelev:2014mda}. The amplitude of the
modulations is further examined in Sec.~\ref{Sec:fd} of this article.

Furthermore, one   notes that the near-side peak
of US and LS $P_2$ correlation functions measured in \pPb\ collisions  is considerably narrower than those observed in $R_2$.  Additionally,  the shape of
the near-side peaks observed in US and LS correlation functions are remarkably different. The US peaks are wider and rounder at the top, while the LS peaks
are very narrow at the top but appear to fan out with relatively longer tails
along both the $\Delta\eta$ and $\Delta\varphi$ axes. Such differences may
arise in part due to Coulomb  and  HBT  effects.
The evolution of the width  of the near side peak of
the $R_2$ and $P_2$ distributions as a function of the multiplicity class are discussed in Sec. \ref{Sec:nearsidepeakwidth}.

In addition, the $R_2$ correlation functions observed  in \pPb\ feature an away-side shape and dependence on $\Delta\eta$ significantly different than those observed in \PbPb. The away-side  of $R_2$ observed in the lowest \pPb\  multiplicity class   is dominated by a structure essentially independent of $\Delta\varphi$ and with a strong concave dependence on $\Delta\eta$.  This structure progressively evolves, with increasing multiplicity, into an elongated, but still concave, $\Delta\eta$ distribution in the  0--20\%  multiplicity class. In contrast,  the away-side of $P_2$ correlations features a much smaller amplitude (relative to the near-side peak) and exhibits a weaker dependence on $\Delta\eta$  than  observed in $R_2$.

Finally, at large multiplicity, one also notes the emergence of flow-like modulations in both the  $R_2$ and $P_2$ correlation functions. A quantitative study of the strength of these modulations is presented in Sec.~\ref{Sec:fd}.


\subsection{Charge-independent correlations}
\label{Sec:ChargeIndependent}

\begin{figure}[h!]
\includegraphics[width=0.46\linewidth]{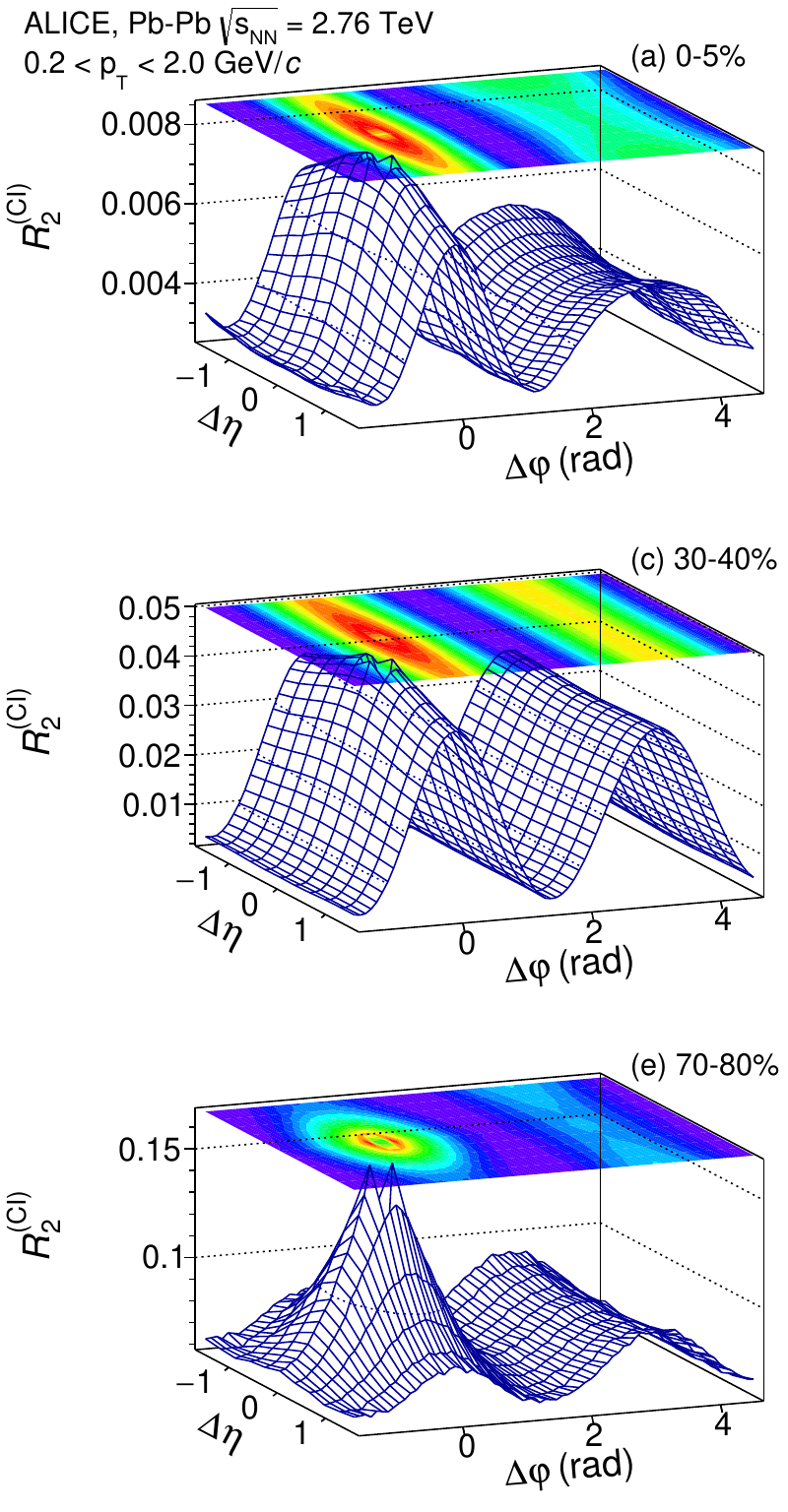}
\includegraphics[width=0.46\linewidth]{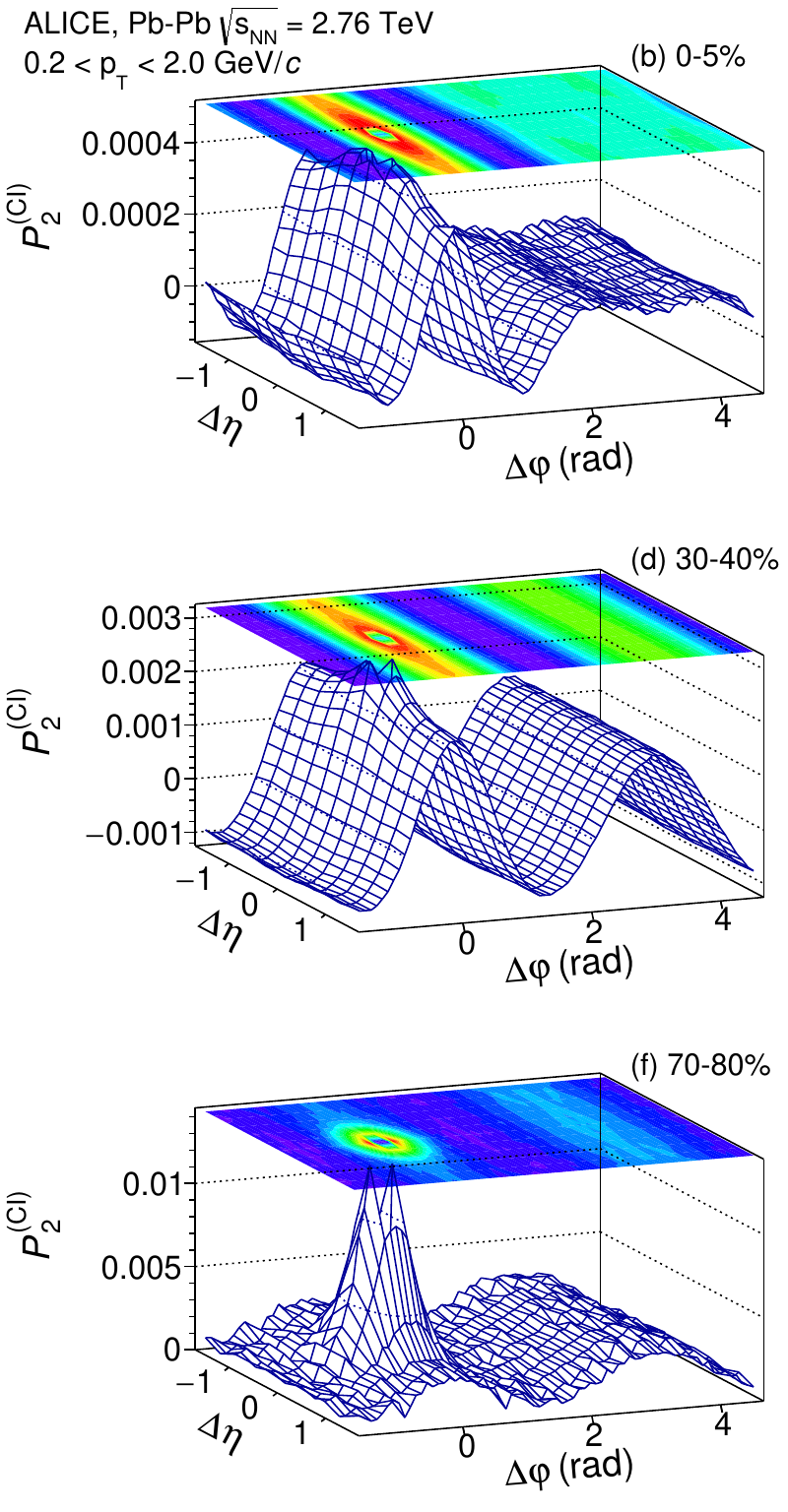}
\caption{Correlation functions  $R_{2}^{\rm (CI)}$  and $P_{2}^{\rm (CI)}$
measured with charged particles in the range $0.2 < \pt < 2.0$
  \gevc\   for selected centrality classes in \PbPb\ collisions.}
\label{Fig:Corr_2D_ChCI_PbPb}
\end{figure}

\begin{figure}[h!]
\includegraphics[width=0.46\linewidth]{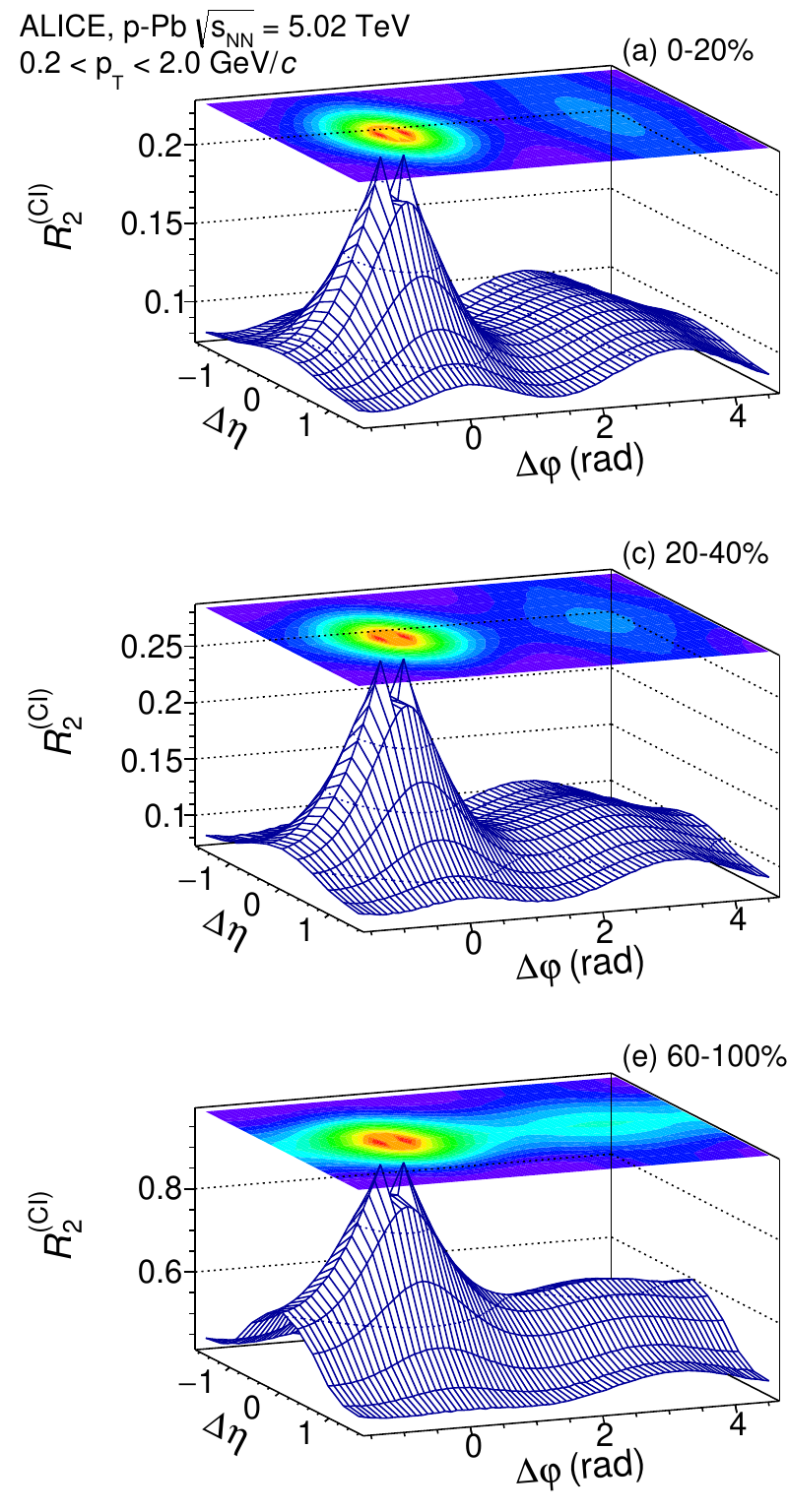}
\includegraphics[width=0.46\linewidth]{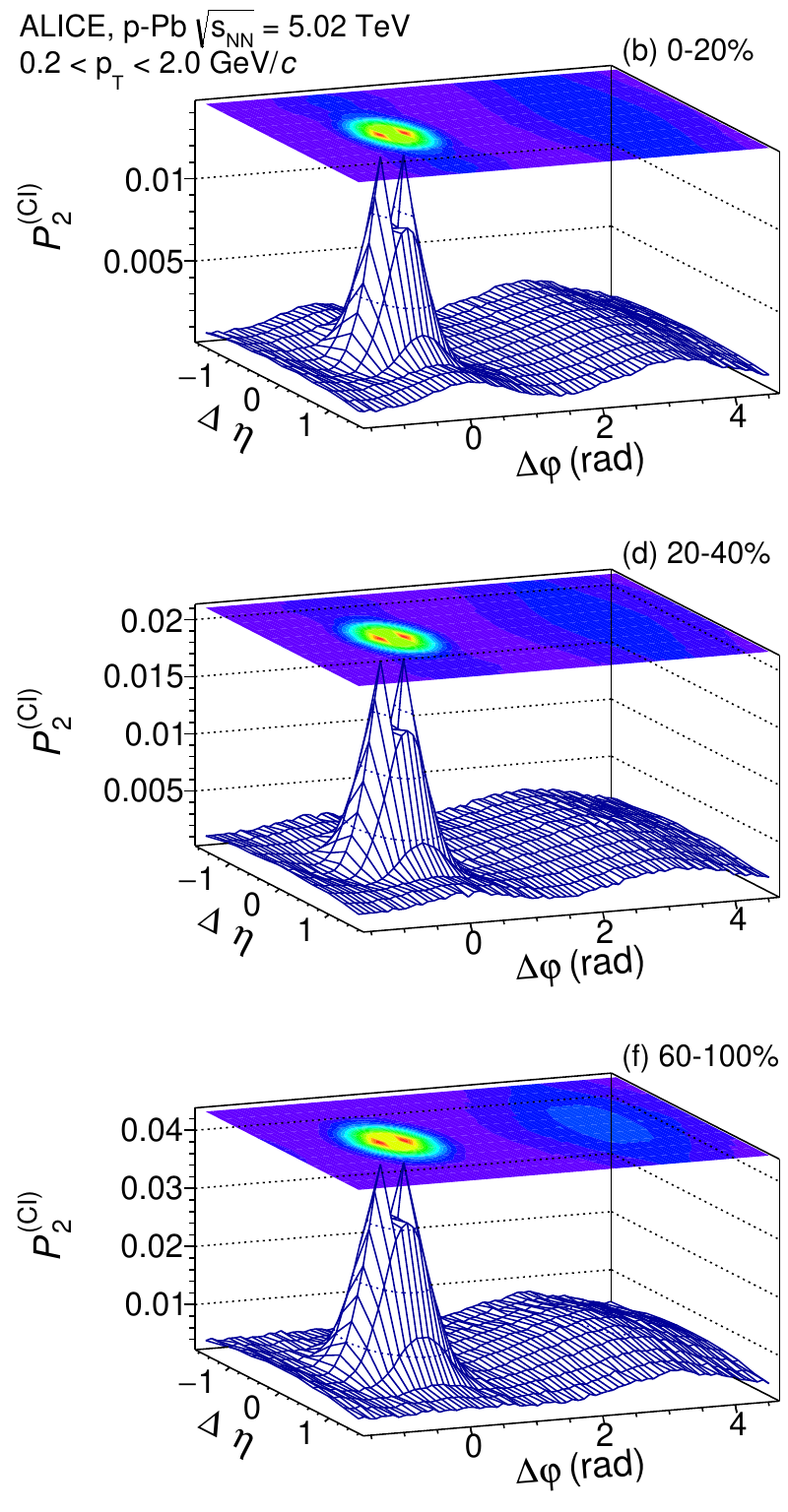}
\caption{Correlation functions  $R_{2}^{\rm (CI)}$  and $P_{2}^{\rm (CI)}$
measured with charged particles in the range $0.2 < \pt < 2.0$
  \gevc\   for selected multiplicity classes in  \pPb\   collisions.}
\label{Fig:Corr_2D_ChCI_pPb}
\end{figure}

Figures~\ref{Fig:Corr_2D_ChCI_PbPb} and \ref{Fig:Corr_2D_ChCI_pPb}  present 
$R_{2}^{\rm (CI)}$    and $P_{2}^{\rm (CI)}$   correlation functions,  determined according to Eq. (\ref{Eq:CI}),  for selected multiplicity classes in  \PbPb\ collisions at $\snn = $ 2.76 TeV and
\pPb\ collisions at $\snn = $ 5.02 TeV, respectively.
The CI correlation functions constitute signatures of the particle
production dynamics and the evolution of the collision system formed
in \PbPb\ and   \pPb\  interactions. As averages of the US and
LS distributions,  these carry essentially the
same features  as these correlation functions. They show the same decreasing
amplitude trend as a function of collision centrality in \PbPb\ collisions and multiplicity classes in \pPb\ collisions, as well as  the emergence of strong  $\Delta \varphi$ modulation in mid-central \PbPb\ collisions. In absence of medium induced effects, the shape of these correlation functions should be independent of the collision centrality and their magnitude should scale with the inverse of the number of binary nucleon-nucleon collisions. From Figs.~\ref{Fig:Corr_2D_ChCI_PbPb}--\ref{Fig:Corr_2D_ChCI_pPb}, one  observes that
the two correlation functions exhibit decreasing amplitude 
with increasing multiplicity in both \PbPb\ and \pPb\ 
collisions. However, both $R_2$ and $P_2$ correlation functions show non-scaling
behaviors: their shapes, i.e., dependences on $\Delta\eta$ and $\Delta\varphi$, significantly
evolve with increasing
multiplicity in both \pPb\ and \PbPb\ collisions. This lack of scaling indicates a different reaction dynamics and collision system evolution with produced
particle multiplicity. The appearance of strong $\Delta\varphi$ modulations,
associated with collective flow, has been observed in several measurements
of  two-particle correlation functions~\cite{PhysRevC.75.034901,Voloshin:2004th,Alver:2009id,Adare:2008ae}.
We find that both the near-side and flow-like feature of $P_2$ and $R_2$
exhibit a somewhat different evolution with  produced particle multiplicity. The near-side peak
of $P_2$ correlations is significantly narrower in $\Delta\eta$
and $\Delta\varphi$ than that observed with $R_2$. One also notes that the
away-side of $P_2$ has a significantly different evolution with collision
centrality than $R_2$, featuring a dip and double hump structure in 5\% most central
\PbPb\  collisions not seen in $R_2$ correlation of the same centrality class.
Clearly, the $P_2$ observable is more sensitive to the presence of higher
harmonics than $R_2$. The flow components of the two observables, however, are not
independent and have been reported to be closely related~\cite{Adam:2017ucq}.
The harmonics coefficients
$v_n$ obtained with the $P_2$ observable, for relative pseudorapidities
$\Delta\eta>0.9$  are successfully predicted by a simple formula, known as
flow ansatz~\cite{monica,Adam:2017ucq}. This ansatz is based on the notion that two-particle correlations observed
in \PbPb\  collisions are predominantly determined by particle emission
relative to a collision's symmetry plane.  The dependences of the harmonic
flow coefficients $v_n$ on charge combination, pseudorapidity
difference $\Delta\eta$, and produced particle multiplicity are presented
in more detail in Sec. \ref{Sec:fd}.

\subsection{Charge-dependent correlations}
\label{Sec:ChargeDependent}

Energy-momentum and quantum number (e.g., charge, strangeness, baryon number) conservation laws
govern
the production of particles  and thus have a strong impact
on correlation functions. Given the very high energy scale
reached in the \PbPb\ and  \pPb\   interactions reported
in this work, it is reasonable to assume that considerations
of energy-momentum conservation may play an equally important role in
the production of LS and US charge pairs. One should then be able to
remove, or at  least suppress, the effect of energy-momentum conservation on particle correlations by considering charge-dependent (CD)
correlation functions. The shape and strength of CD  correlation
functions should thus be predominantly driven by  processes of
creation of charge pairs, their transport,   and the fact that electric charge is a conserved quantity.

\begin{figure}[h!]
\includegraphics[width=0.46\linewidth]{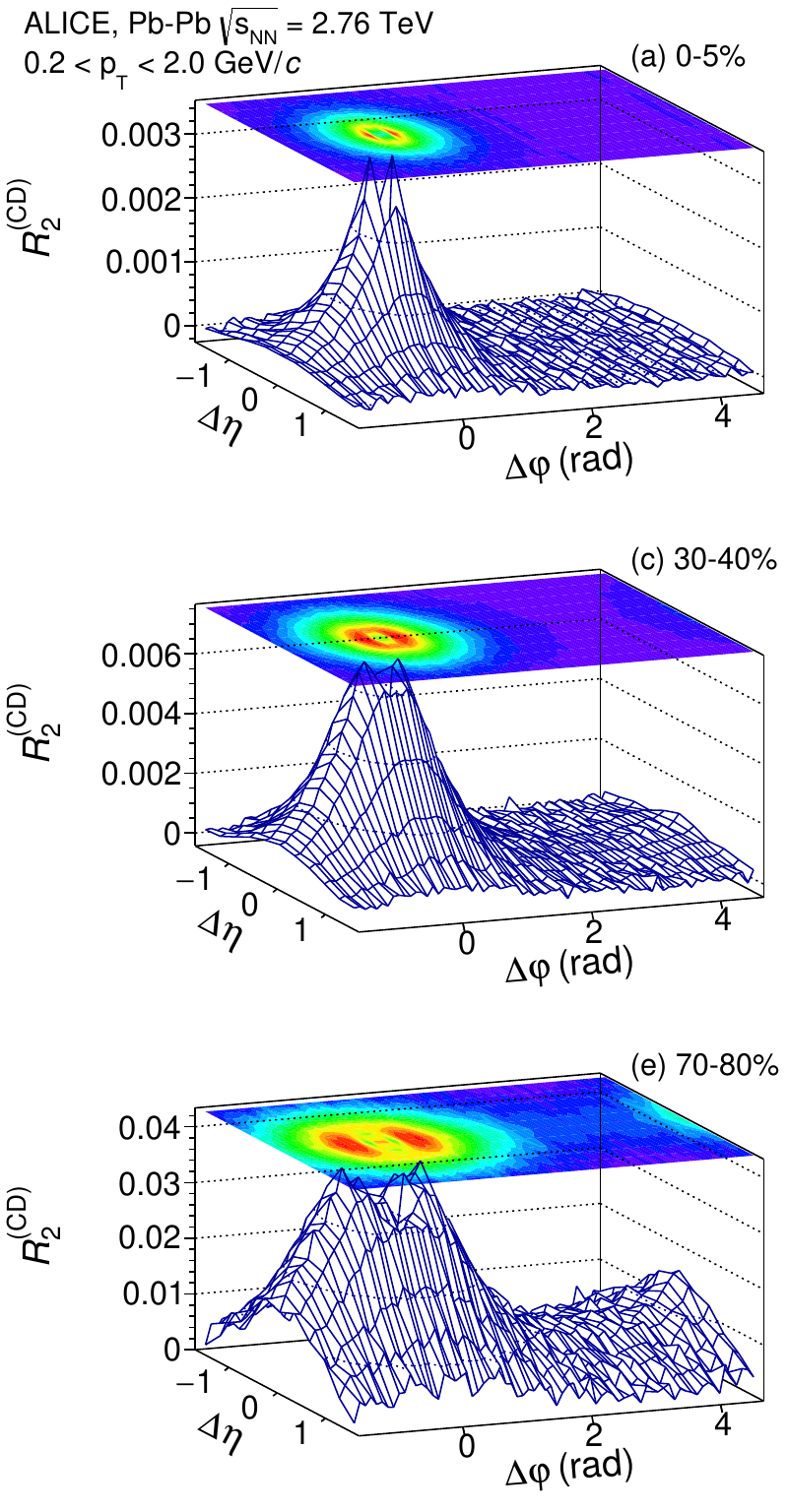}
\includegraphics[width=0.46\linewidth]{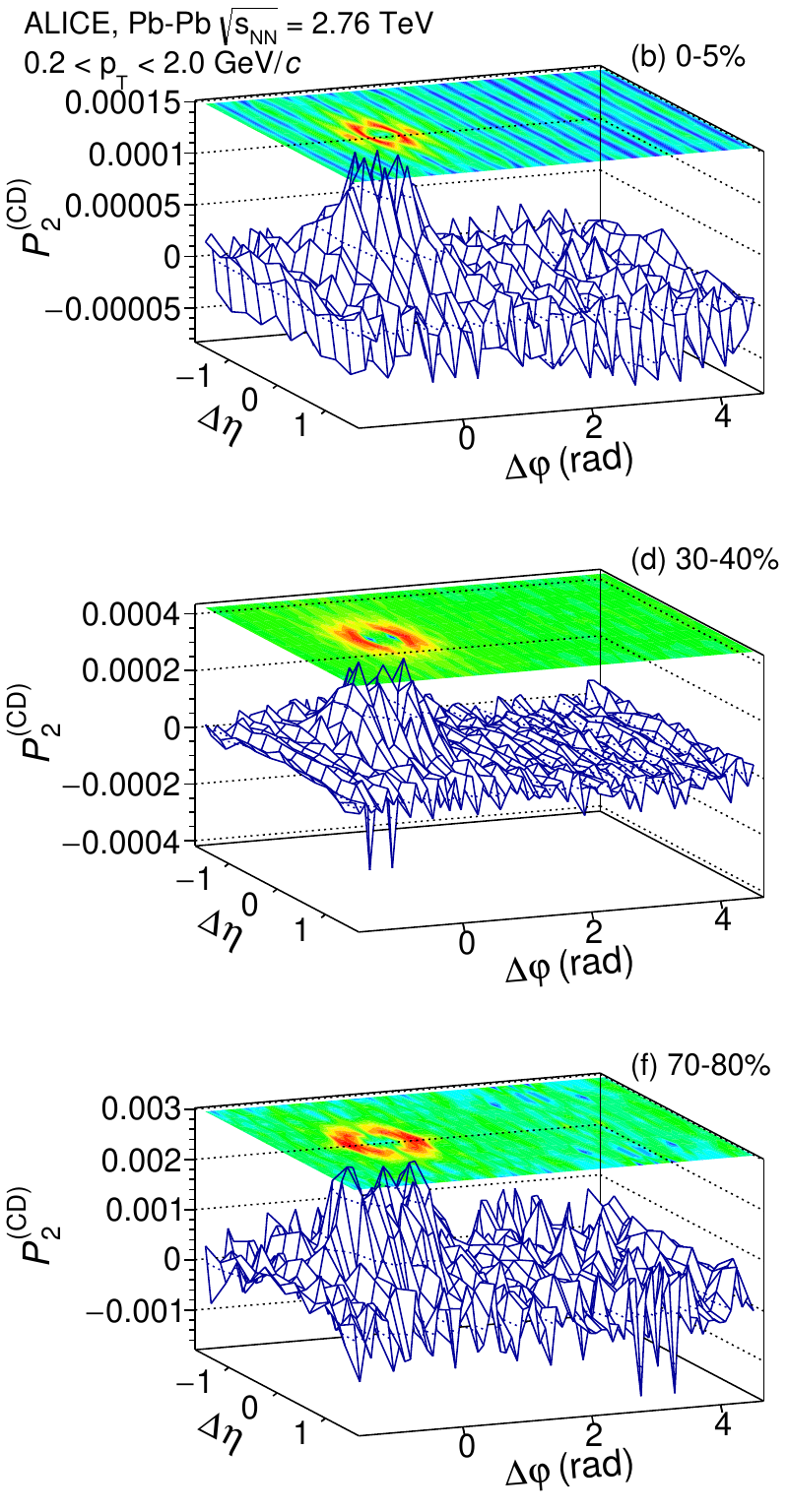}
\caption{Correlation functions  $R_{2}^{\rm (CD)}$   and $P_{2}^{\rm (CD)}$   measured with charged particles in the range $0.2 < \pt < 2.0$   \gevc\ for selected centrality classes in  \PbPb\   collisions.}
\label{Fig:Corr_2D_ChCD_PbPb}
\end{figure}

\begin{figure}[h!]
\includegraphics[width=0.46\linewidth]{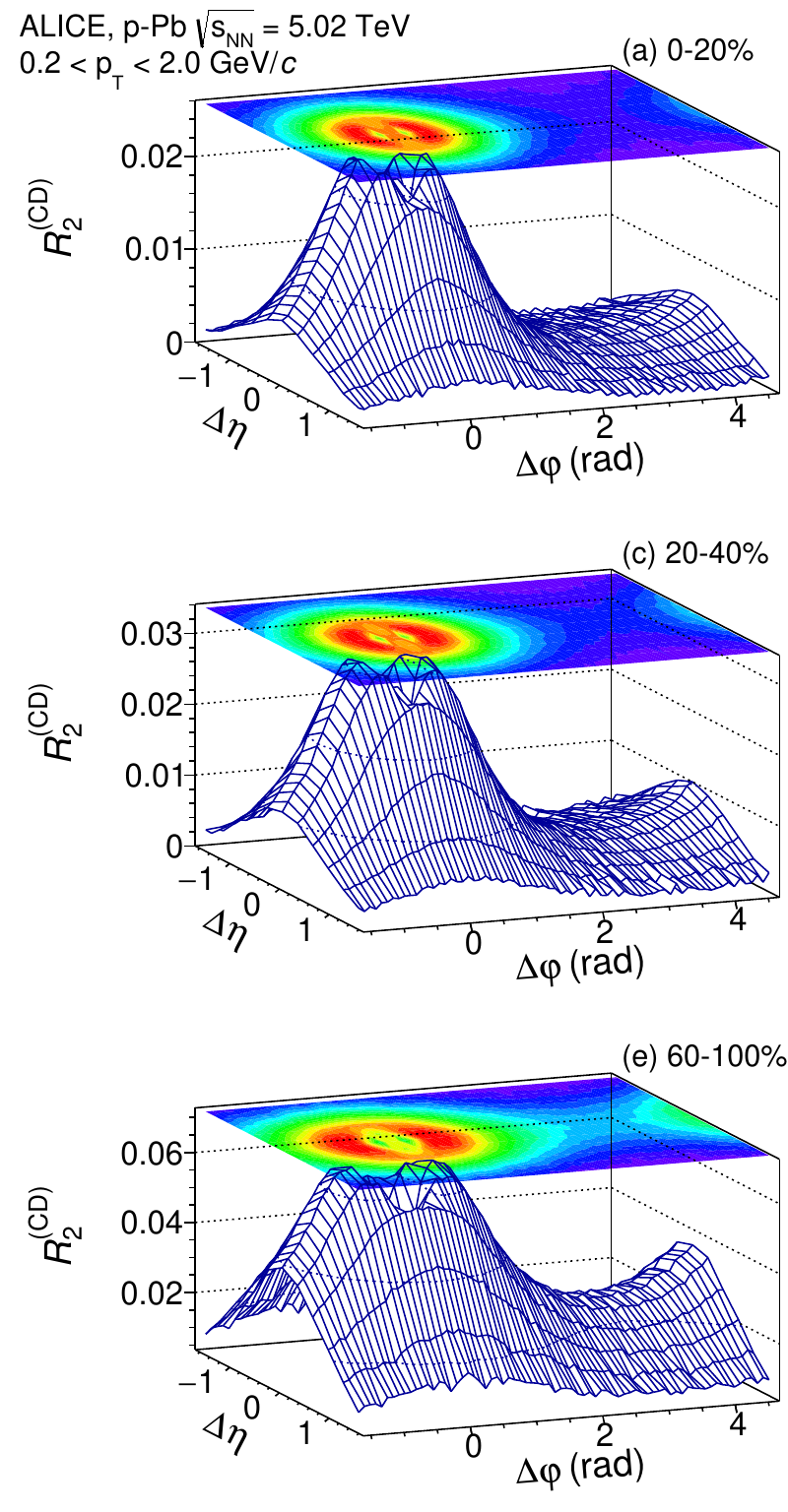}
\includegraphics[width=0.46\linewidth]{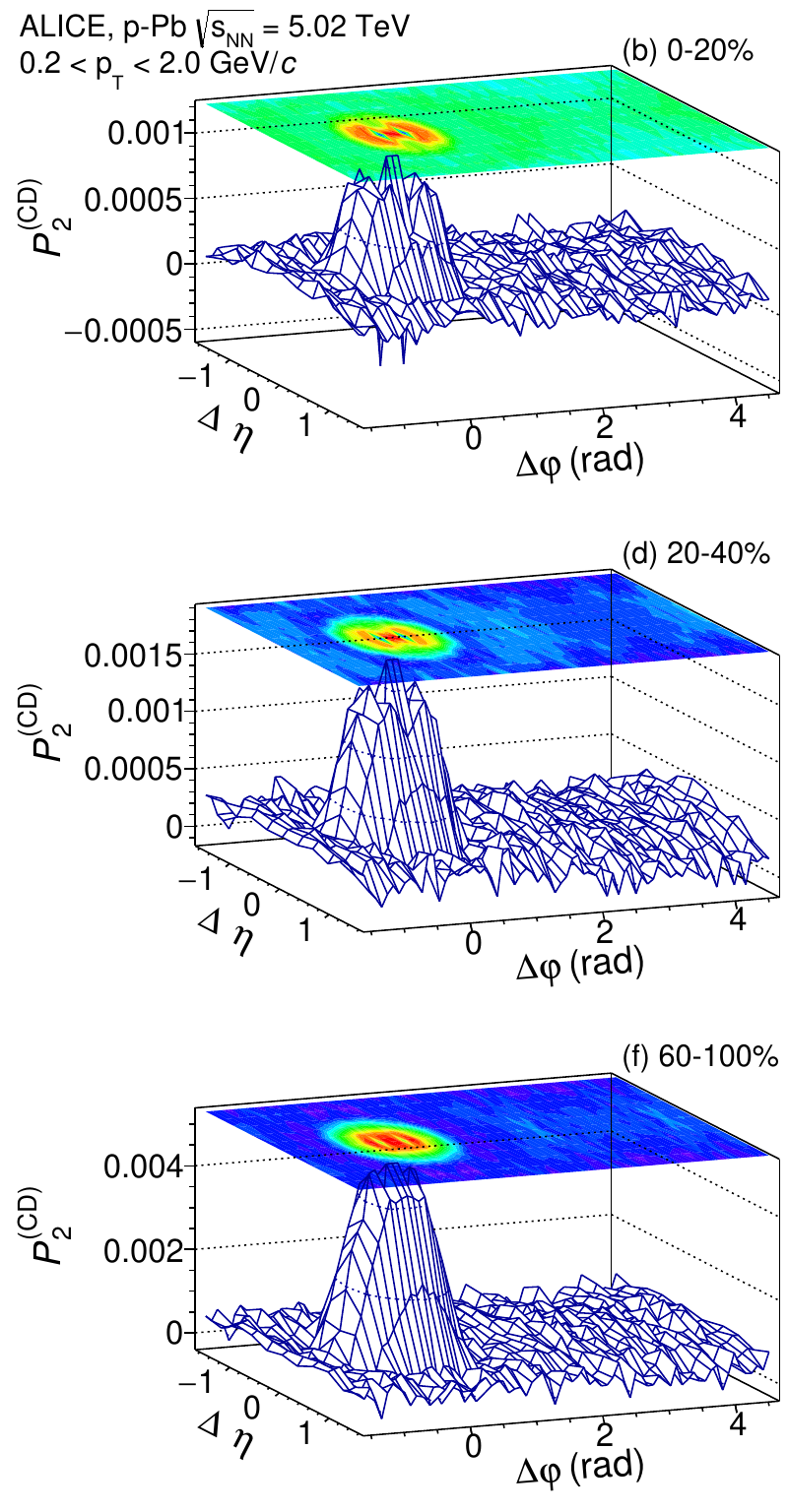}
\caption{Correlation functions  $R_{2}^{\rm (CD)}$   and $P_{2}^{\rm (CD)}$   measured with charged particles in the range $0.2 < \pt < 2.0$   \gevc\ for selected multiplicity classes in \pPb\   collisions.}
\label{Fig:Corr_2D_ChCD_pPb}
\end{figure}

The CD correlation functions $R_2^{\rm (CD)}$ and $P_2^{\rm (CD)}$, shown
in Figs.~\ref{Fig:Corr_2D_ChCD_PbPb} and \ref{Fig:Corr_2D_ChCD_pPb}, respectively, were obtained  according to Eq.~(\ref{Eq:CD})  based on US and LS correlation functions presented in Sec. \ref{Sec:ChargeExclusiveResults}.
In 70--80\% central  \PbPb\  collisions, the $R_{2}^{\rm (CD)}$ correlation function features a very
strong and relatively  broad near-side peak that
extends to  $\Delta\varphi\sim \pi$ and slowly decreases in amplitude
for large values of $|\Delta \eta|$.
The width of the near-side peak  narrows in the centrality range 30--40\% and even more  
in the 0--5\% range. One notes, in particular, that
the away side of these two correlation functions is essentially
flat and nearly vanishing, except for minor and incompletely corrected
detector effects -- most noticeable in the case of the $P_2^{\rm (CD)}$ observable in Fig.~\ref{Fig:Corr_2D_ChCD_PbPb}.
The low-amplitude, high-frequency modulations seen on
the away-side of  $P_2^{\rm (CD)}$ in  the 0--5\% collisions are due to instrumental
effects near the boundaries between TPC sectors. Although these
effects are very much suppressed by the weight-based analysis
used in this work, they could not be completely eliminated.
 The presence of narrow near-side peaks as well as  flat and essentially vanishing away-side in $R_2^{\rm (CD)}$ and  $P_2^{\rm (CD)}$  indicate  that the  US pair production on the
near and away sides (seen in $R_2^{\rm (CD)}$ and  $P_2^{\rm (CD)}$)  are uncorrelated and causally disconnected.
By contrast, the finite away-side amplitude observed in the charge-independent  correlation functions $R_2^{\rm (CI)}$, shown
in Fig.~\ref{Fig:Corr_2D_ChCI_PbPb}, indicates that the corresponding charged-particle correlations
must arise from particle production mechanisms insensitive to charge
conservation.

The  narrowing of  $R_2^{\rm (CD)}$  observed with increasing produced particle multiplicity in Figs.~\ref{Fig:Corr_2D_ChCD_PbPb}--\ref{Fig:Corr_2D_ChCD_pPb}  is qualitatively similar
to the
narrowing of the balance function (BF) reported by the ALICE
collaboration~\cite{Abelev2013267,Adam:2015gda}.
A quantitative comparison of the widths obtained
from $R_2^{\rm (CD)}$ correlations and those already reported for the BF  is presented in Sec.~\ref{Sec:nearsidepeakwidth}.

The strength of $P_2^{\rm (CD)}$ in \PbPb\ collisions is approximately one order of
magnitude weaker than that of $R_2^{\rm (CD)}$. One finds that the
away-side of $P_2^{\rm (CD)}$ is essentially flat, i.e., independent of
$\Delta\eta$ and $\Delta\varphi$, in all centrality classes.
The salient feature of $P_2^{\rm (CD)}$ is a   near-side peak
significantly narrower than the near-side peak observed
in $R_2^{\rm (CD)}$.  This is an interesting result given that both $R_2$
and $P_2$ are derived from the same two-particle density
$\rho_2(\vec p_1,\vec p_2)$. It provides indications that the
product  $\Delta \textit{p}_{\rm T,1} \Delta \textit{p}_{\rm T,2}$  has a significant
dependence on $\Delta\eta$ and $\Delta\varphi$ for correlated US pairs. Also note, in Fig.~\ref{Fig:Corr_2D_ChCD_pPb},  that the near-side peak of $P_2$  observed in   \pPb\  collisions   exhibits
a circular and narrow undershoot ring at  $\sqrt{|\Delta\eta|^2 + |\Delta\varphi|^2} \sim 0.75$. For larger particle separations,
the product $\Delta \textit{p}_{\rm T,1} \Delta \textit{p}_{\rm T,2}$ is approximately constant and  averages to a small positive value,
whereas for smaller separations, it forms  a clear peak. In the undershoot region, the strength of the correlation dips to zero or even below zero.  The  origin of the  very narrow width of the $P_2$ near-side peak and the presence
of the undershoot is discussed in  Sec.~\ref{Sec:discussion}.

It is  interesting to compare the $R_2^{\rm (CD)}$ and $P_2^{\rm (CD)}$ correlation functions obtained in  \pPb\   collisions, shown in Fig. \ref{Fig:Corr_2D_ChCD_pPb}, with  those obtained in \PbPb\  collisions discussed above. The $R_2^{\rm (CD)}$ correlation functions feature
strong and broad near-side peaks similar to that observed in the 70--80\% centrality range. However, the latter has an amplitude smaller than those featured in Fig.~\ref{Fig:Corr_2D_ChCD_pPb}, consistent with the notion that collisions in that centrality range involve a significant geometrical overlap yielding a larger number of binary collisions, on average, than  \pPb\   collisions. One also notes that the near-side peak observed in 0--20\% collisions remains fairly broad and features an amplitude nearly half of that observed in 60--100\% collisions.  Finally, one also observes that all three multiplicity classes feature finite correlation  amplitudes at $\Delta\eta \approx 0$, $\Delta\varphi \approx \pi$, much like the $R_2^{\rm (CD)}$ distribution observed in 70--80\%  \PbPb\  collisions.  These features have already been reported in~\cite{Adam:2015gda}. Remarkably, all three  \pPb\   $P_2^{\rm (CD)}$ shown in Fig. \ref{Fig:Corr_2D_ChCD_pPb} exhibit essentially uniform, but non-vanishing, correlation amplitudes on the away-side. This indicates that $P_2$ correlations manifest a different sensitivity to particle production than  number correlations $R_2$. Note that such a conclusion could not be readily established based on the 70--80\% centrality range in \PbPb\  collisions for the $P_2^{\rm (CD)}$ distribution because of finite residual sector boundary effects observed  in that distribution. One additionally notes that all    $P_2^{\rm (CD)}$ correlation functions measured in \pPb\ exhibit a rather sharp and narrow near-side peak. The width of these peaks is quantified more precisely in the next section, but it is visually rather obvious that the $P_2^{\rm (CD)}$ near-side peaks are much narrower than those observed in $R_2^{\rm (CD)}$ correlations. It is also interesting to notice that the amplitude of the near-side reduces by about a factor of five from 60--100\% to 0--20\% multiplicity classes, while the amplitude  of the $R_2^{\rm (CD)}$ correlation decreases by a factor of two only. Clearly, the $P_2^{\rm (CD)}$  correlation  has a rather different sensitivity to charge creation than the $R_2^{\rm (CD)}$ correlation.

\subsection{Near-side peak widths}
\label{Sec:nearsidepeakwidth}

The presence of a relatively narrow near-side peak in $R_2$ and $P_2$ correlation functions indicates that the production of
two particles (or more) at small relative azimuthal angle and
pseudorapidity is substantially more probable than large angle emission. Such narrowly focused emission may in principle be produced by in flight decays of highly boosted resonances or clusters, jet fragmentation, or  string (or color tube) fragmentation~\cite{Voloshin:2004th,Voloshin:2005qj,Pruneau:2007ua,Dumitru:2008wn,Gavin:2008ev,Dusling:2009ar,Romatschke:2006bb,Shuryak:2007fu}.
However, these different production mechanisms  feature distinct $\pt$ dependences and may thus produce noticeable differences in the structures of the   $R_2$ and $P_2$ correlation functions. Comprehensive particle production models should in principle enable  detailed calculations of the shape and strength of $R_2^{\rm (CI)}$,  $R_2^{\rm (CD)}$, $P_2^{\rm (CI)}$, and $P_2^{\rm (CD)}$ to be compared to two-dimensional distributions presented in this work. It is interesting, nonetheless, to extract simple characterizations of  these distributions and consider their  multiplicity dependence in \PbPb\  and \pPb\ collisions.

\begin{figure}[h!]
\includegraphics[width=0.46\linewidth]{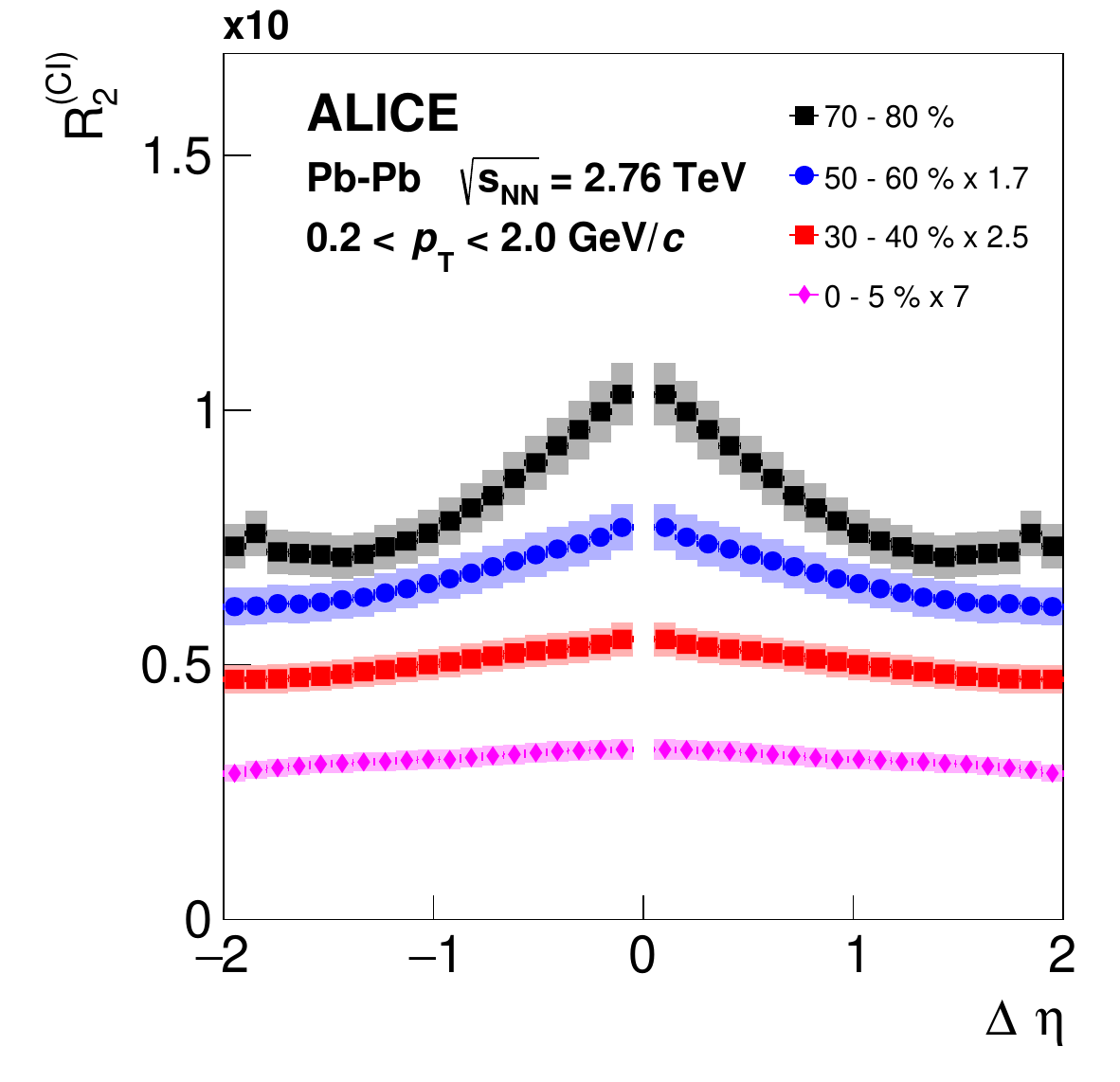}
\includegraphics[width=0.46\linewidth]{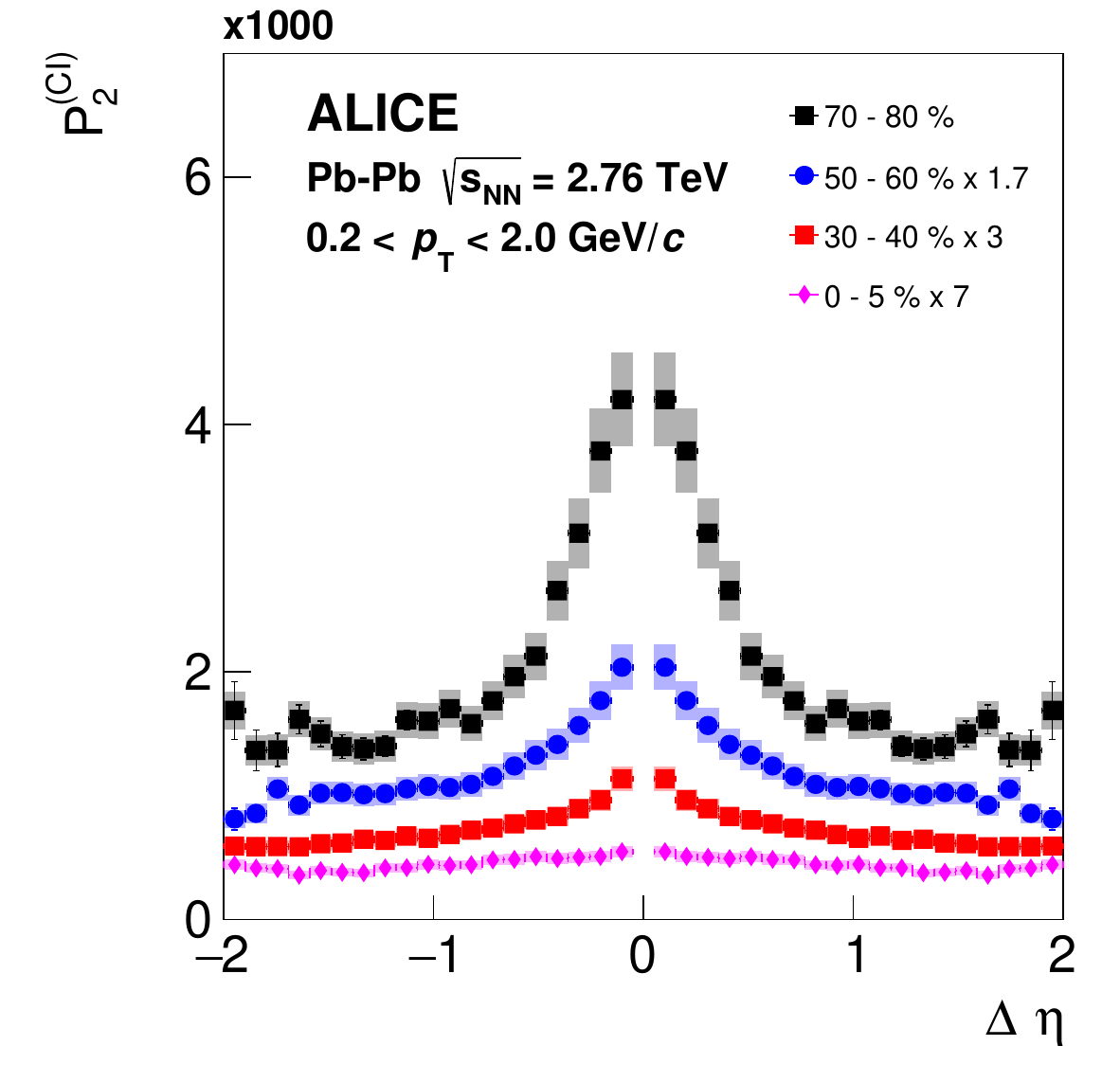}
\caption{Projections of $R_2^{\rm (CI)}$ and $P_2^{\rm (CI)}$ correlation functions, measured in \PbPb\  collision at $\sqrt{s_{NN}} = 2.76$ TeV, for selected ranges of collision
centrality. Projections onto the $\Delta\eta$  axis are calculated as averages of the two-dimensional correlations in the range  $|\Delta \varphi| \le \pi$.  Vertical bars and shaded areas represent statistical and systematic  uncertainties, respectively.}
\label{Fig:PbPb_R2CI_P2_projectionsOntoEta}
\end{figure}

\begin{figure}[h!]
\includegraphics[width=0.46\linewidth]{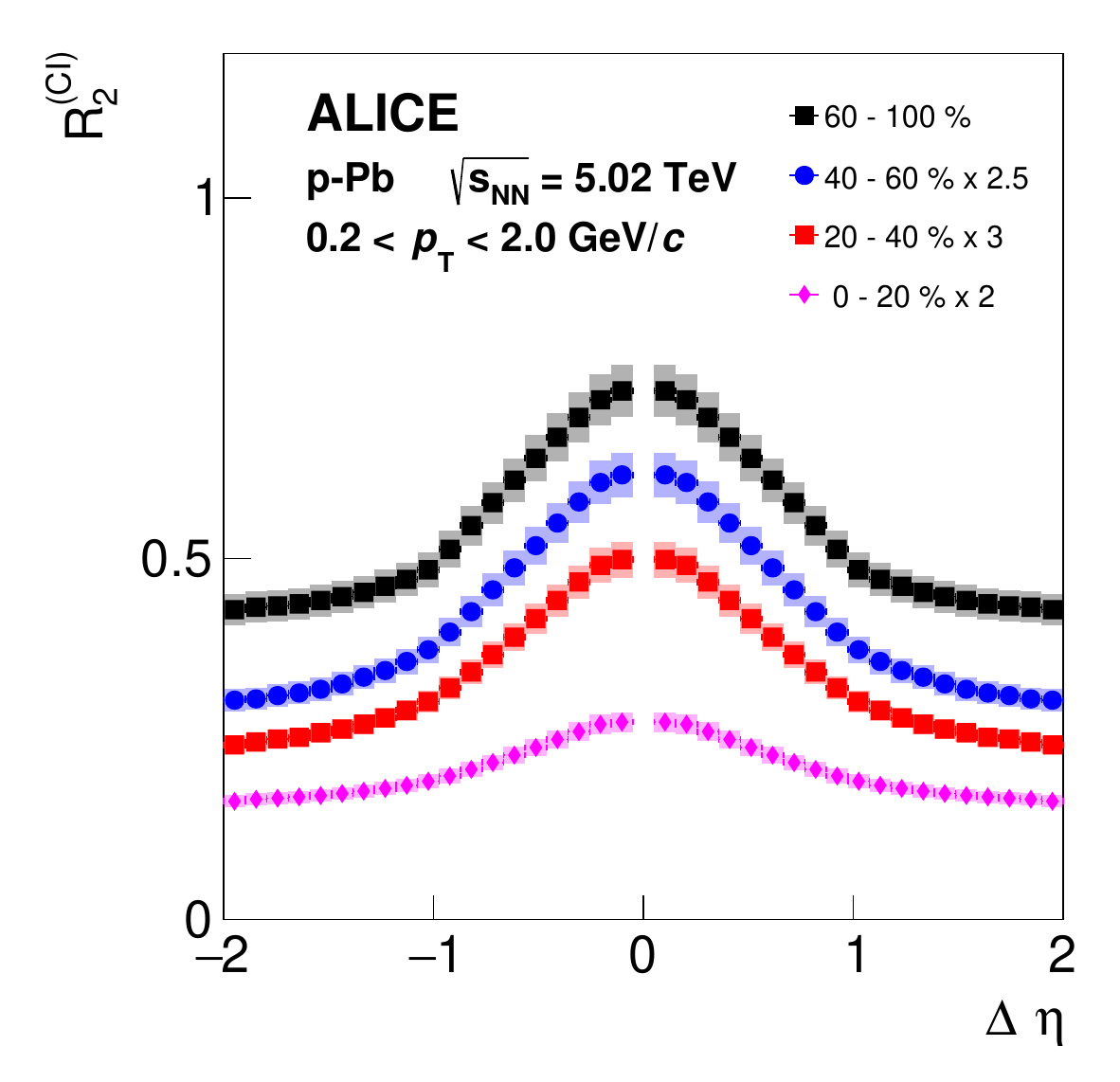}
\includegraphics[width=0.46\linewidth]{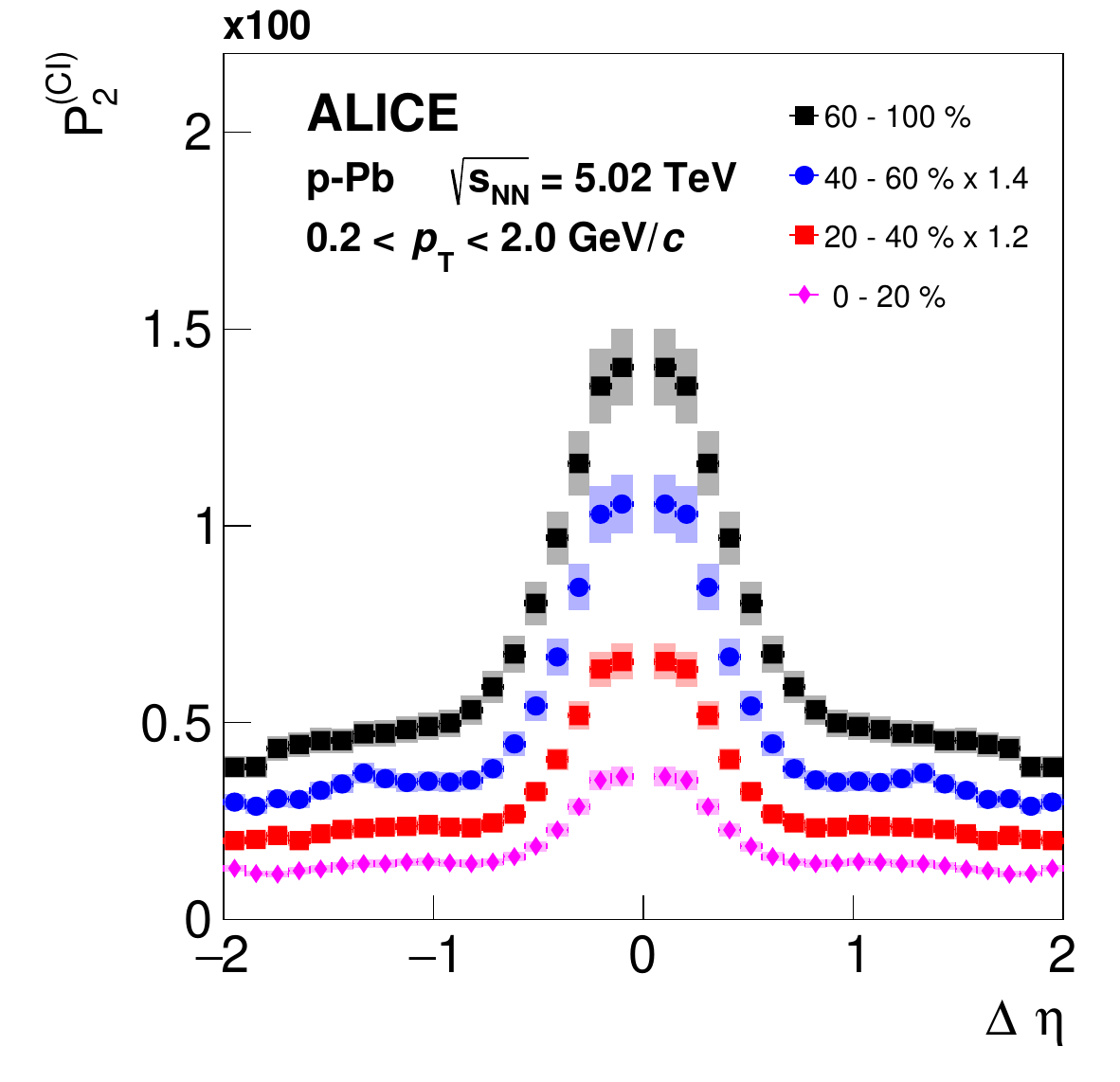}
\caption{Projections of $R_2^{\rm (CI)}$ and $P_2^{\rm (CI)}$ correlation functions, measured in  \pPb\   collisions at $\sqrt{s_{NN}} = 5.02$ TeV, for selected multiplicity classes. Projections onto the $\Delta\eta$  axis are calculated as averages of the two-dimensional correlations in the range  $|\Delta \varphi| \le \pi$. Vertical bars and shaded areas represent statistical and systematic  uncertainties, respectively.   }
\label{Fig:pPb_R2CI_P2_projectionsOntoEta}
\end{figure}

\begin{figure}[h!]
\includegraphics[width=0.46\linewidth]{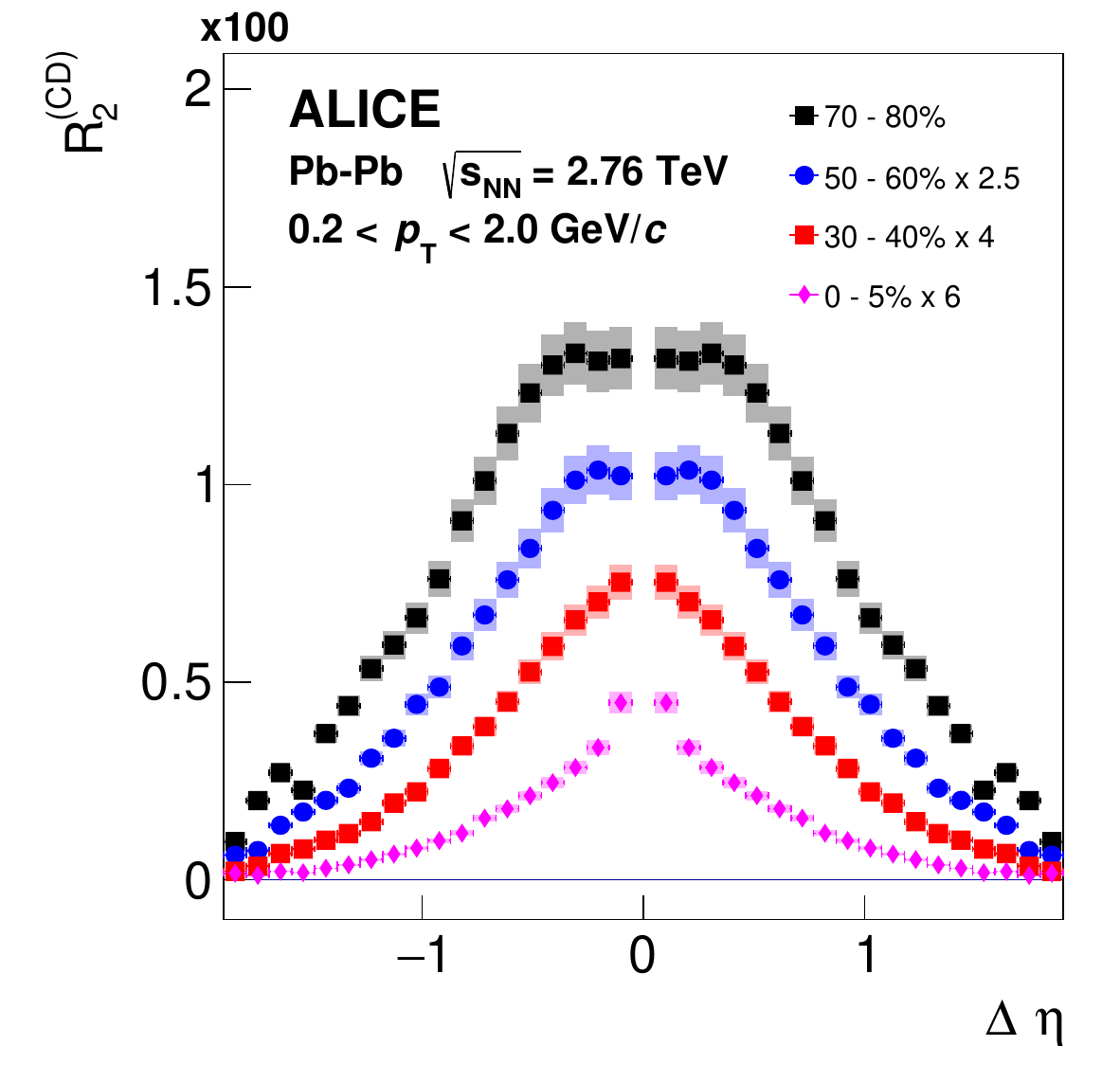}
\includegraphics[width=0.46\linewidth]{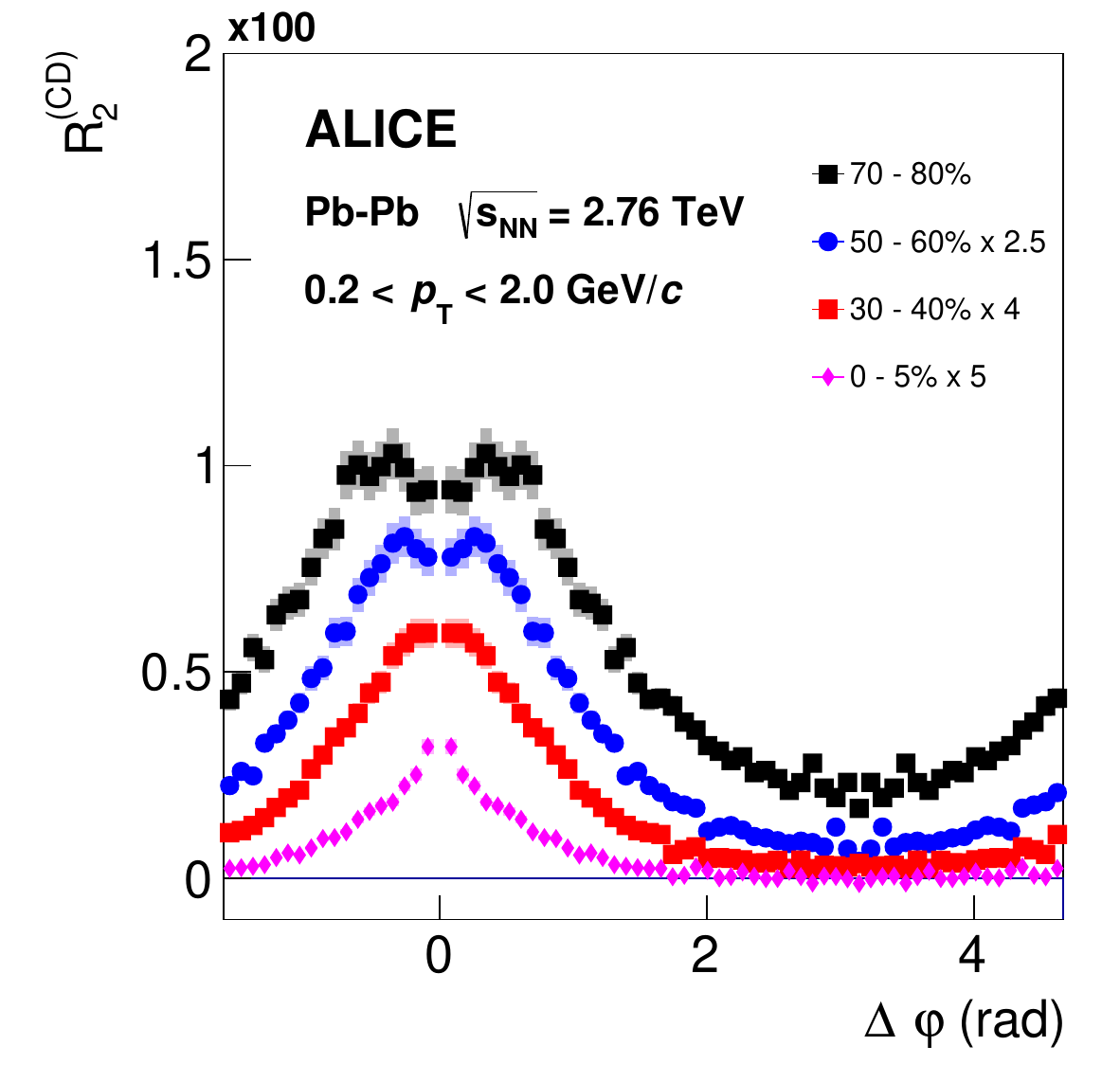}
\caption{Projections of $R_2^{\rm (CD)}$ correlation functions, measured in \PbPb\  collision at $\sqrt{s_{NN}} = 2.76$ TeV, for selected ranges of collision
centrality.  The $\Delta\eta$  and $\Delta\varphi$ projections are calculated as averages of the two-dimensional correlations in the ranges $|\Delta \varphi| \le \pi$  and $|\Delta \eta| \le 1.8$, respectively. Vertical bars and shaded areas represent statistical and systematic  uncertainties, respectively. }
\label{Fig:PbPb_R2_CD_projections}
\end{figure}

\begin{figure}[h!]
\includegraphics[width=0.46\linewidth]{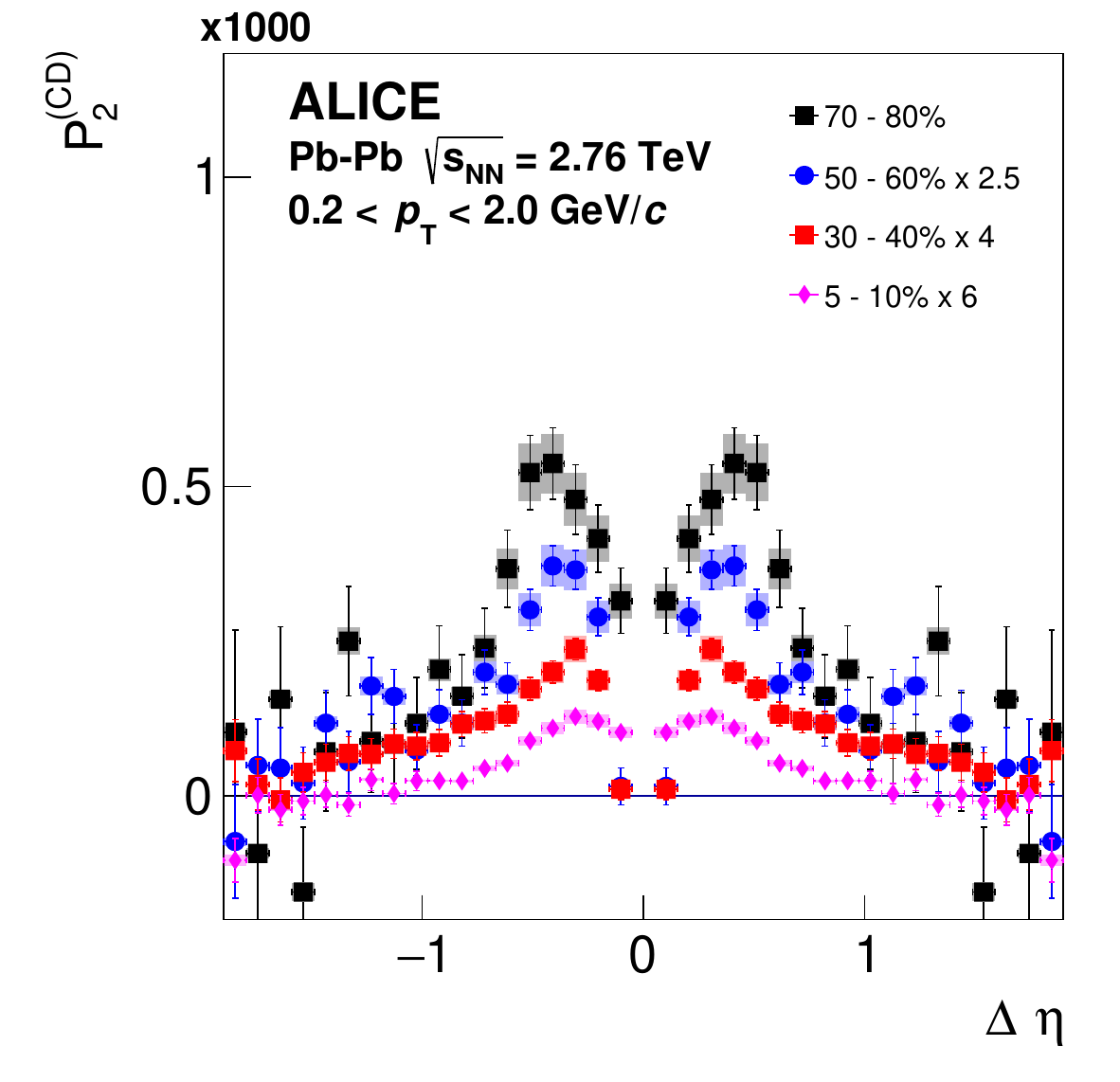}
\includegraphics[width=0.46\linewidth]{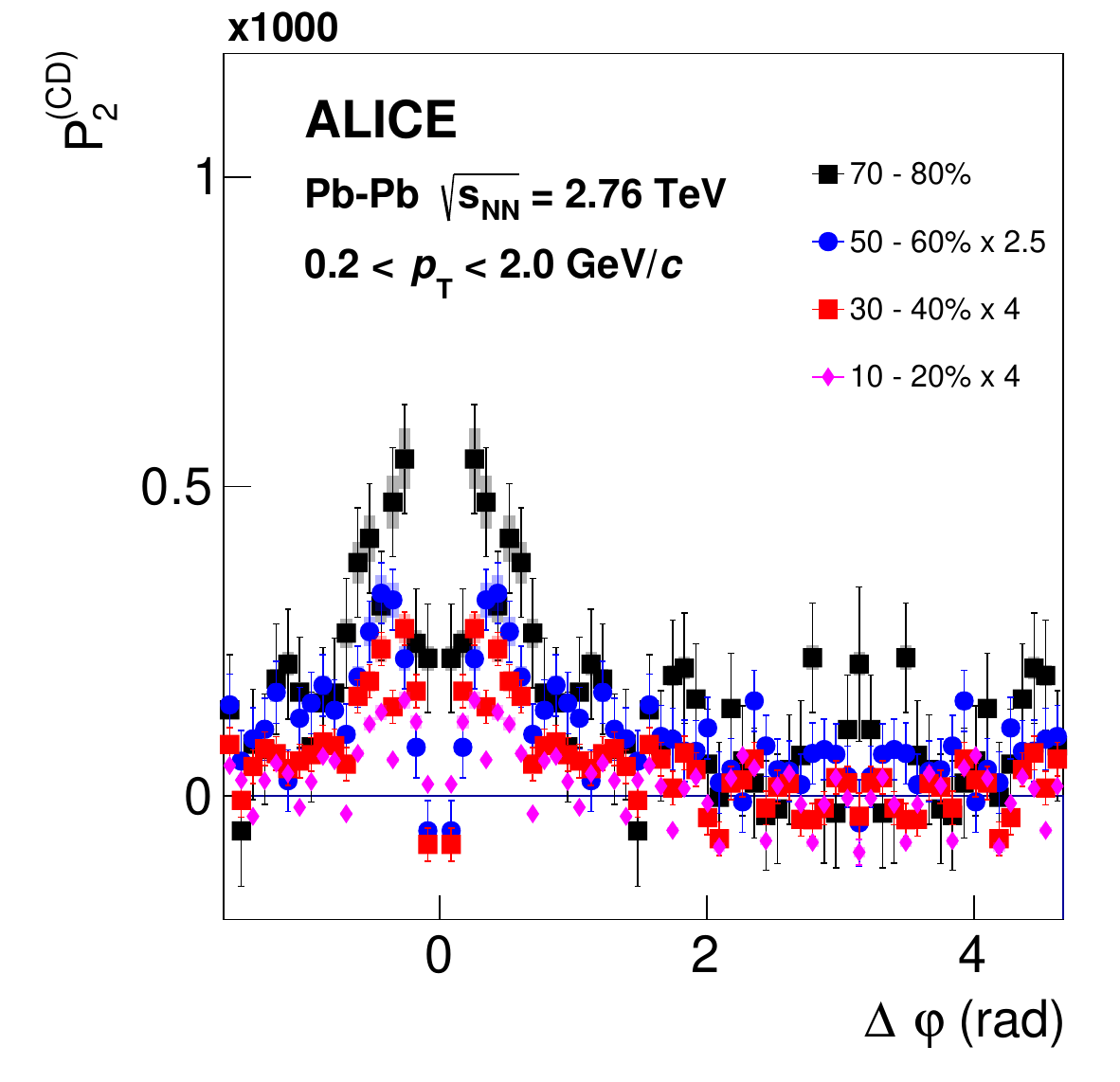}
\caption{Projections of  $P_2^{\rm (CD)}$ correlation functions, measured in \PbPb\  collision at $\sqrt{s_{NN}} = 2.76$ TeV, for selected ranges of collision
centrality.  The $\Delta\eta$  and $\Delta\varphi$ projections are calculated as averages of the two-dimensional correlations in the ranges $|\Delta \varphi| \le \pi$  and $|\Delta \eta| \le 1.8$, respectively. Vertical bars and shaded areas represent statistical and systematic  uncertainties, respectively. }
\label{Fig:PbPb_P2_CD_projections}
\end{figure}

\begin{figure}[h!]
\includegraphics[width=0.46\linewidth]{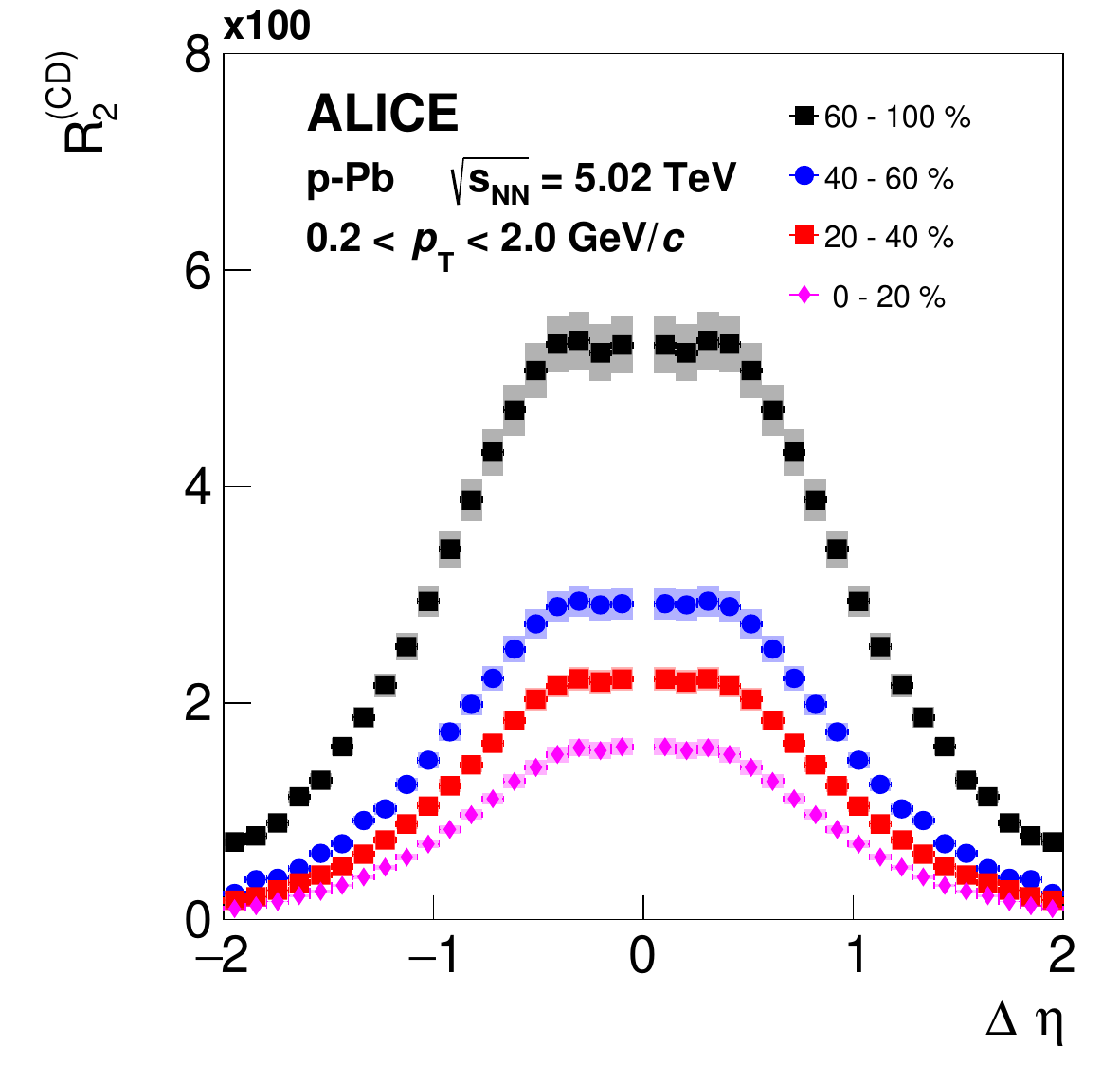}
\includegraphics[width=0.46\linewidth]{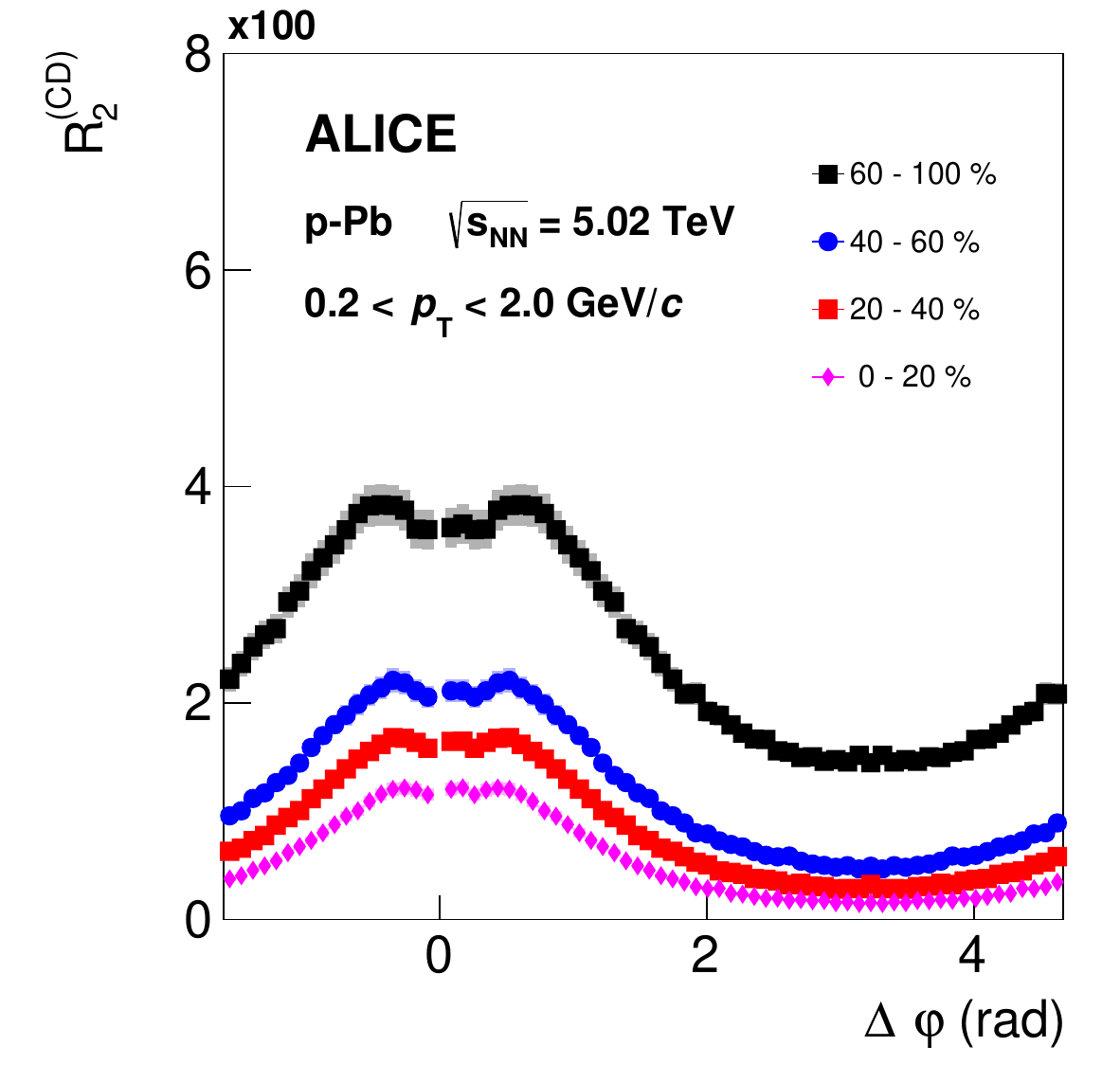}
\caption{Projections of $R_2^{\rm (CD)}$ correlation functions, measured in \pPb\  collision at $\sqrt{s_{NN}} = 5.02$ TeV, for selected multiplicity classes.  The $\Delta\eta$  and $\Delta\varphi$ projections are calculated as averages of the two-dimensional correlations in the ranges $|\Delta \varphi| \le \pi$  and $|\Delta \eta| \le 1.8$, respectively. Vertical bars and shaded areas represent statistical and systematic  uncertainties, respectively.}
\label{Fig:pPb_R2_CD_projections}
\end{figure}

\begin{figure}[h!]
\includegraphics[width=0.46\linewidth]{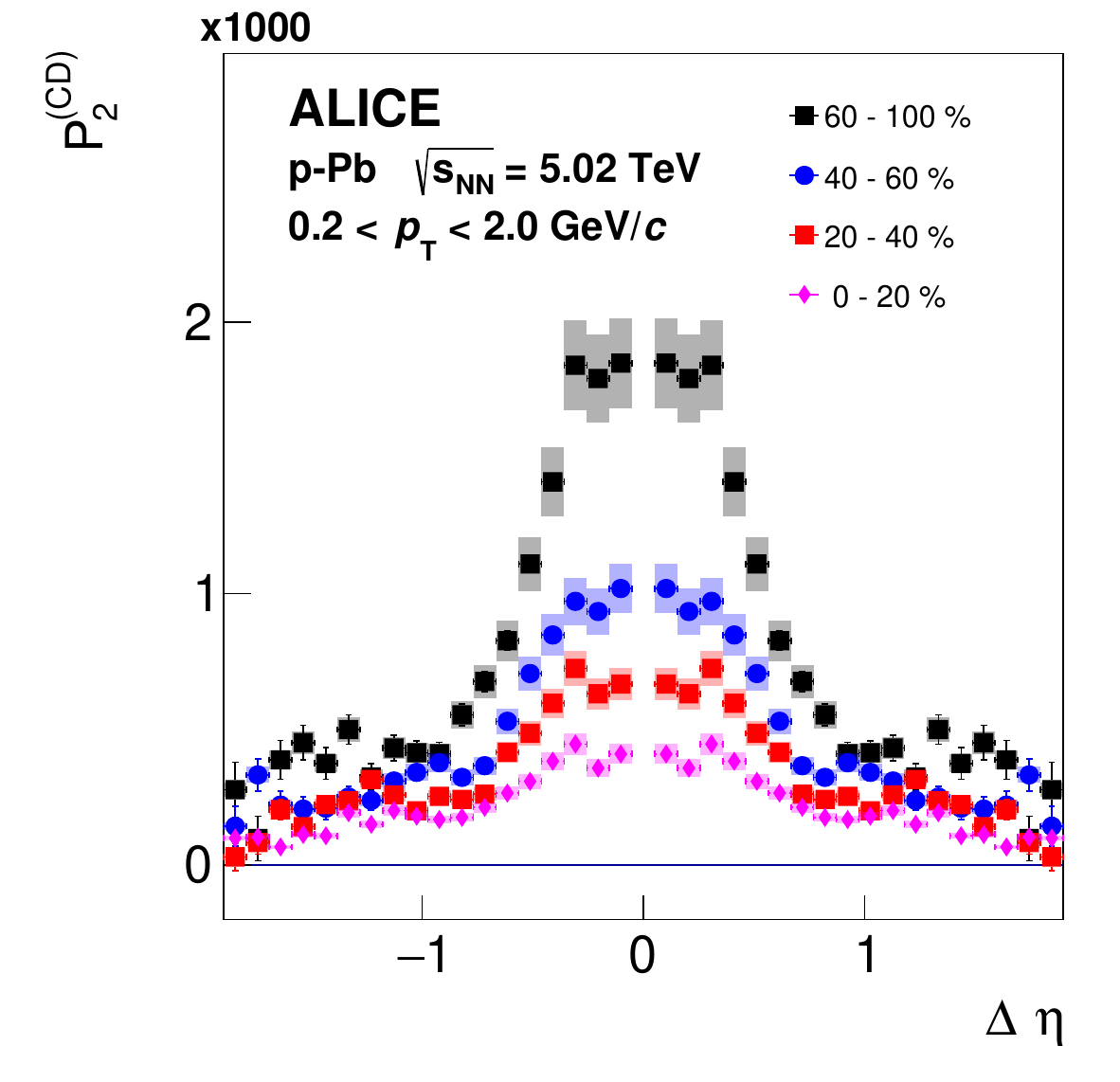}
\includegraphics[width=0.46\linewidth]{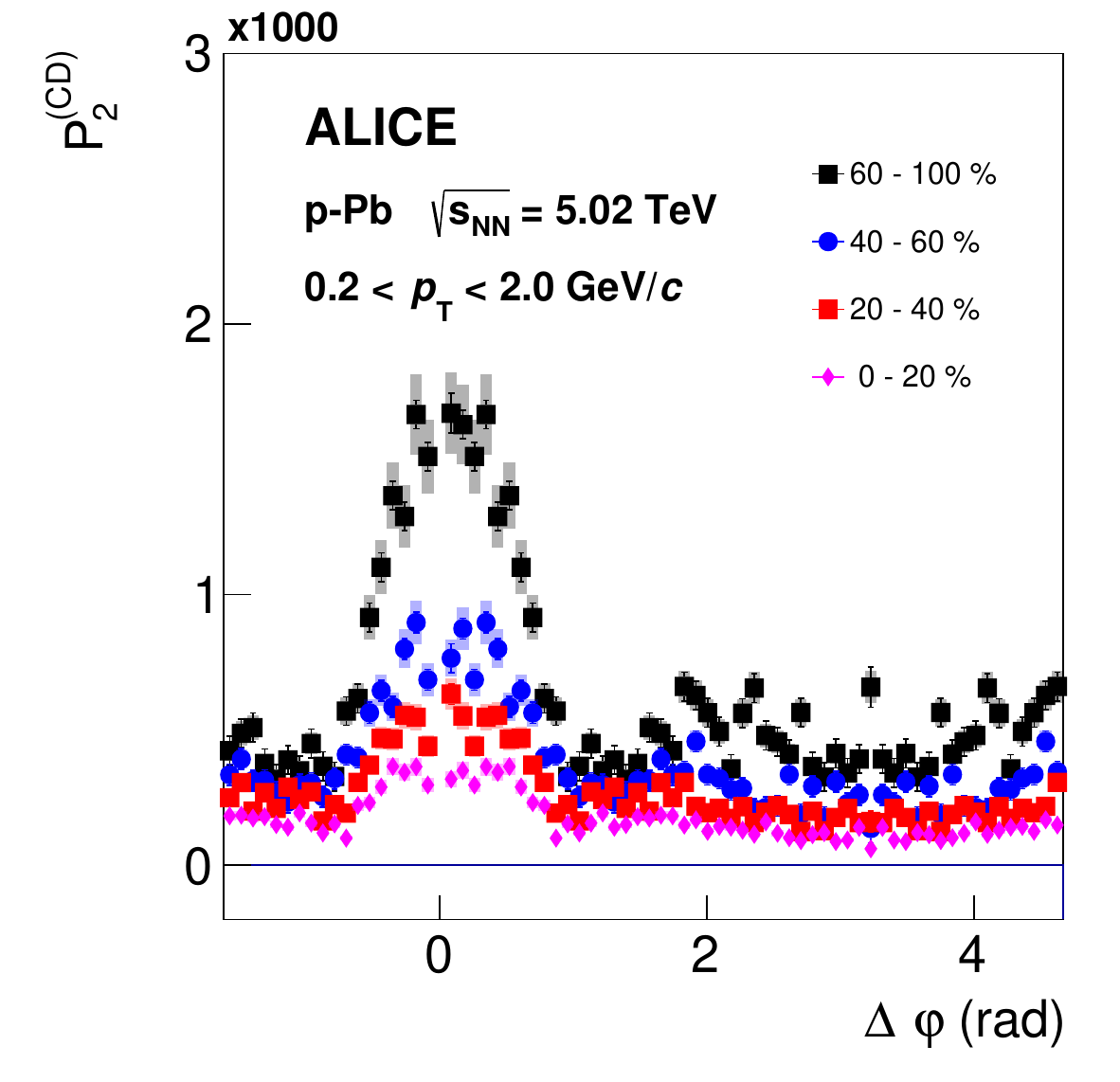}
\caption{Projections of $P_2^{\rm (CD)}$ correlation functions, measured in \pPb\  collision at $\sqrt{s_{NN}} = 5.02$ TeV, for selected multiplicity classes.  The $\Delta\eta$  and $\Delta\varphi$ projections are calculated as averages of the two-dimensional correlations in the ranges $|\Delta \varphi| \le \pi$  and $|\Delta \eta| \le 1.8$, respectively. Vertical bars and shaded areas represent statistical and systematic  uncertainties, respectively.}
\label{Fig:pPb_P2_CD_projections}
\end{figure}

Measurements of the evolution of the width of the distributions with increasing  multiplicity, in particular, are of interest  given that variations of the widths might reflect important changes in the underlying particle production mechanisms~\cite{Pratt:2010zn,Bass:2000az,Jeon:2001ue}.  In order to enable comparisons with previous works (e.g., balance function) \cite{Aggarwal:2010ya,Abelev:2013csa}, we proceed to determine the longitudinal and azimuthal means as well as the RMS  widths of the measured  correlation functions in terms of the moments $\la\Delta \eta^k\ra$ and  $\la\Delta \varphi^k\ra$, with $k=1,2$, calculated according to
\be\label{Eq:widthR2Eta}
\la \Delta\eta^k \ra = \frac{\sum\limits_{\Delta\eta_{\min}}^{\Delta\eta_{\max}}\left[O(\Delta\eta_{i},\Delta\varphi_{i})- O_{\rm off}\right]\Delta\eta_{i}^k}{\sum\limits_{\Delta\eta_{\min}}^{\Delta\eta_{\max}}\left[O(\Delta\eta_{i},\Delta\varphi_{i})- O_{\rm off}\right]} \\ \label{Eq:widthR2Phi}
\la \Delta\varphi^k \ra = \frac{\sum\limits_{\Delta\varphi_{\min}}^{\Delta\varphi_{\max}}\left[O(\Delta\varphi_{i},\Delta\varphi_{i}^k)- O_{\rm off}\right]\Delta\varphi_{i}^k}{\sum\limits_{\Delta\varphi_{\min}}^{\Delta\varphi_{\max}}\left[O(\Delta\varphi_{i},\Delta\varphi_{i})- O_{\rm off}\right]},
\ee
where $O(\Delta\eta_{i},\Delta\varphi_{i})$  represents  values of the correlation functions  $R_2^{\rm (CI)}$,  $R_2^{\rm (CD)}$, $P_2^{\rm (CI)}$, or $P_2^{\rm (CD)}$ for
the relative pseudorapidity bin $\Delta\eta_{i}$ (azimuthal angle $\Delta\varphi_{i}$).  For $k=1$, the summations are
carried out  one-sided, i.e., from  $\Delta\eta_{\min}= 0$ ($\Delta\varphi_{\min}=0$)  to maximum values  $\Delta\eta_{\max}$ ($\Delta\varphi_{\max}$), while for $k=2$,
the summations are carried  two-sided, i.e.,  in the range   $-\Delta \eta_{\max} \le \Delta\eta \le \Delta \eta_{\max}$ ($-\Delta \varphi_{\max} \le \Delta\varphi \le \Delta \varphi_{\max}$). For $\la \Delta\eta^k \ra$ calculations, $\Delta\eta_{\max}$ is  chosen either at the edge of the acceptance or at  $\Delta\eta$
values where the correlation functions reach a plateau (most particularly in the case of CD correlations) to avoid undue accumulation of noise in the calculation of the moments. For $\la \Delta\varphi^k \ra$ calculations, the upper edge of the range  is set to $\Delta\varphi_{\max} = \pi$ for $R_2^{\rm (CD)}$ correlations and whichever values the $\Delta\varphi$ projections reach  a minimum, in the case of $P_2^{\rm (CD)}$ correlations.   Offsets  $O_{\rm off}$ are nominally used to eliminate trivial dependences of
the averages  on the  width of the experimental acceptance. For   calculations of $\la\Delta\varphi^k\ra$, offsets $O_{\rm off}$ are determined by taking a 3 bin average near  $\Delta \varphi = \pi$, while for calculations of $\la\Delta\eta^k\ra$,  offsets $O_{\rm off}$ are evaluated near
the edge of the acceptance $\Delta\eta\sim 2$. However, in the case of $R_2^{\rm (CD)}$, since  the correlation is vanishing for large $|\Delta\eta|$ values, one uses a null offset. In this case, contributions to $\la\Delta\eta^k\ra^{1/k}$ from the unobserved part of $R_2^{\rm (CD)}$, i.e. beyond the  acceptance, are then  neglected. Moments $\la\Delta\eta^k\ra$ and $\la\Delta\varphi^k\ra$ are determined on the basis of projections of the
$R_2^{\rm (CI)}$,  $R_2^{\rm (CD)}$, $P_2^{\rm (CI)}$, or $P_2^{\rm (CD)}$ correlation functions onto the $\Delta\eta$ and $\Delta\varphi$ axes, respectively.
Projections onto $\Delta\eta$ are calculated in the range $|\Delta\varphi|\le \pi$, whereas the projections onto $\Delta\varphi$
are determined in the range $|\Delta\eta|\le 1.8$ for $R_2$ correlations and $|\Delta\eta|\le 1$ for $P_2$ correlations, also to suppress accumulation of statistical noise.
Only $\Delta\eta$ projections and the corresponding moments $\la \Delta\eta^k \ra$ are considered in the case of $R_2^{\rm (CI)}$ and  $P_2^{\rm (CI)}$ given that  these correlation functions feature strong azimuthal modulations. Projections of the $R_2^{\rm (CI)}$,  $R_2^{\rm (CD)}$, $P_2^{\rm (CI)}$, or $P_2^{\rm (CD)}$ correlation functions are shown in Figs.~\ref{Fig:PbPb_R2CI_P2_projectionsOntoEta} --\ref{Fig:pPb_P2_CD_projections}. They have been divided by the number of integrated bins and scaled for ease of comparison.

\begin{figure}[h!]
\includegraphics[width=0.46\linewidth]{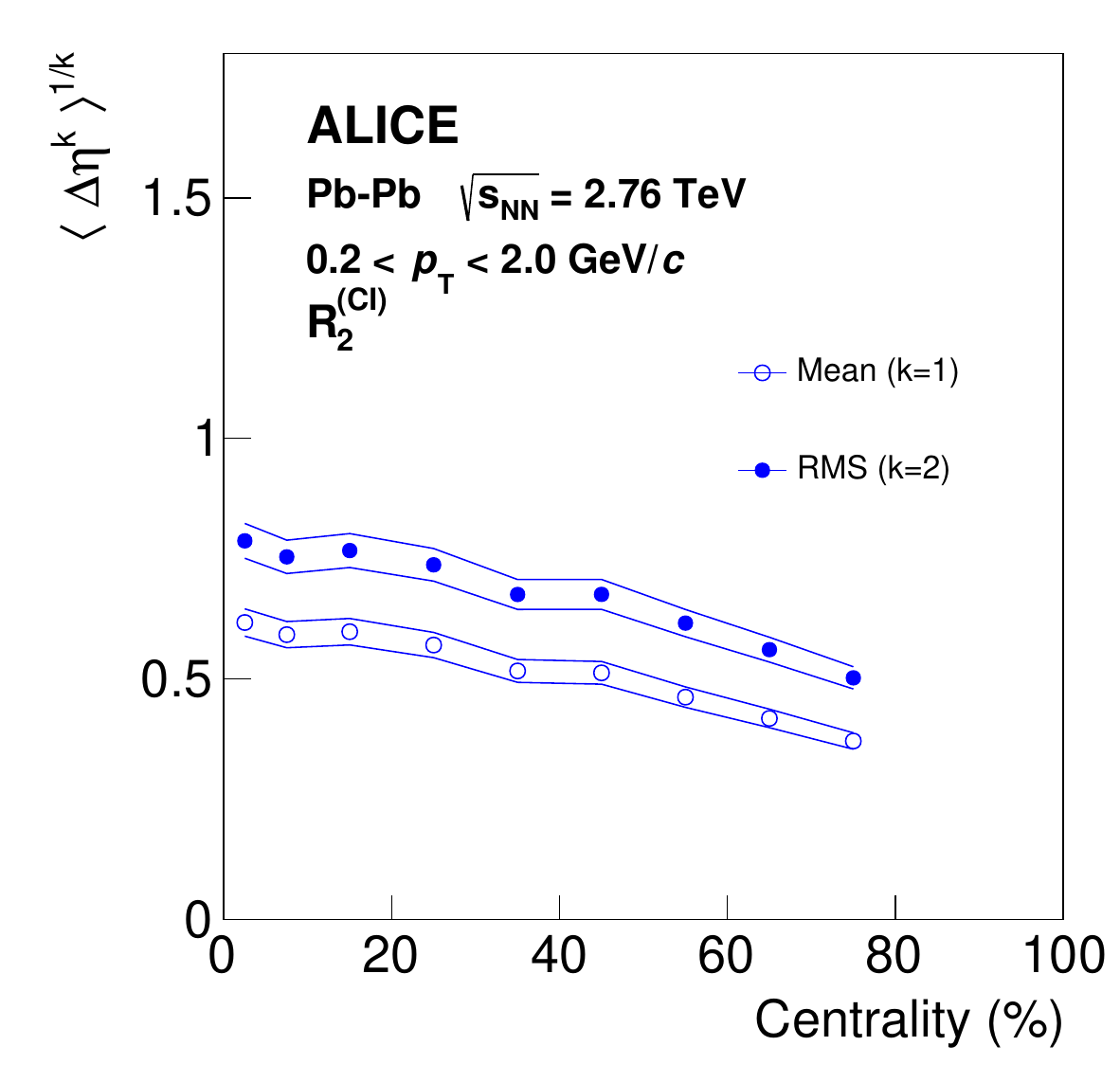}
\includegraphics[width=0.46\linewidth]{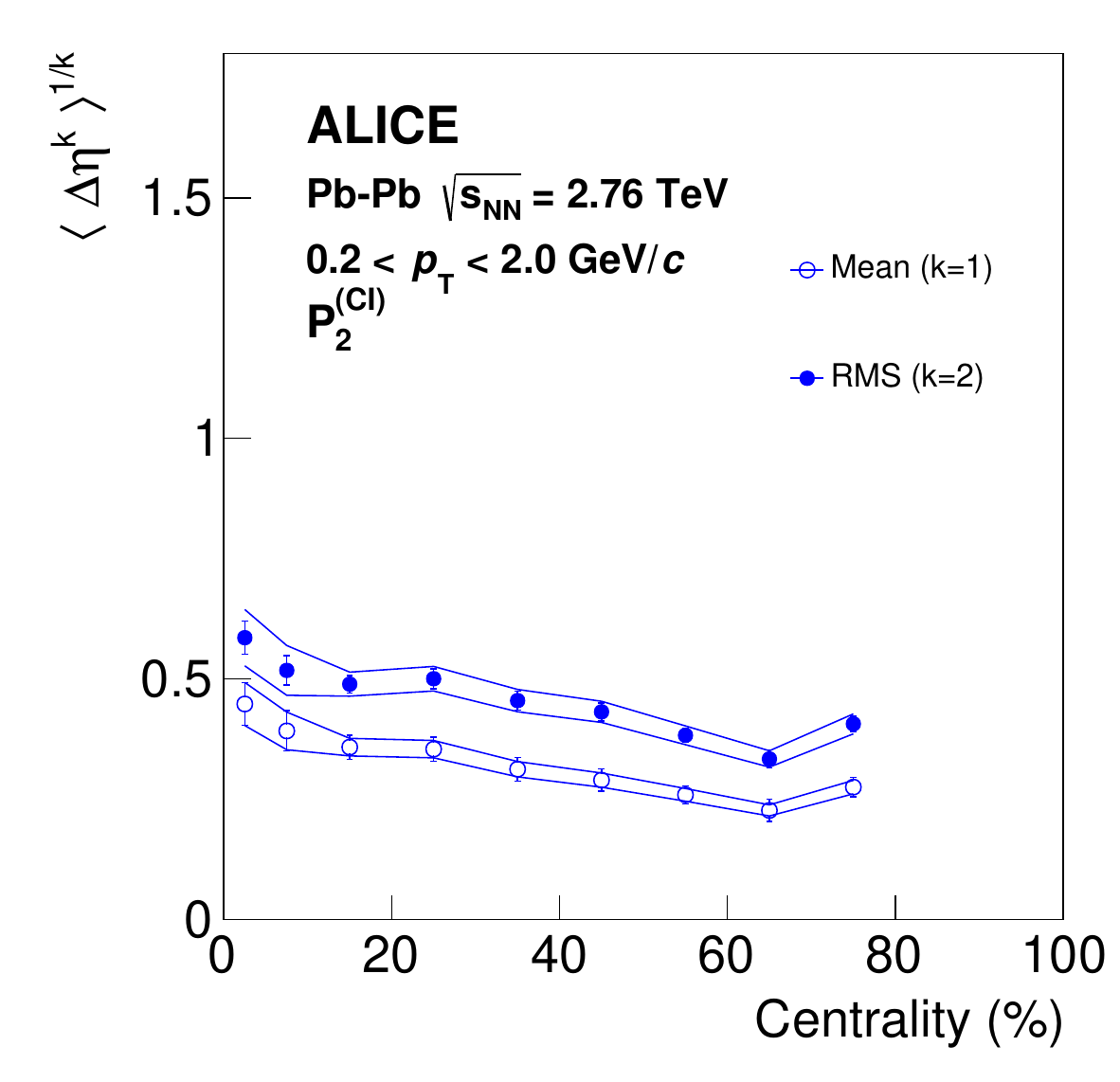}
\caption{Width of the near-side peak of $R_2^{\rm (CI)}$ (left)  and $P_2^{\rm (CI)}$ (right) correlation functions  along $ \Delta \eta$  measured in
\PbPb\   collisions  as a function of the collision centrality class. Vertical bars and solid lines represent statistical and systematic  uncertainties, respectively. }
\label{Fig:PbPb_R2CI_P2CI_WidthVsCent}
\end{figure}

\begin{figure}[h!]
\includegraphics[width=0.46\linewidth]{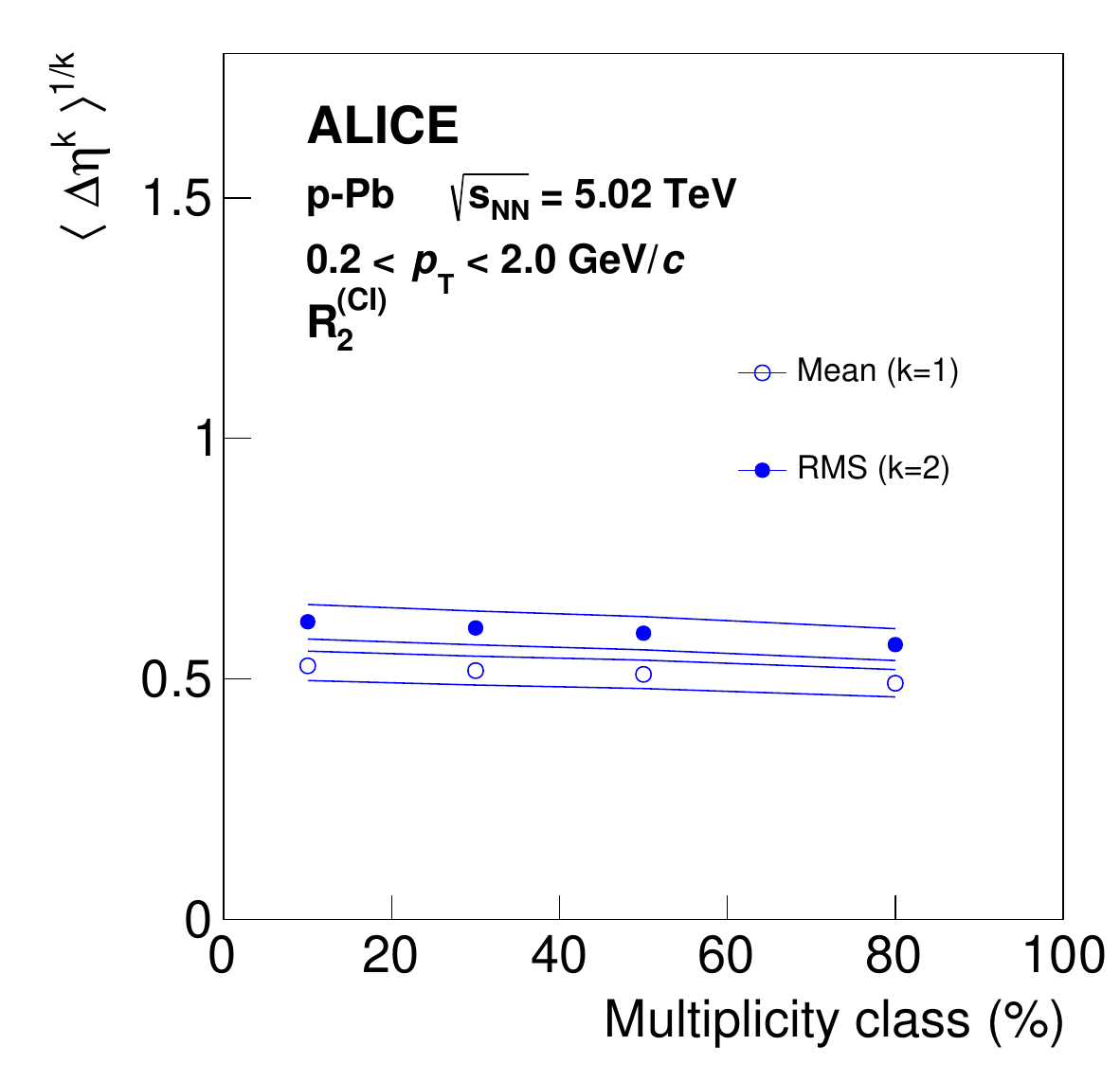}
\includegraphics[width=0.46\linewidth]{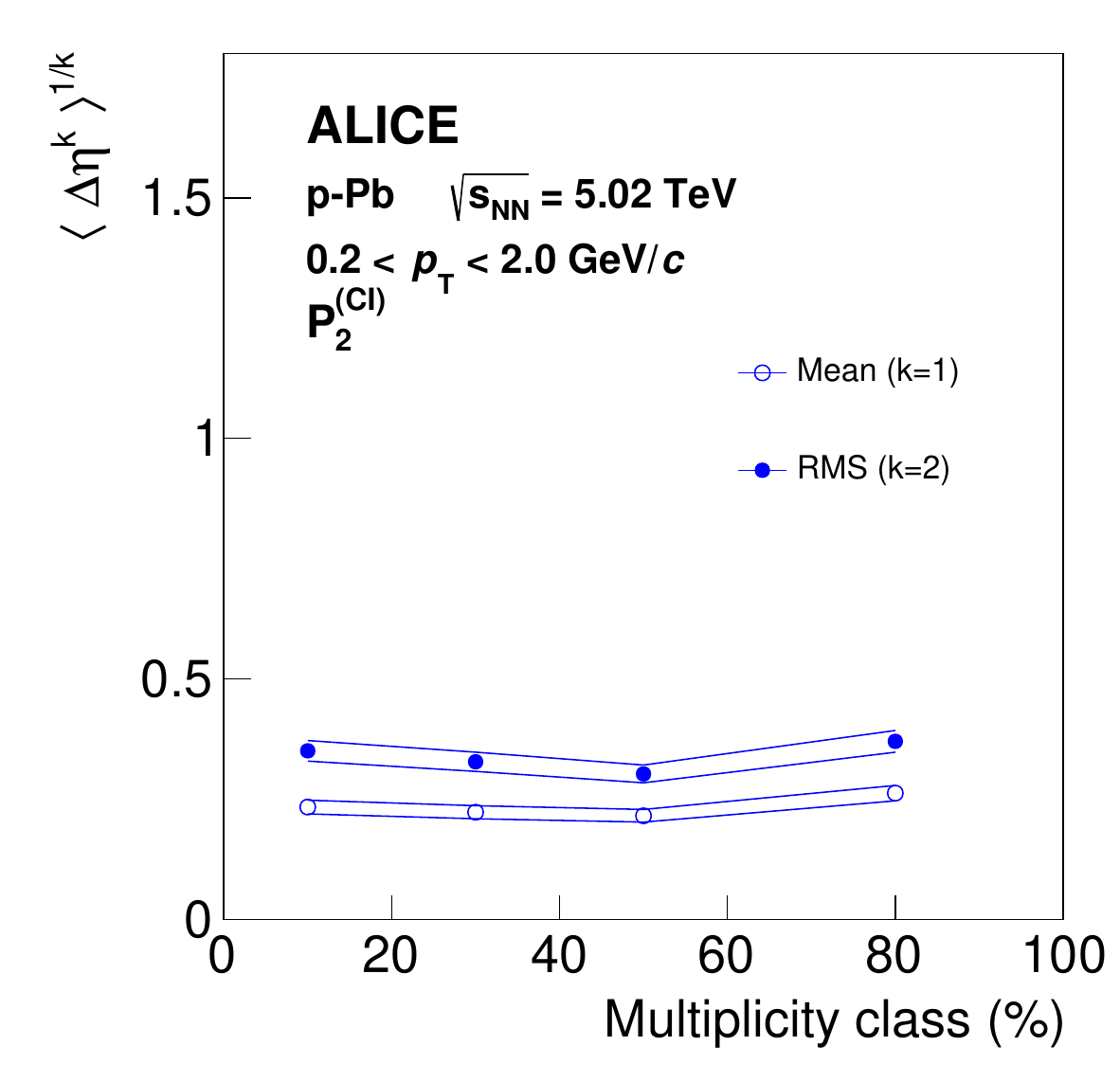}
\caption{Width of the near-side peak of $R_2^{\rm (CI)}$ (left)  and $P_2^{\rm (CI)}$ (right) correlation functions  along $ \Delta \eta $  measured in \pPb\  collisions  as a function of  produced particle multiplicity class. Vertical bars and solid lines represent statistical and systematic  uncertainties, respectively. }
\label{Fig:pPb_R2CI_P2CI_WidthVsCent}
\end{figure}

\begin{figure}[h!]
\includegraphics[width=0.46\linewidth]{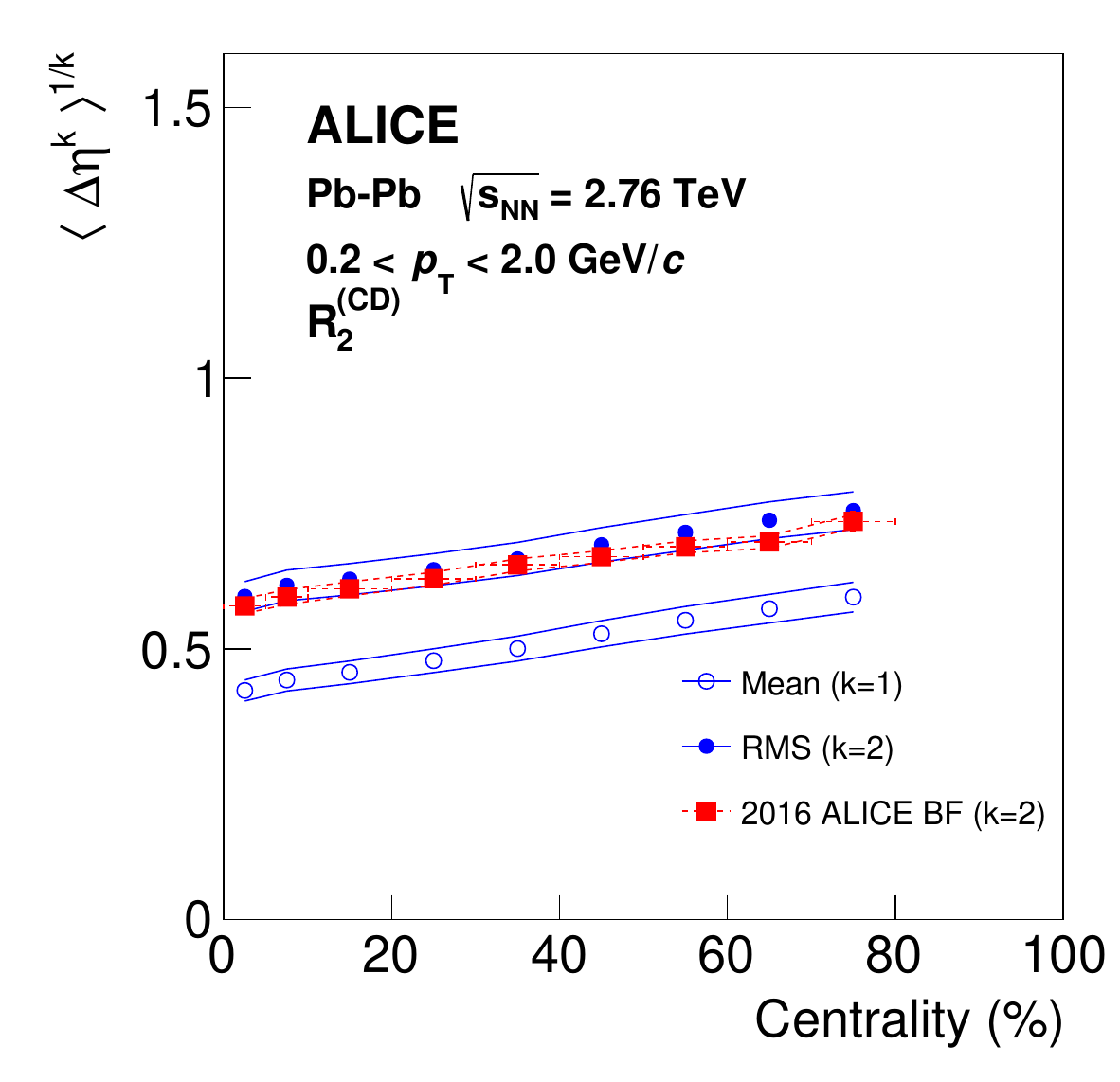}
\includegraphics[width=0.46\linewidth]{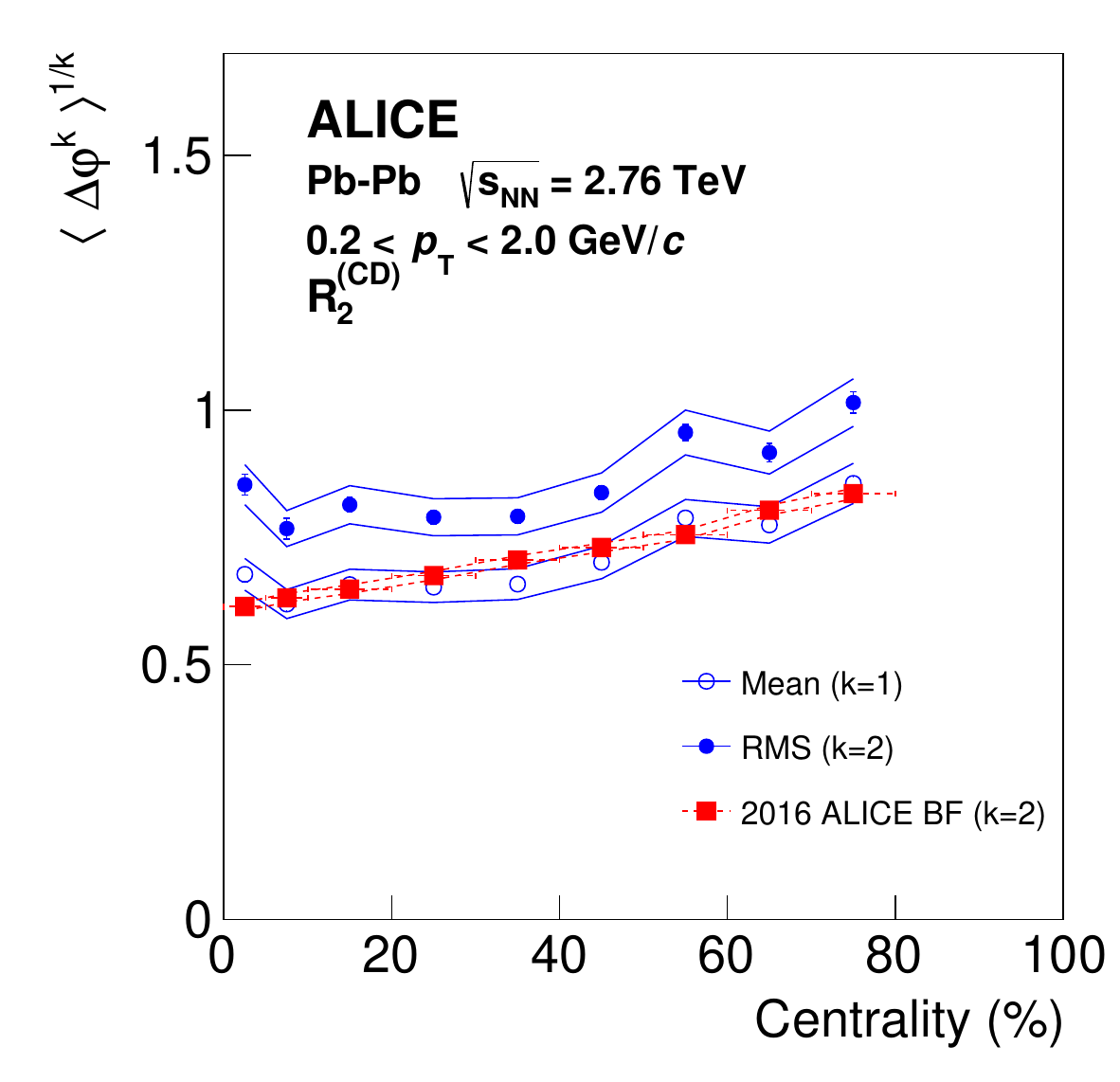}
\caption{Width of the near-side peak of $R_2^{\rm (CD)}$  correlation functions  along $ \Delta \eta $ (left) and   $\Delta \varphi$ (right) measured in \PbPb\   collisions  as a function of collision centrality class. Vertical bars and solid lines represent statistical and systematic  uncertainties, respectively.  Mean and RMS $\Delta \varphi$  widths (right: blue circles) were computed in the range $-\pi \le \Delta\varphi \le \pi$ with an offset according to Eq.~(\ref{Eq:widthR2Phi}).  Red symbols   show RMS $\Delta\eta$ and $\Delta\varphi$ widths (systematic uncertainties shown as red dashed lines) reported by a prior ALICE analysis based on measurements of balance functions~\cite{Abelev:2013csa}. The $\Delta\varphi$ widths reported in this earlier work were computed in the range  $-\pi/2 \le \Delta\varphi \le \pi/2$.  }
\label{Fig:PbPb_R2CD_WidthVsCent}
\end{figure}

\begin{figure}[h!]
\includegraphics[width=0.46\linewidth]{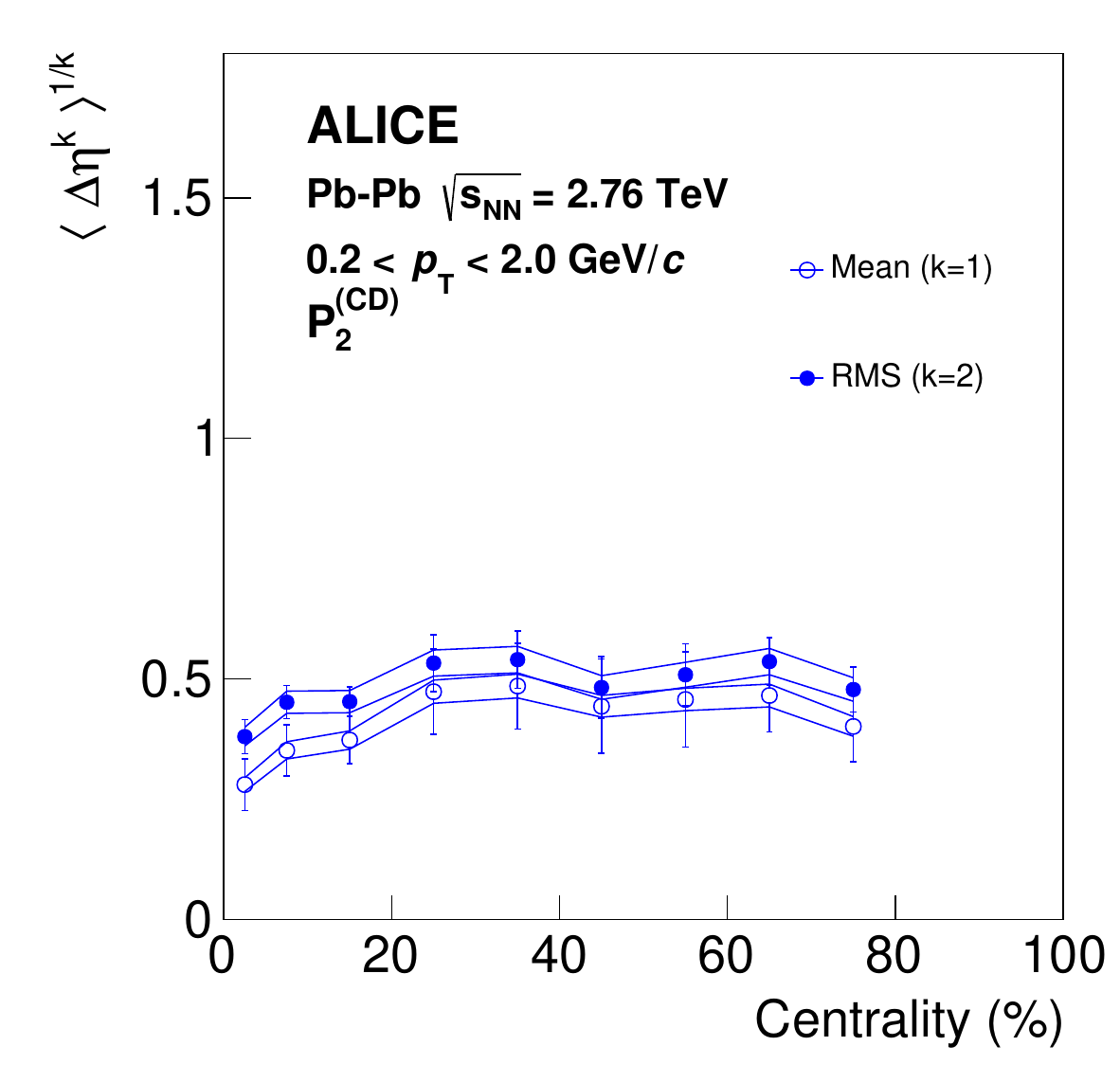}
\includegraphics[width=0.46\linewidth]{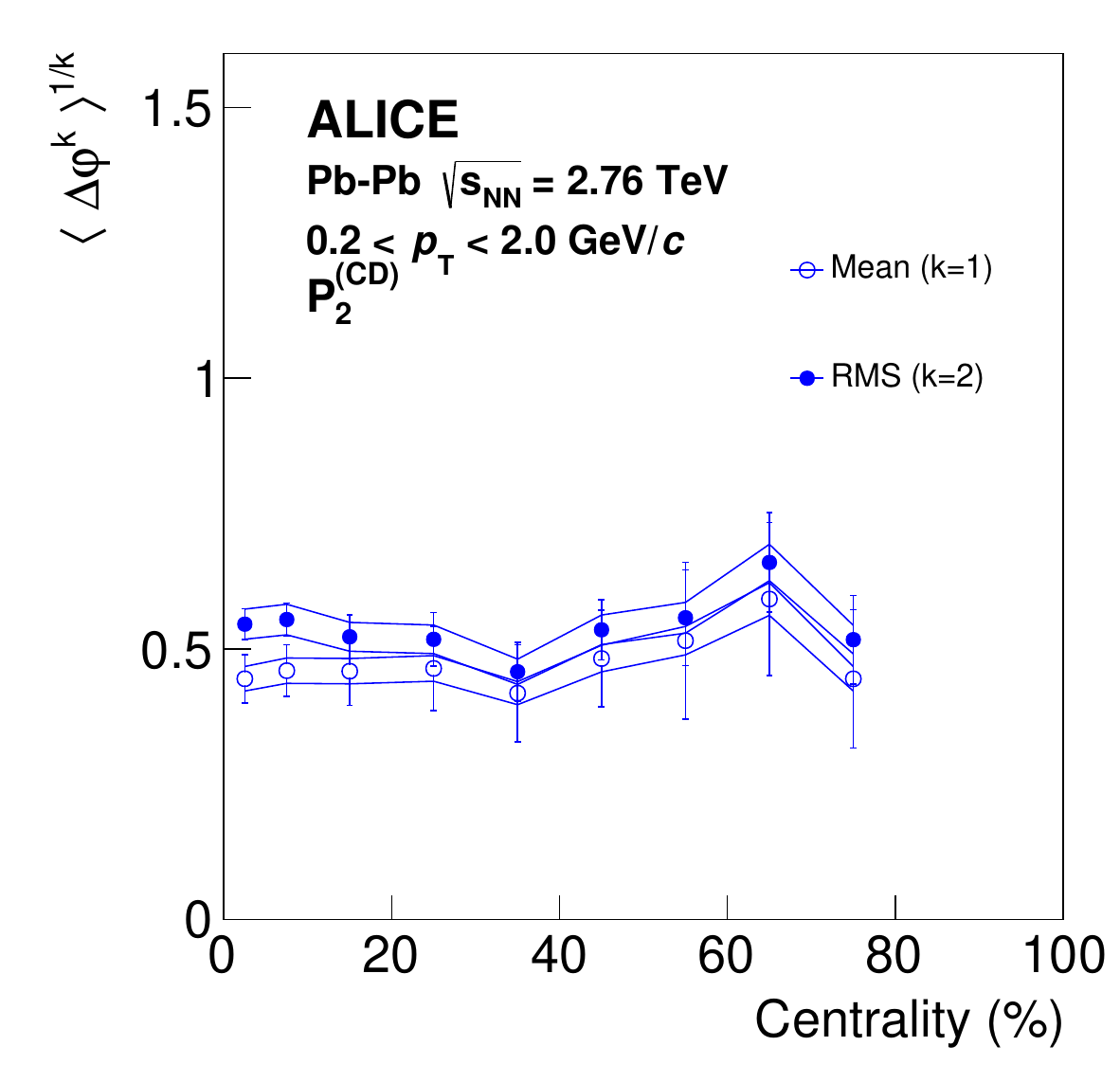}
\caption{Width of the near-side peak of $P_2^{\rm (CD)}$  correlation functions  along $\Delta \eta$ (left) and   $\Delta \varphi$  (right) measured in \PbPb\    collisions  as a function of collision centrality class. Vertical bars and solid lines represent statistical and systematic  uncertainties, respectively.  }
\label{Fig:PbPb_P2CD_WidthVsCent}
\end{figure}

\begin{figure}[h!]
\includegraphics[width=0.46\linewidth]{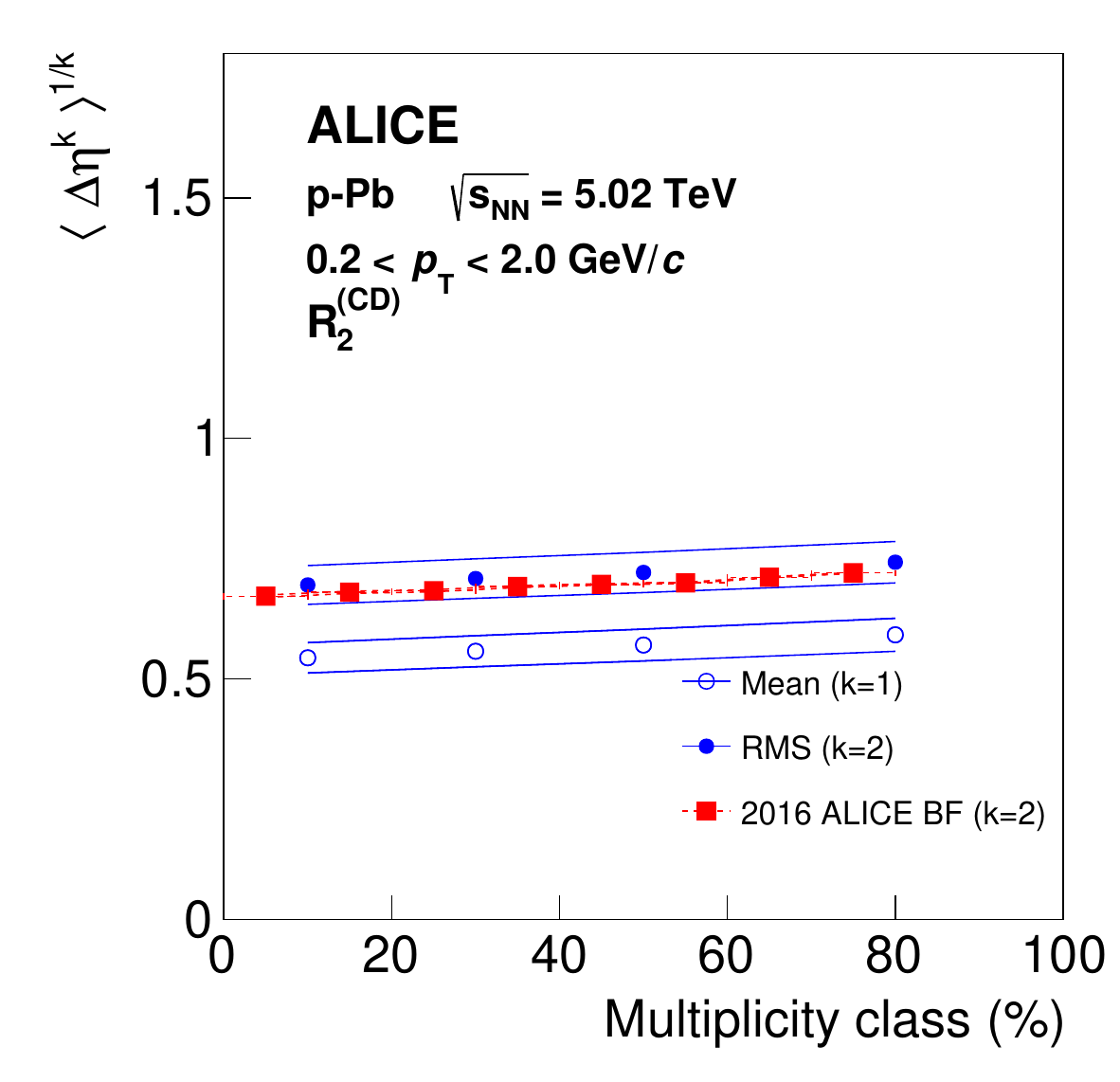}
\includegraphics[width=0.46\linewidth]{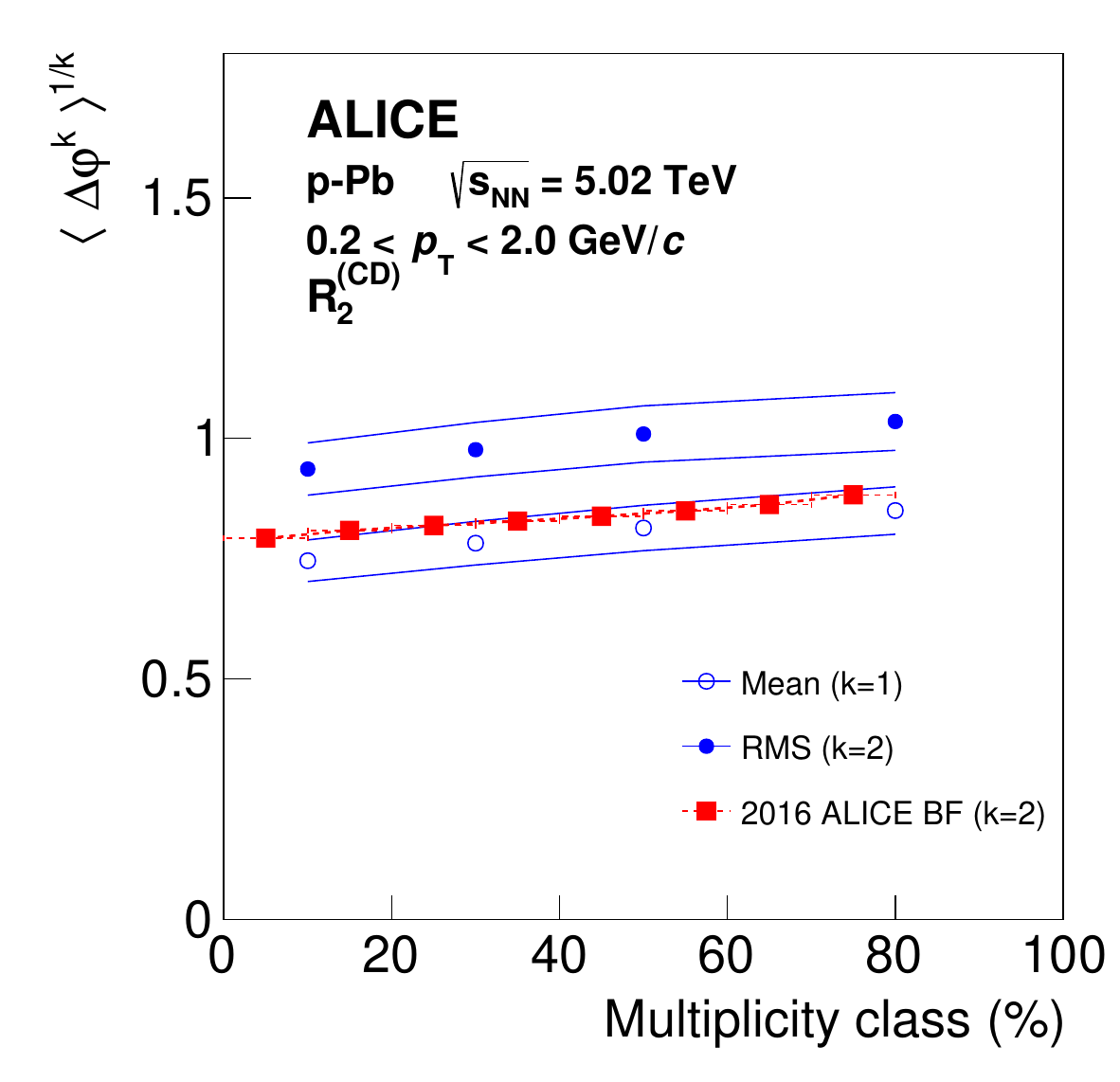}
\caption{Width of the near-side peak of $R_2^{\rm (CD)}$  correlation functions  along $\Delta \eta$ (left) and   $\Delta \varphi$ (right)   measured in    \pPb\     collisions  as a function of produced particle multiplicity class. Vertical bars and solid lines represent statistical and systematic  uncertainties, respectively. Mean and RMS $\Delta \varphi$  widths (right: blue circles) were computed in the range $-\pi \le \Delta\varphi \le \pi$ with an offset according to Eq.~(\ref{Eq:widthR2Phi}).  Red symbols   show RMS $\Delta\eta$ and $\Delta\varphi$ widths (systematic uncertainties shown as red dashed lines) reported by a prior ALICE analysis based on measurements of balance functions~\cite{Abelev:2013csa}. The $\Delta\varphi$ widths reported in this earlier work were computed in the range  $-\pi/2 \le \Delta\varphi \le \pi/2$.}
\label{Fig:pPb_R2CD_WidthVsCent}
\end{figure}

\begin{figure}[h!]
\includegraphics[width=0.46\linewidth]{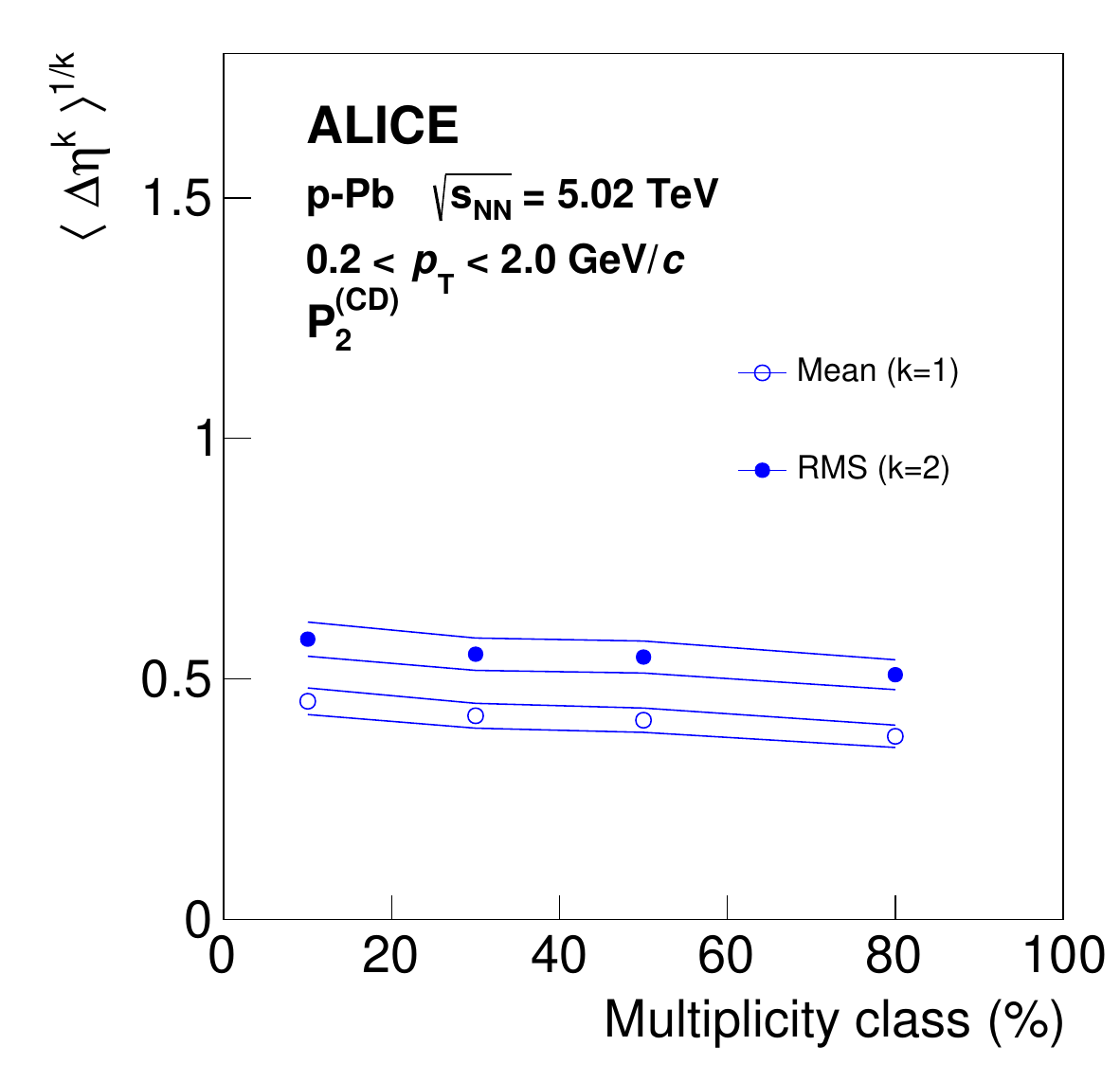}
\includegraphics[width=0.46\linewidth]{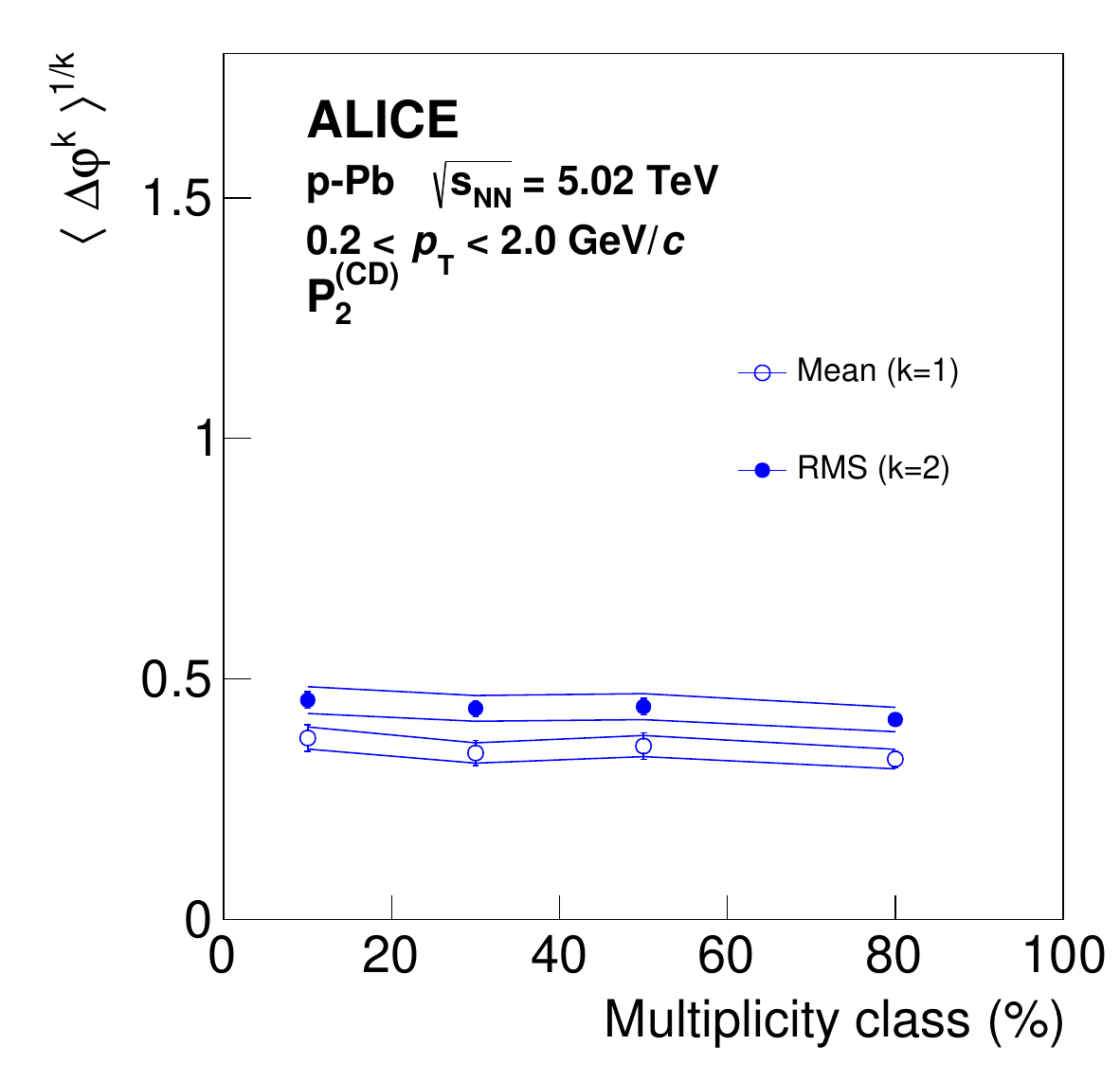}
\caption{Width of the near-side peak of $P_2^{\rm (CD)}$  correlation functions  along $\Delta \eta$ (left) and   $\Delta \varphi$ (right)  measured in    \pPb\    collisions  as a function of produced particle multiplicity class. Vertical bars and solid lines represent statistical and systematic  uncertainties, respectively. }
\label{Fig:pPb_P2CD_WidthVsCent}
\end{figure}

The longitudinal widths of the near-side peaks of $R_2^{\rm (CI)}$ and $P_2^{\rm (CI)}$ correlation functions are presented in Figs. \ref{Fig:PbPb_R2CI_P2CI_WidthVsCent} --\ref{Fig:pPb_R2CI_P2CI_WidthVsCent} as a function of collision centrality and multiplicity class, respectively, while the longitudinal and azimuthal widths of near-side peak of $R_2^{\rm (CD)}$ and $P_2^{\rm (CD)}$ are displayed in Figs. \ref{Fig:PbPb_R2CD_WidthVsCent}--\ref{Fig:pPb_P2CD_WidthVsCent}.

The widths  $\la\Delta\eta^k\ra^{1/k}$  of $R_2^{\rm (CI)}$  (Fig.~\ref{Fig:PbPb_R2CI_P2CI_WidthVsCent}) grow monotonically in
\PbPb\ collisions from 70--80\%  to 0--5\% multipicity classes, reaching a maximum in the 5\% most central collisions. A similar monotonic increase is observed for $P_2$, except for the 70--80\% multiplicity class.
By contrast, in    \pPb\    collisions (Fig.~\ref{Fig:pPb_R2CI_P2CI_WidthVsCent}), the longitudinal widths of the near-side peak of $R_2^{\rm (CI)}$
and $P_2^{\rm (CI)}$ have rather weak dependence, if any, on multiplicity. These different dependences may in part be attributed to diffusion processes, expected  to play a larger role in the longer lived systems created in  more central  \PbPb\  collisions~\cite{AbdelAziz:2006jv,Gavin:2016jfw}. However,   the formation of long-range color tubes or strings  compounded with radial flow may also play an important role in the observed longitudinal broadening of the near-side peak of the $R_{2}^{\rm (CI)}$ and $P_{2}^{\rm (CI)}$ correlation functions~\cite{Moschelli:2009bk}.
Interestingly, the longitudinal widths  $\la\Delta\eta^k\ra^{1/k}$ observed for  $P_2^{\rm (CI)}$ are significantly smaller than those observed for $R_2^{\rm (CI)}$ in both \PbPb\ and     \pPb\    collisions. Charge-dependent correlation functions are expected to have a different sensitivity to particle correlations than charge-independent correlations. This is readily verified in Figs.~\ref{Fig:PbPb_R2CD_WidthVsCent}--\ref{Fig:pPb_P2CD_WidthVsCent}, that display the near-side peak width of  CD  correlations, measured in both \PbPb\  and  \pPb,  as a function of produced particle multiplicity classes. One finds that in contrast to CI correlations whose near-side peaks width increase  with produced particle multiplicity, the widths of the near-side peak of  $R_2^{\rm (CD)}$ correlation functions  monotonically decrease with increasing multiplicity.

The widths measured in this work, shown with solid blue circles for $k=2$ (RMS) and open blue circles for $k=1$ (one-sided mean) in Figs.~\ref{Fig:PbPb_R2CD_WidthVsCent} and~\ref{Fig:pPb_R2CD_WidthVsCent}, are compared with RMS values of the longitudinal
and azimuthal widths, shown in red, of the balance function reported by the ALICE collaboration~\cite{Abelev:2013csa}. One observes that the RMS widths, $\la\Delta\eta^2\ra^{1/2}$, obtained in this work are in very good agreement with the longitudinal RMS  values reported for the balance function. A similar trend with collision centrality is
observed for the RMS width, $\la\Delta\varphi^2\ra^{1/2}$, albeit with a finite offset owing to differences in the RMS calculation methods used in this and the prior work. In this work, an offset, evaluated at the minimum of the $\Delta\varphi$ projection is used and the RMS calculation is performed in the range $-\pi \le \Delta\varphi \le \pi$, whereas the widths reported in~\cite{Abelev:2013csa} were evaluated without the use of an offset and in the range $-\pi/2 \le \Delta\varphi \le \pi/2$.

The $R_{2}^{\rm (CD)}$ distributions measured in \PbPb\  exhibit a
strong reduction  from peripheral to central while the widths
measured in  \pPb\   show a weaker but nonetheless noticeable
reduction with increased charged-particle production. Multiplicity class  dependences of the widths of the near-side peak of $P_2^{\rm (CD)}$ correlations are more difficult to assess owing to larger statistical and systematic uncertainties: measurements in \PbPb\  are consistent with a modest decrease with increasing collision centrality, whereas those in   \pPb\   suggest a reverse trend.

\begin{figure}[h!]
\includegraphics[width=0.46\linewidth]{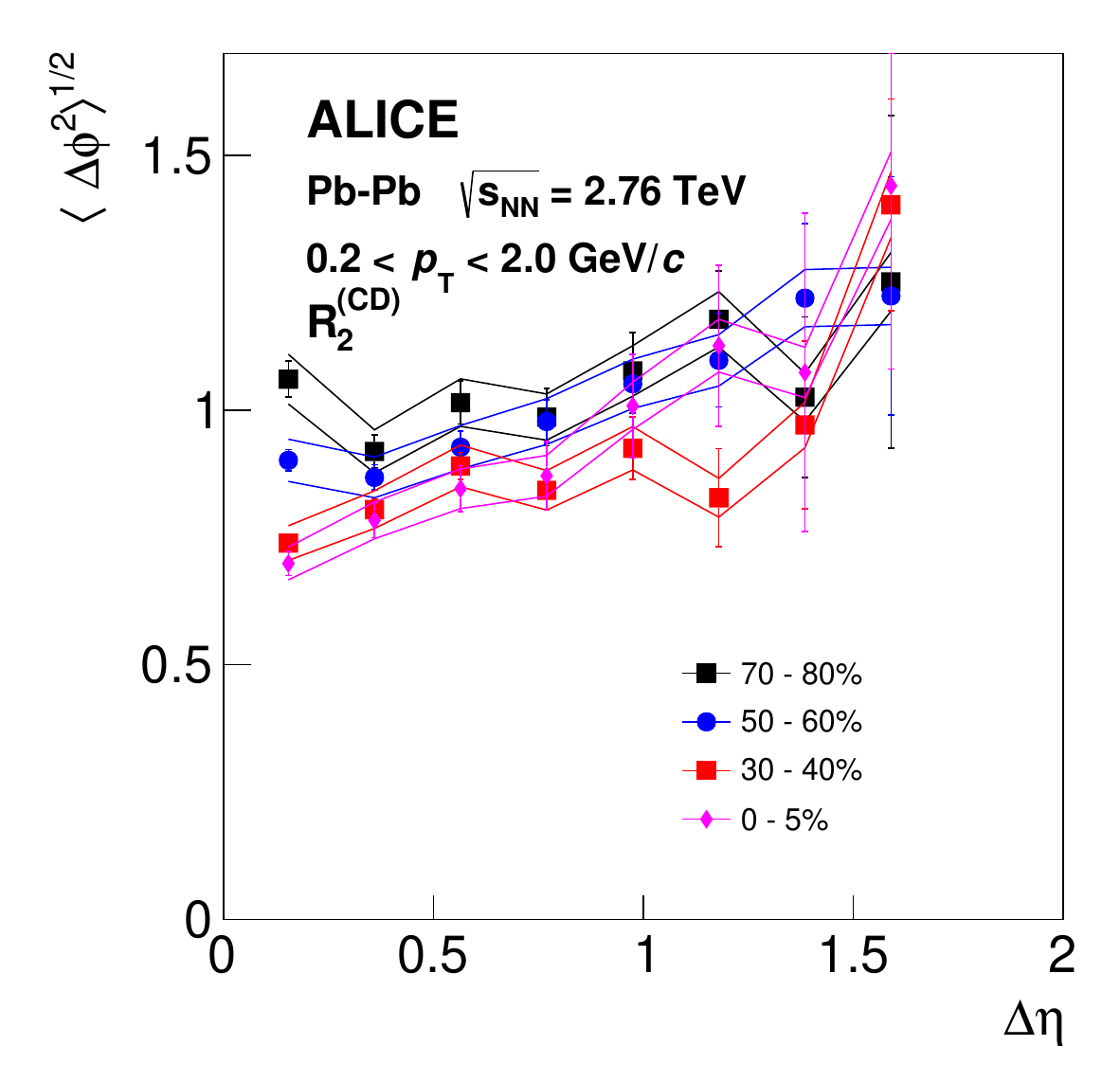}
\includegraphics[width=0.46\linewidth]{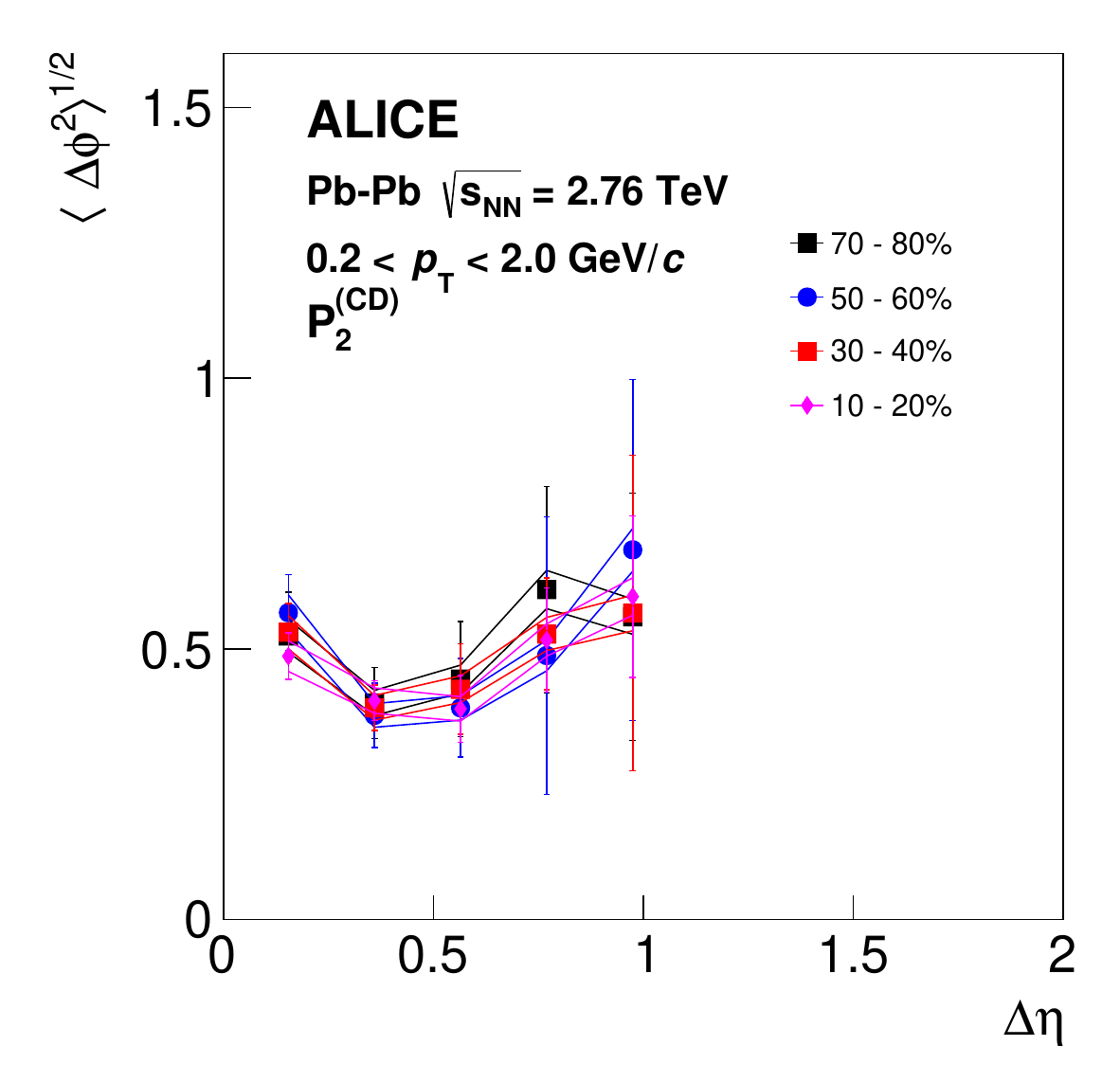}
\caption{Width of the near-side peak of $R_2^{\rm (CD)}$  (left) and  $P_2^{\rm (CD)}$  (right) correlation functions  along $\Delta \varphi$  measured in
\PbPb\   collisions  as a function of the $\Delta \eta$ pair separation for selected collision centralities. Vertical bars and solid lines represent statistical and systematic  uncertainties, respectively. }
\label{Fig:PbPb_R2_P2_WidthVsEtaVsCent}
\end{figure}

\begin{figure}[h!]
\includegraphics[width=0.46\linewidth]{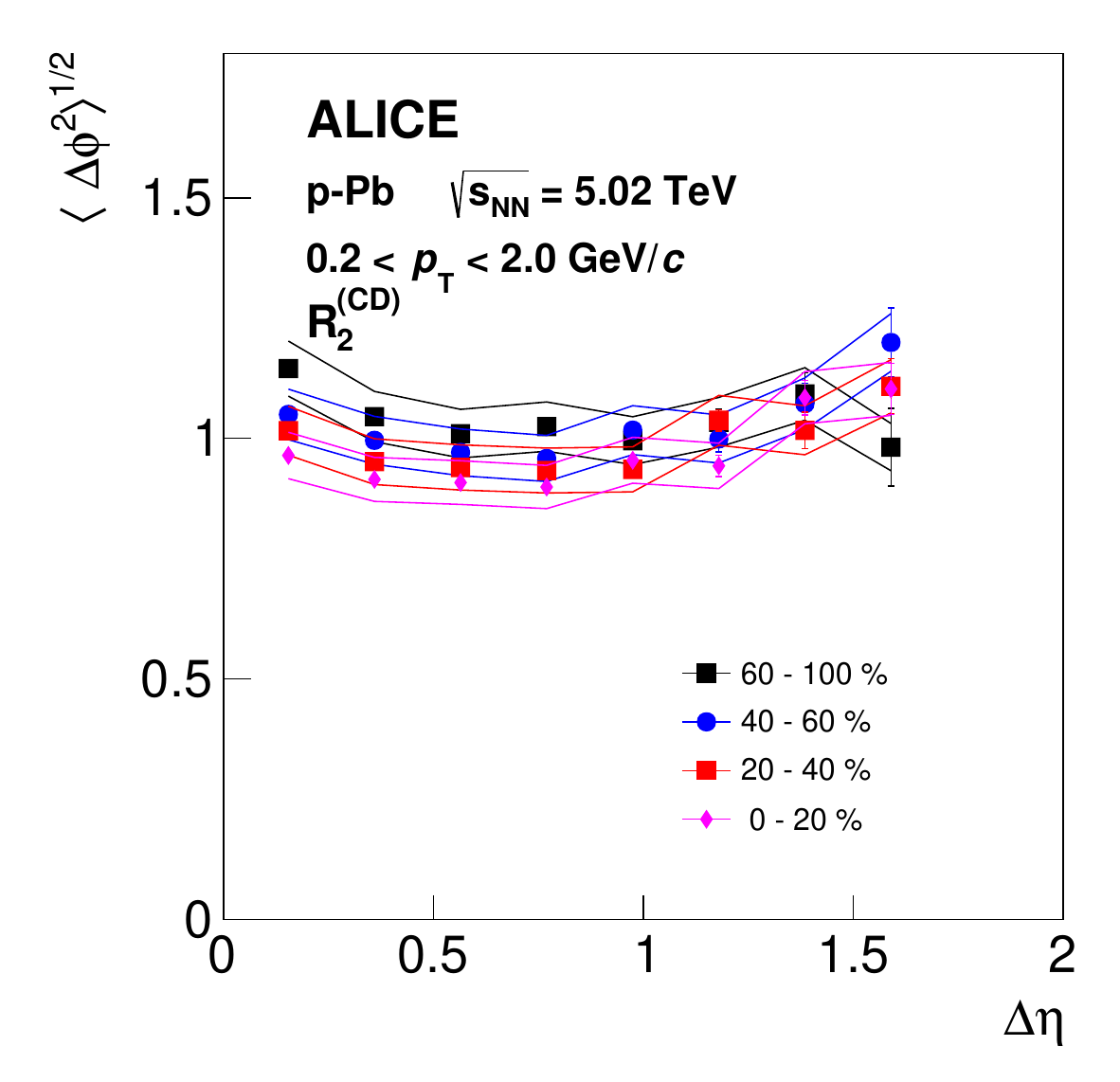}
\includegraphics[width=0.46\linewidth]{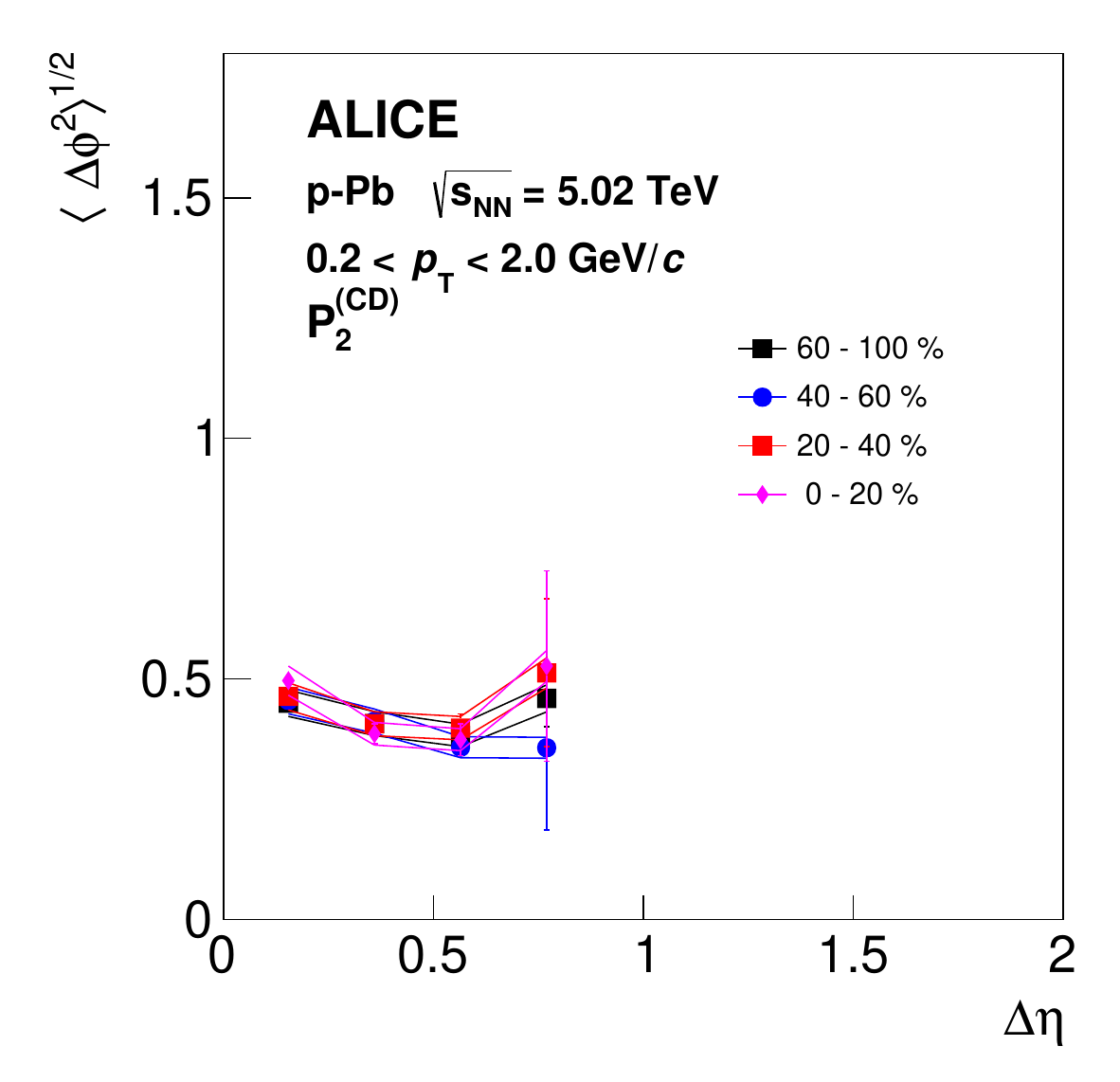}
\caption{Width of the near-side peak of $R_2^{\rm (CD)}$ (left) and  $P_2^{\rm (CD)}$ (right) correlation functions  along $\Delta \varphi$   measured in  \pPb\    collisions  as a function of the $\Delta \eta$ pair separation for selected ranges of produced multiplicities. Vertical bars and solid lines represent statistical and systematic  uncertainties, respectively.}
\label{Fig:pPb_R2_P2_WidthVsEtaVsCent}
\end{figure}

The reductions of the longitudinal and azimuthal widths of the near-side peak of  $R_2^{\rm (CD)}$ observed in \PbPb\  and  \pPb\   collisions are in agreement with prior measurements  (both at RHIC and LHC) and are qualitatively consistent  with the presence of strong radial flow and the existence of two-stage emission in these collisions, particularly, in \PbPb\  collisions. However, one must also consider the role of diffusion processes, which for longer system lifetimes, would produce a broadening of the $R_2^{\rm (CD)}$
correlations. Traditional collision centrality dependent analyses of the width of balance functions or $R_2^{\rm (CD)}$ do not readily enable separation of the diffusion process, radial flow, and two-stage hadronization. However, the longitudinal (rapidity) expansion of the system might provide a useful clock towards the evaluation of azimuthal diffusion processes. As the system expands  longitudinally,  scatterings within the QGP phase would produce a progressive broadening of the
CD correlation functions in $\Delta\varphi$. It thus becomes of interest to study whether there is evidence for larger diffusion at progressively wider $\Delta\eta$ separations. Figures \ref{Fig:PbPb_R2_P2_WidthVsEtaVsCent}--\ref{Fig:pPb_R2_P2_WidthVsEtaVsCent} display the azimuthal RMS width, $\la \Delta\varphi^2 \ra^{1/2}$, measured in selected collision centrality ranges (\PbPb) and multiplicity classes (\pPb), as a function of the pair separation $\Delta\eta$. First note that in both  \pPb\   and peripheral \PbPb\  collisions, the presence of a strong HBT component leads to small or even negative values of $R_2^{\rm (CD)}$ and
$P_2^{\rm (CD)}$ at short pair separations in $\Delta\eta$ thereby creating a  depletion  near $\Delta\eta,\Delta\varphi=0$ in plots of this correlator  vs. $\Delta\eta, \Delta\varphi$. This depression effectively pushes outward, in $\Delta\varphi$, the value of the azimuthal width of the distribution thereby leading to enhanced values of $\la \Delta\varphi \ra$
for short pair separations (i.e., $\Delta\eta<0.5$). However, the HBT contribution to $R_2^{\rm (CD)}$ is   very narrow and not resolved by this measurement in mid to central \PbPb\  collisions. It consequently does not appreciably contribute to the calculation of the $\Delta\varphi$
widths in the  0--50\% centrality interval. The  width of $R_2^{\rm (CD)}$ in these mid-to-central collisions is thus believed to be dominated by charge conserving particle production processes and the evolution dynamics of the collision systems. In the 0--5\% and  30--40\% collision centralities, one finds that the RMS  width  $\la \Delta\varphi^2 \ra^{1/2}$ is in fact smallest at shortest pair separation and essentially monotonically grows with increasing pair separation. The growth for 0--5\% collisions can be approximately described with a function of the form $a + b \Delta\eta^{1/2}$ which suggests that the observed width dependence is compatible with a naive model of the diffusion process. Indeed, the azimuthal width of the correlation peak should qualitatively grow as the  power $1/2$ of the lifetime of the system, i.e., $\tau^{1/2}$, which in turn, should be roughly proportional to $\Delta \eta^{1/2}$ for sufficiently large separations. However, a fit with a linear function $a{'} + b{'} \Delta\eta$
produces a $\chi^2/dof$ of similar magnitude as the  $\Delta\eta^{1/2}$ fit. It is thus not possible, with this measurement, to precisely assess the
$\Delta\varphi$ broadening dependence on the pair separation in  $\Delta \eta$.  While the measured evolution   of the $R_2^{\rm (CD)}$ $\Delta\varphi$ width  with pair separation might indicate  the presence of diffusion processes, it might also be attributable to  radial flow effects~\cite{pruneauInPrep}.
Hydrodynamic models of the evolution of heavy-ion collisions and blast-wave fits of \AuAu\  and \PbPb\  data reveal the presence of significant radial flow with  velocity profiles
dependent on the point of origin of the produced particles~\cite{Kolb:2003dz,Retiere:2003kf}. Given balanced charged-particle pairs originate from a common production mechanism such as  resonance decays or string fragmentation, the pair separation in $\Delta \eta$ and $\Delta \varphi$  is thus expected to decrease with the outward radial
velocity of the source. Slow sources shall produce large pair separations in $\Delta\eta$ and $\Delta\varphi$, on average, while larger radial velocity will produce significantly smaller $\Delta\eta$ and $\Delta\varphi$ separations. In effect, differential flow profiles shall yield, overall, $\Delta\varphi$ widths  that
increase with the pair separation in  $\Delta\eta$. The observed dependence of $\Delta\varphi$ widths on pair separation might then in part result from
radial flow, diffusion, and possibly other effects~\cite{pruneauInPrep}.  A proper assessment of these contributions shall thus require  model studies beyond the scope of this work.

\subsection{Fourier decompositions of $R_2$ and $P_{2}$ correlation functions}
\label{Sec:fd}


Correlation analyses based on multi-particle cumulants, including the scalar-product,  $Q$-distribution, Lee-Yang Zeros, and Fourier-Bessel Transforms methods, have established the presence of strong collective anisotropic flow in \AuAu\ and \PbPb\ collisions~\cite{Adams:2005dq,Adcox:2004mh,Back:2004je,Voloshin:2008dg,Aamodt:2010pa}, and recent multi-particle correlation analyses suggest that collective behavior  might also play an important role in \pPb\ and \pp\ collisions~\cite{Aad:2012gla,Khachatryan:2010gv,Chatrchyan:2013nka,Aad:2013fja,Bozek:2010pb,Bozek:2012gr,Bozek:2013uha,dEnterria:2010xip,Abelev:2014mda}. However, non-collective particle production
mechanisms, including resonance decays, jets, and other non-flow effects, are also known to contribute to
correlation functions, particularly at small particle pair separation in (pseudo)rapidity  and in small collision systems. One studies the interplay of flow and non-flow effects by carrying out Fourier  decomposition of the $\Delta\varphi$ dependence of $R_2(\Delta\eta,\Delta\varphi)$ and $P_{2}(\Delta\eta,\Delta\varphi)$ as a function of the pair separation $|\Delta\eta|$. Flow coefficients $v_n[R_2]$ and $v_n[P_2]$, calculated according to Eqs.~(10,11) are reported for \PbPb\ collisions, whereas harmonic coefficients $b_n[R_2]$ and $b_n[P_2]$ are reported for \pPb\ collisions.

\begin{figure}[h!]
\includegraphics[width=0.96\linewidth]{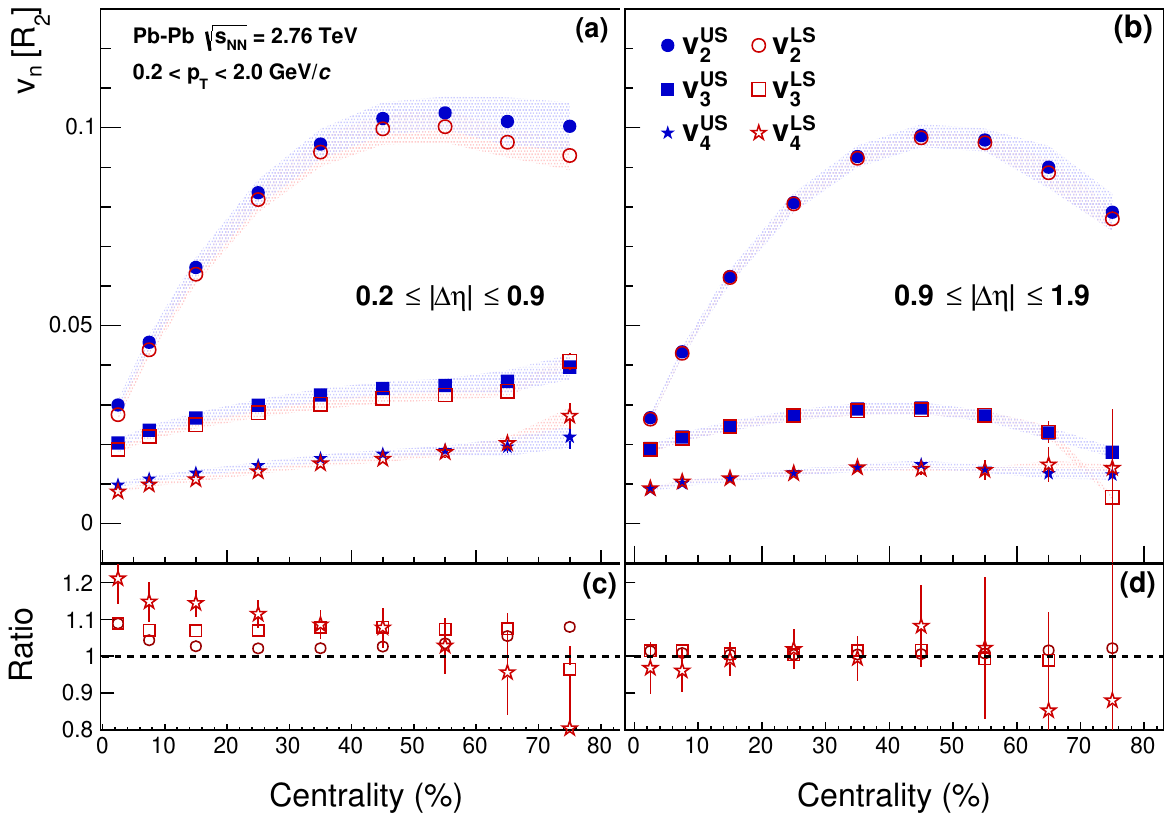}
\caption{Fourier coefficients $v_{n}$, with $n=2,3,4$,
extracted from  US and LS  $R_{2}$ correlation functions in the range
$0.2 \leq |\Delta\eta| \leq 0.9$ and $0.9 \leq |\Delta\eta| \leq 1.9$ in panels
(a) and (b), respectively. The ratios between US and LS $v_{n}$
coefficients are shown in panels (c) and (d). Vertical bars and shaded areas represent statistical and systematic  uncertainties, respectively.}
\label{Fig:vnR2_lsus_PbPb}
\end{figure}

\begin{figure}[h!]
\includegraphics[width=0.96\linewidth]{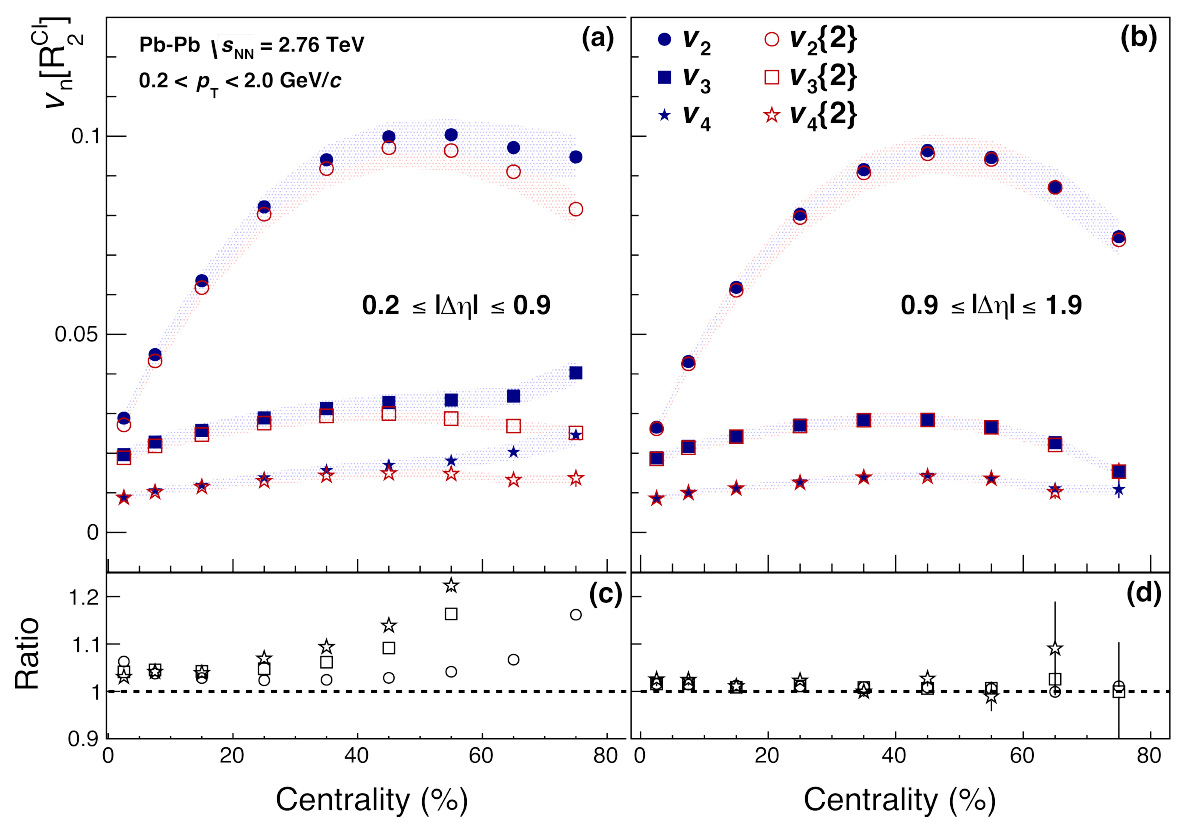}
\caption{Solid symbols: coefficients $v_{n}$,   $n=2,3,4$,
obtained from Fourier decompositions of  charge-independent correlators, $R_{2}^{\rm (CI)}$,
in the ranges $0.2 \leq |\Delta\eta| \leq 0.9$ (left) and $0.9 \leq |\Delta\eta| \leq 1.9$ (right). Open symbols;  flow coefficients $v_{n}$ obtained with the scalar-product  method according to Eq.~(\ref{Eq:mth_sp}). Panels  (c) and (d): Ratios of the coefficients $v_{n}$  values obtained from $R_{2}^{\rm (CI)}$ to those obtained with  the scalar-product method. Vertical  bars and shaded areas indicate statistical and systematic uncertainties, respectively.}
\label{Fig:vnR2_Vs_EP}
\end{figure}

\begin{figure}[h!]
\includegraphics[width=0.96\linewidth]{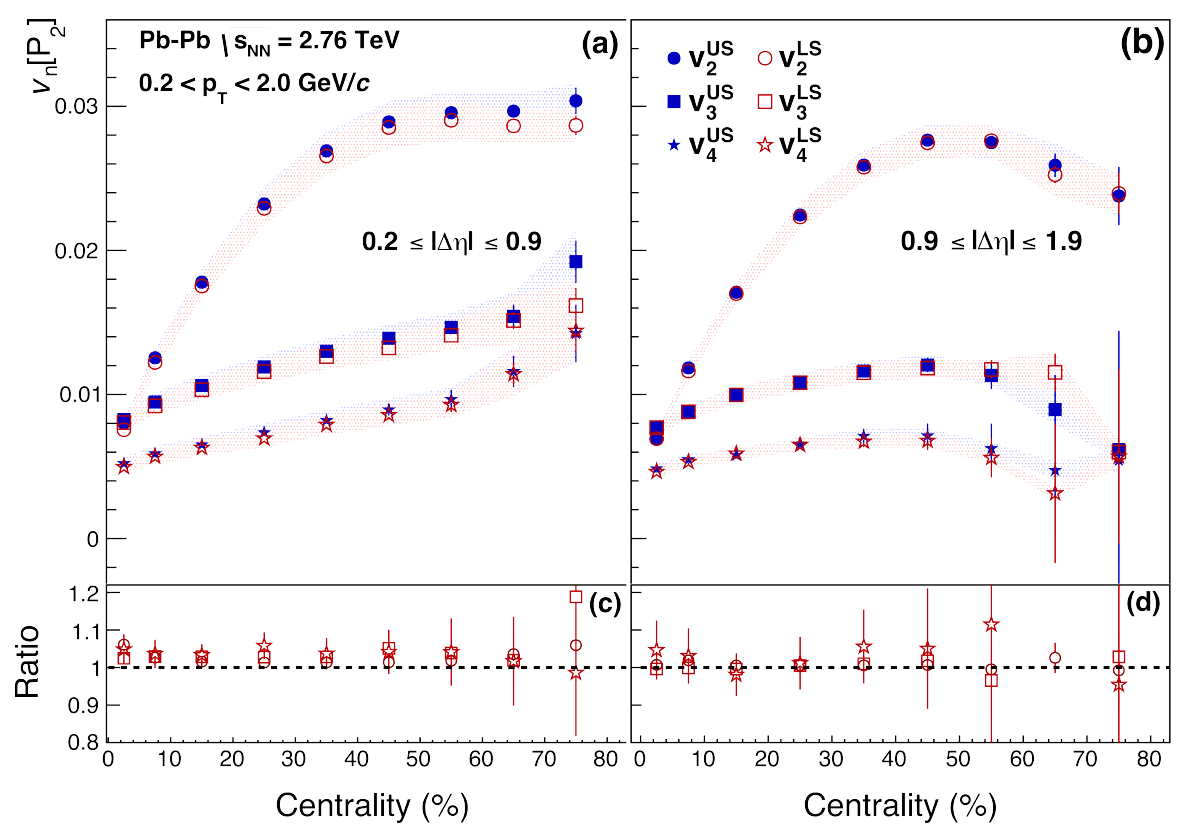}
\caption{Fourier  coefficients with $n=2,3,4$
obtained in $P_{2}$ for US and LS charge-correlations in the ranges
$0.2 \leq |\Delta\eta| \leq 0.9$ and $0.9 \leq |\Delta\eta| \leq 1.9$ in panels
(a) and (b), respectively. The ratios between US and LS $v_{n}$
coefficients are shown in panels (c) and (d). Vertical  bars and shaded areas indicate statistical and systematic uncertainties, respectively.}
\label{Fig:vnP2_lsus_PbPb}
\end{figure}
Figure~\ref{Fig:vnR2_lsus_PbPb} presents the $v_{n}$ coefficients,  $n=2,3,4$, (defined in Sec.~\ref{Sec:scalarProduct}) plotted as a function of  \PbPb\  collision centrality,
obtained from projections of  $R_{2}^{\rm (US)}$ and $R_{2}^{\rm (LS)}$, in ranges  $0.2 \leq |\Delta\eta| \leq 0.9$ and  $0.9 \leq |\Delta\eta| \leq 1.9$. One observes that the $v_{n}[R_{2}]$ coefficients obtained from US and LS correlations are
essentially identical at ``large'' $|\Delta\eta|$ (i.e., $|\Delta\eta| \geq 0.9$). Aside from weak Coulomb distortions~\cite{Pratt:2003gh}, one expects that two-particle correlations determined by collective behavior to be essentially independent of the charge of the particles. The near perfect agreement between LS and US Fourier coefficients of order 2, 3, and 4 is thus an indication that non-flow effects, which might exhibit explicit dependences on charges, are rather weak for pair separations in excess of $|\Delta\eta| = 0.9$.
The observed azimuthal coefficients at $|\Delta\eta| > 0.9$ are thus  consistent with the dominance of
collective flow effects in this range. The US and LS coefficients obtained for pairs with $0.2 \le |\Delta\eta| \le 0.9$, on the other hand, exhibit systematic discrepancies at all collision centralities. Considering the ratio of US and LS coefficients plotted in the lower panel of  Fig.~\ref{Fig:vnR2_lsus_PbPb}, one observes that US $v_n$ coefficients are systematically larger than those of LS pairs. One also finds that the $v_2$ coefficients exhibit the smallest differences, while the $v_4$ coefficients have the largest. This behavior is largely driven by the presence of the stronger near-side peak observed in US $R_2$ correlations, and is thus a result of non-flow effects associated with the creation of charge particle pairs.

Figure~\ref{Fig:vnR2_Vs_EP}  compares  $v_n[R_2^{\rm (CI)}]$ coefficients,  $n=2,3,4$ (solid symbols),   extracted from $R_{2}^{\rm (CI)}$ correlation functions  with flow coefficients $v_2\{2\}$ (open symbols) obtained with the  scalar-product method  according to Eq.~(\ref{Eq:mth_sp}). The comparison is carried out in panel (a) and (b) for charged-particle pairs with   pseudorapidity separations of $0.2\le |\Delta \eta|\le 0.9$  and
$0.9\le |\Delta \eta|\le 1.9$, respectively. Panel (c) and (d) display the ratio of coefficients obtained with the two methods. One observes that the deviations between the
$v_2\{2\}$ and $v_n[R_2^{\rm (CI)}]$ for $0.9\le |\Delta \eta|\le 1.9$ are typically smaller than 2\%, irrespective of collision centrality. Such small deviations are expected given  $v_2\{2\}$ coefficients were determined with a minimal $|\Delta\eta|$ of 0.9 units of pseudorapidity.
The coefficients $v_n$ obtained from $R_2^{\rm (CI)}$, for  pair separation in excess of 0.9, are thus equivalent to those obtained with the SP method. However,
the deviations for pair separations in the range $0.2\le |\Delta \eta|\le 0.9$ are finite in all centrality classes in \PbPb\ collisions. They are smallest in central to mid-central collisions but rise in  excess of  10\% in more  peripheral collisions,  owing to the presence of the near-side peak that dominates the $R_2$ correlations in this collision centrality range.

Similarly to Fig.~\ref{Fig:vnR2_lsus_PbPb}, Fig.~\ref{Fig:vnP2_lsus_PbPb}~presents the $v_{n}$ coefficients,  $n=2,3,4$, plotted as a function of  \PbPb\  collision centrality,
obtained from projections of  $P_{2}^{\rm (US)}$ and $P_{2}^{\rm (LS)}$, in ranges  $0.2 \leq |\Delta\eta| \leq 0.9$ and  $0.9 \leq |\Delta\eta| \leq 1.9$. In this case also, one observes that US and LS $v_n$ coefficients measured for pairs in the range
$0.9 \leq |\Delta\eta| \leq 1.9$ are essentially identical, whereas coefficients for US  pairs in the range $0.2 \leq |\Delta\eta| \leq 0.9$ uniformly exceed those of LS  by about 5\% for $n=$2, 3, and 4, and at all observed centralities.

Comparing the left and right panels of  Figs.~\ref{Fig:vnR2_lsus_PbPb} and \ref{Fig:vnP2_lsus_PbPb}, one concludes that  $v_n[R_2]$ and  $v_n[P_2]$ coefficients exhibit a rather large dependence on the relative pseudorapidity of the pair.   These deviations evidently arise because of non-flow effects manifested by the presence of the strong near-side
peak centered at $\Delta\eta =0$, $\Delta\varphi=0$ observed in $R_2$ and $P_2$ correlations.  One expects  the impact of such non-flow effects on the magnitude of the $v_n$ coefficients to weaken with pair separation.

\begin{figure}[h!]
\includegraphics[width=0.96\linewidth,trim=0pt 6pt 0pt 0pt]{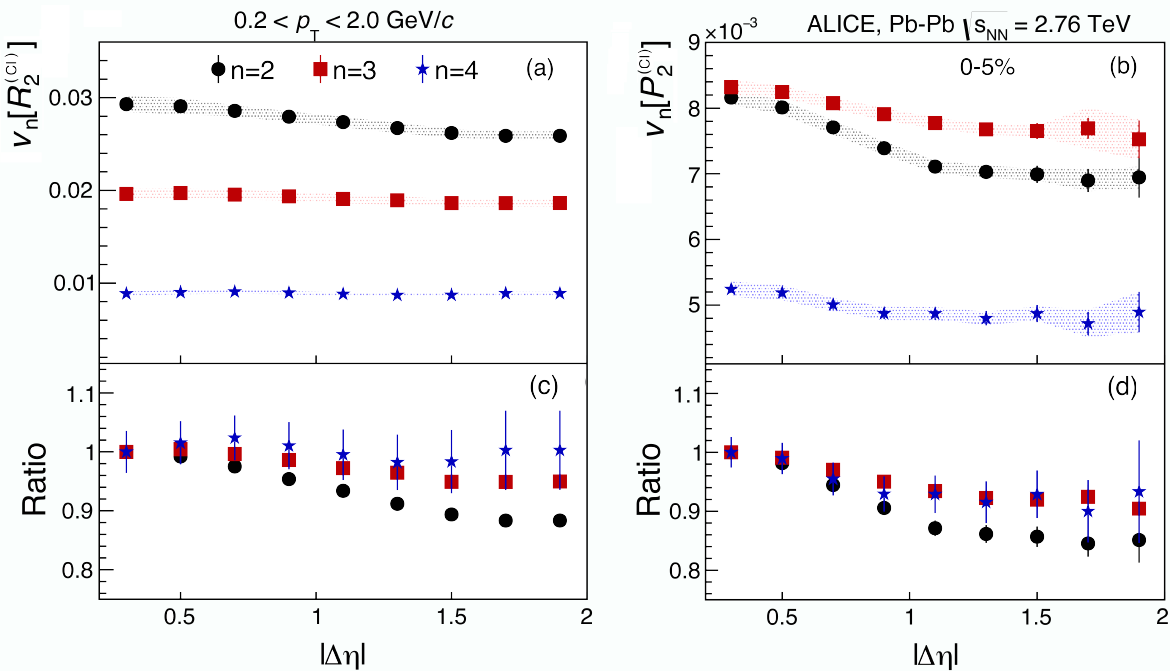}
\caption{Pair separation, $|\Delta\eta|$, dependence of Fourier coefficients
$v_{n}[R_2^{\rm (CI)}]$ (panel a) and  $v_{n}[P_2^{\rm (CI)}]$ (panel b), with $n=2,3,4$,
obtained from $R_2^{\rm (CI)}$ and  $P_2^{\rm (CI)}$ correlation functions in \PbPb\ 5\% most central collisions. Panels (c,d) display ratios of the coefficients to $v_n(|\Delta\eta|)$ to their respective values at $|\Delta\eta|=0.3$. Vertical  bars and shaded areas indicate statistical and systematic uncertainties, respectively. Reproduced from~\cite{Adam:2017ucq}.}
\label{Fig:PRL_fig3}
\end{figure}

\begin{figure}[h!]
\includegraphics[width=0.46\linewidth]{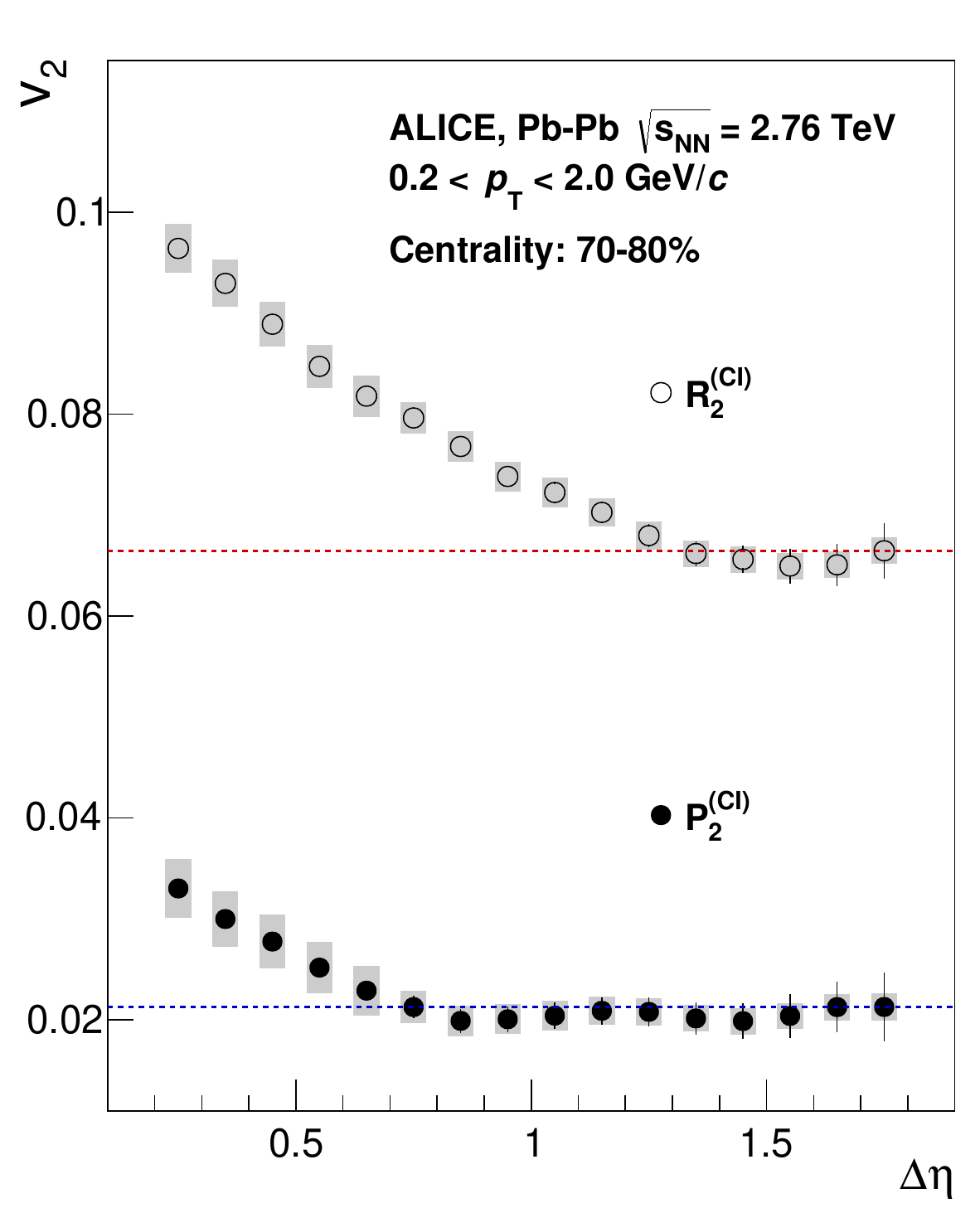}
\includegraphics[width=0.46\linewidth]{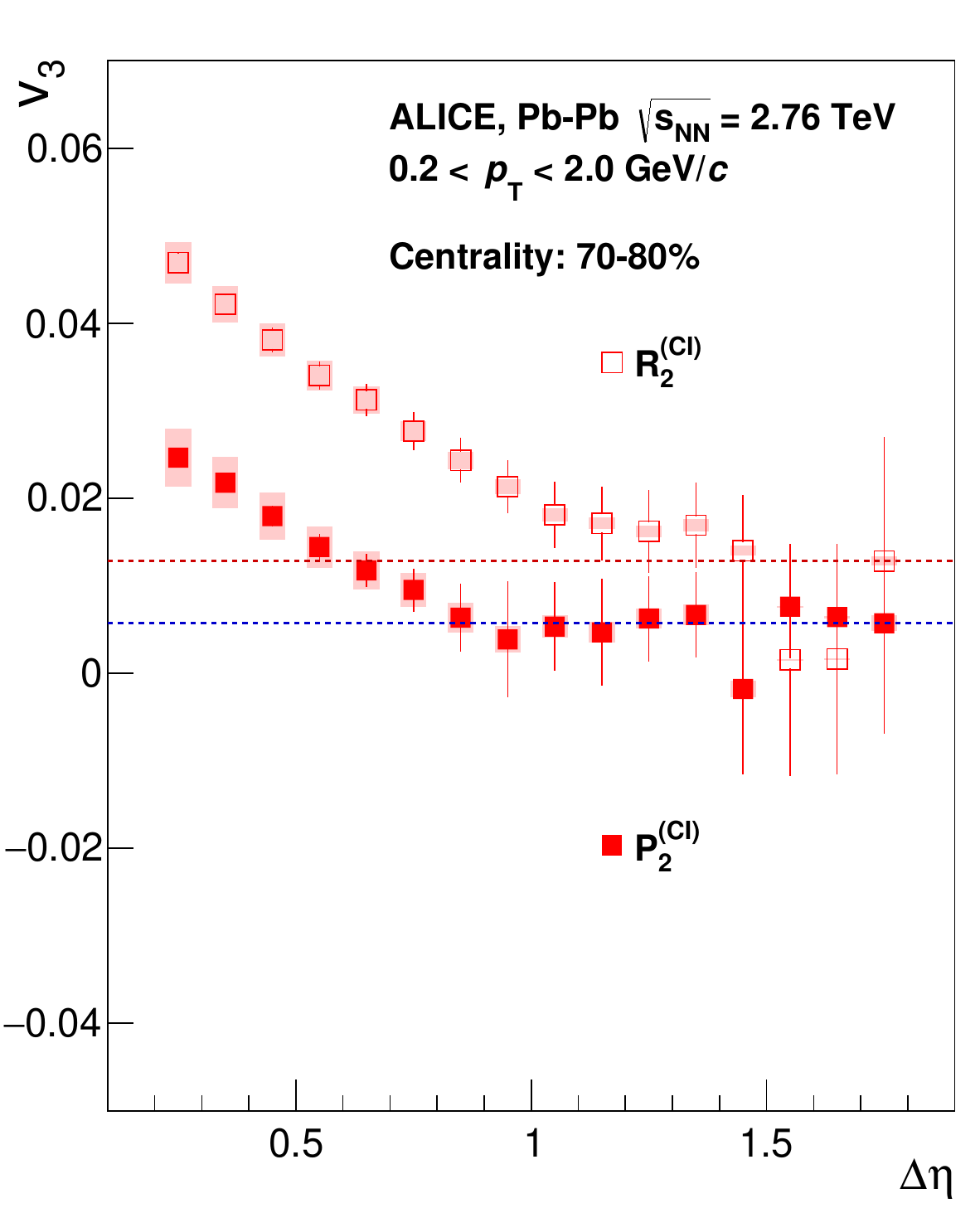}
\caption{Coefficients $v_2$ (left) and $v_3$ (right) as a function of $|\Delta\eta|$ obtained from $P_2^{\rm CI}$ and $R_2^{\rm CI}$ correlation functions in the 70--80\% centrality interval in Pb--Pb collisions. Dotted lines show baselines drawn at $v_n$($|\Delta\eta|=1.75$). Vertical bars and shaded areas indicate statistical and systematic uncertainties, respectively.}
\label{Fig:v2v3_r2VsP2_ci_7080}
\end{figure}

This is explicitly verified
by studying the magnitude of the coefficients as a function of pair separation, shown in Figs.~\ref{Fig:PRL_fig3}--\ref{Fig:v2v3_r2VsP2_ci_7080} for 0--5\% and  70--80\% \PbPb\  collisions, respectively. One observes similar trends for $v_n[R_2]$ and $v_n[P_2]$, coefficients with $n=2,3$. The coefficient amplitudes are largest at
$|\Delta\eta| \sim 0.2$ and decrease approximately linearly with increasing $|\Delta\eta|$ until they seemingly reach  plateaus.
Interestingly, one observes  that the $v_n[P_2]$ coefficients reach their plateau at $|\Delta\eta|\sim 0.7$ in peripheral collisions ($|\Delta\eta|\sim 1$ in central collisions), while the $v_n[R_2]$ coefficients do not reach a plateau until $|\Delta\eta|\sim 1.2-1.3$ ($|\Delta\eta|\sim 1.5$ in central collisions). This numerical difference is evidently due to the fact that the near-side of $P_2$ distributions are significantly
narrower than those of $R_2$ distributions, but it also shows that $P_2$ somehow  features a smaller sensitivity to non-flow. Indeed, non-flow effects in $P_2$ appear to be limited to a narrower range of $\Delta\eta$. Were it not for the fact that high-precision analyses of $P_2$ require a  larger dataset  than those of $R_2$, the suppression of non-flow effects in flow studies might be better achieved  using $\Delta \pt  \Delta \pt $ weighted observables rather than correlators simply based on the number of particles.  The difference between the $P_2$ and $R_2$ coefficients evidently also provides a new perspective and tool to investigate the near-side peak of correlation functions and the nature and origin of non-flow effects.

\begin{figure}[!ht]
\centering
\includegraphics[width=.45\linewidth]{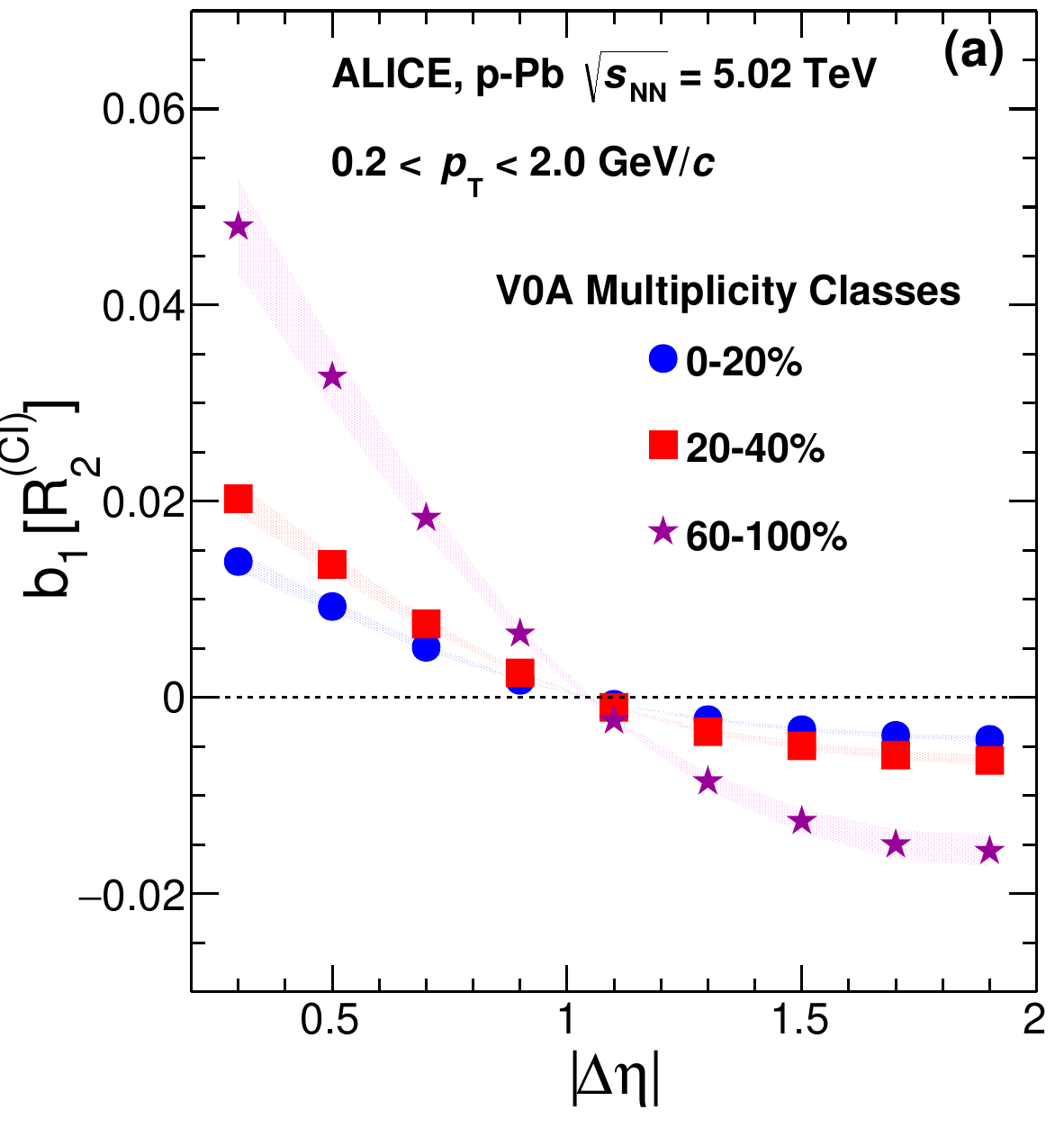}
\includegraphics[width=.45\linewidth]{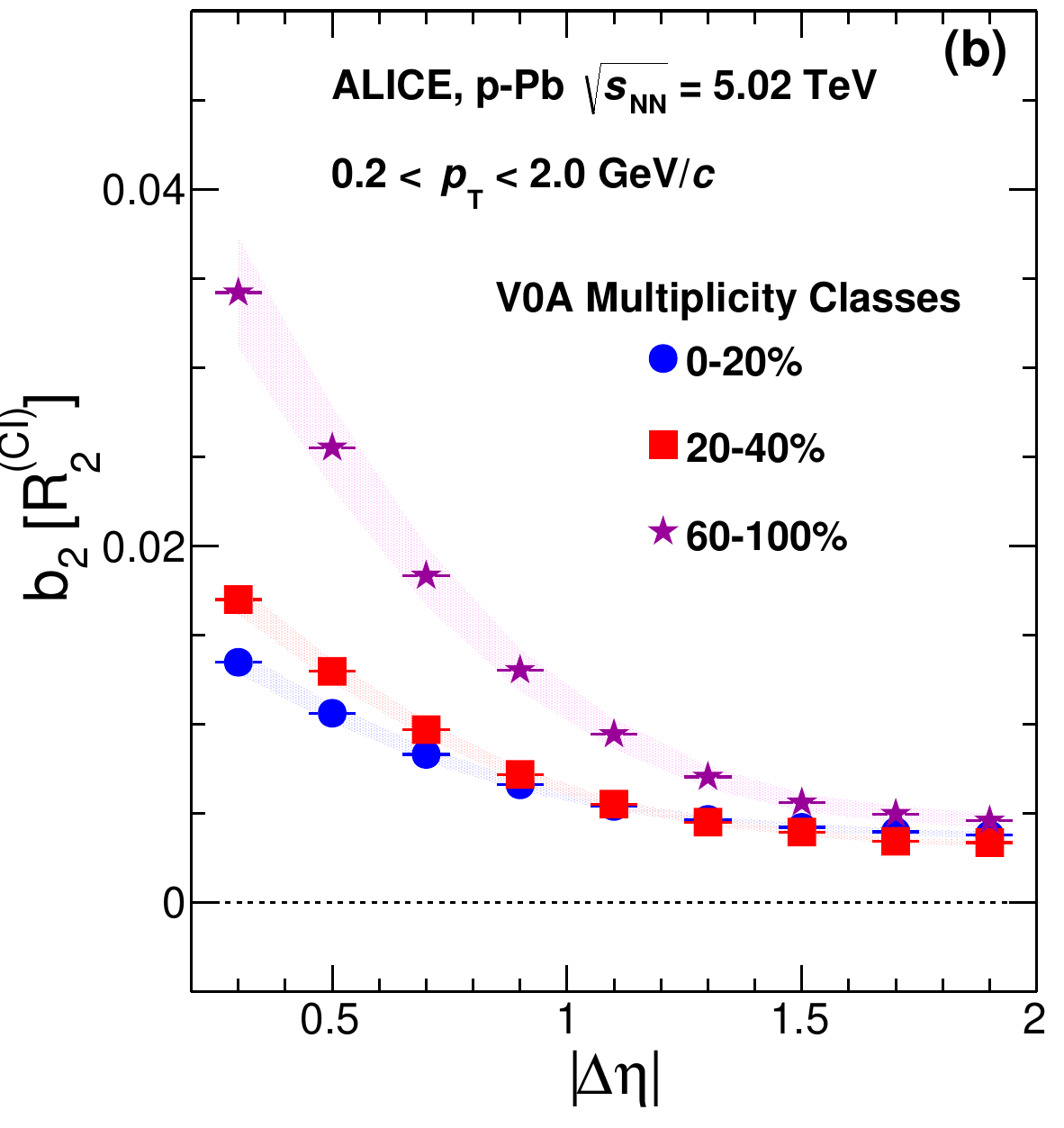}\\
\includegraphics[width=.45\linewidth]{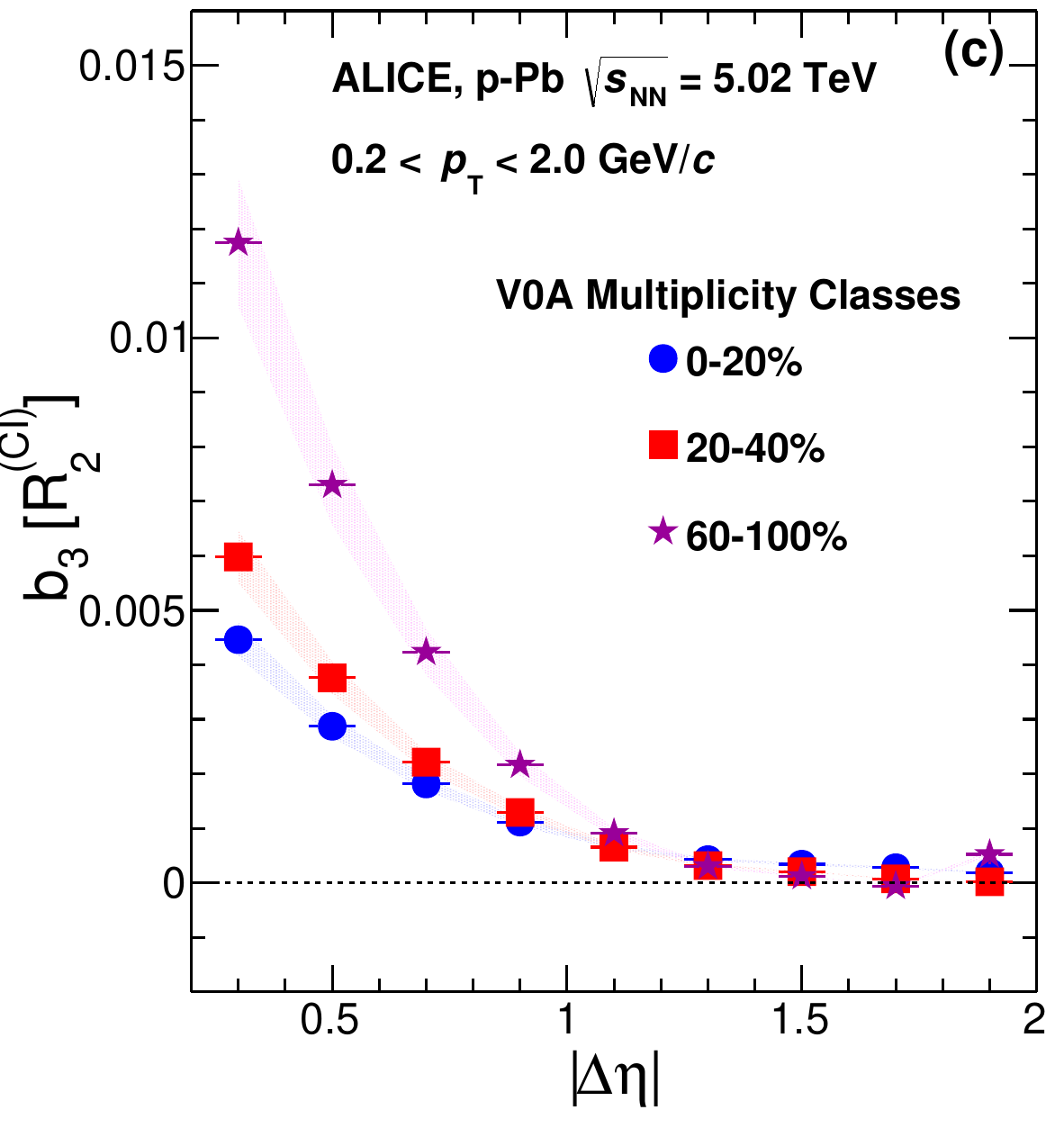}
\includegraphics[width=.45\linewidth]{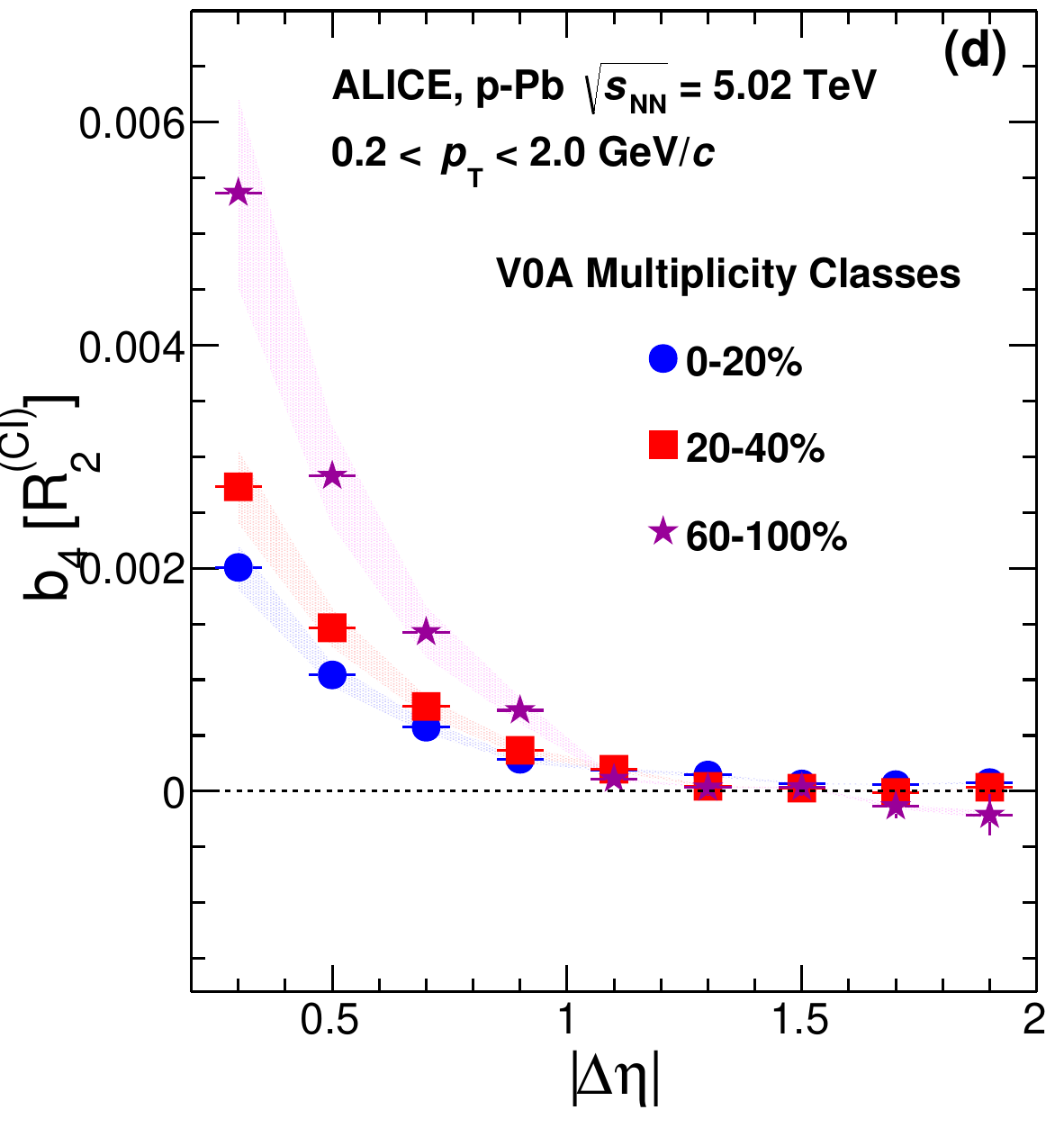}
\caption{Fourier coefficients, $b_{n}$, $n = 1,...,4$, extracted from  $R_{2}^{\rm (CI)}$ correlation functions
measured  in    \pPb\  collisions at
$\snn = $ 5.02 TeV using three multiplicity classes. Vertical bars and shaded areas represent statistical and systematic  uncertainties, respectively.}
\label{Fig:bn_r2_ci_pPb}
\end{figure}

\begin{figure}[!ht]
\centering
\includegraphics[width=.45\linewidth]{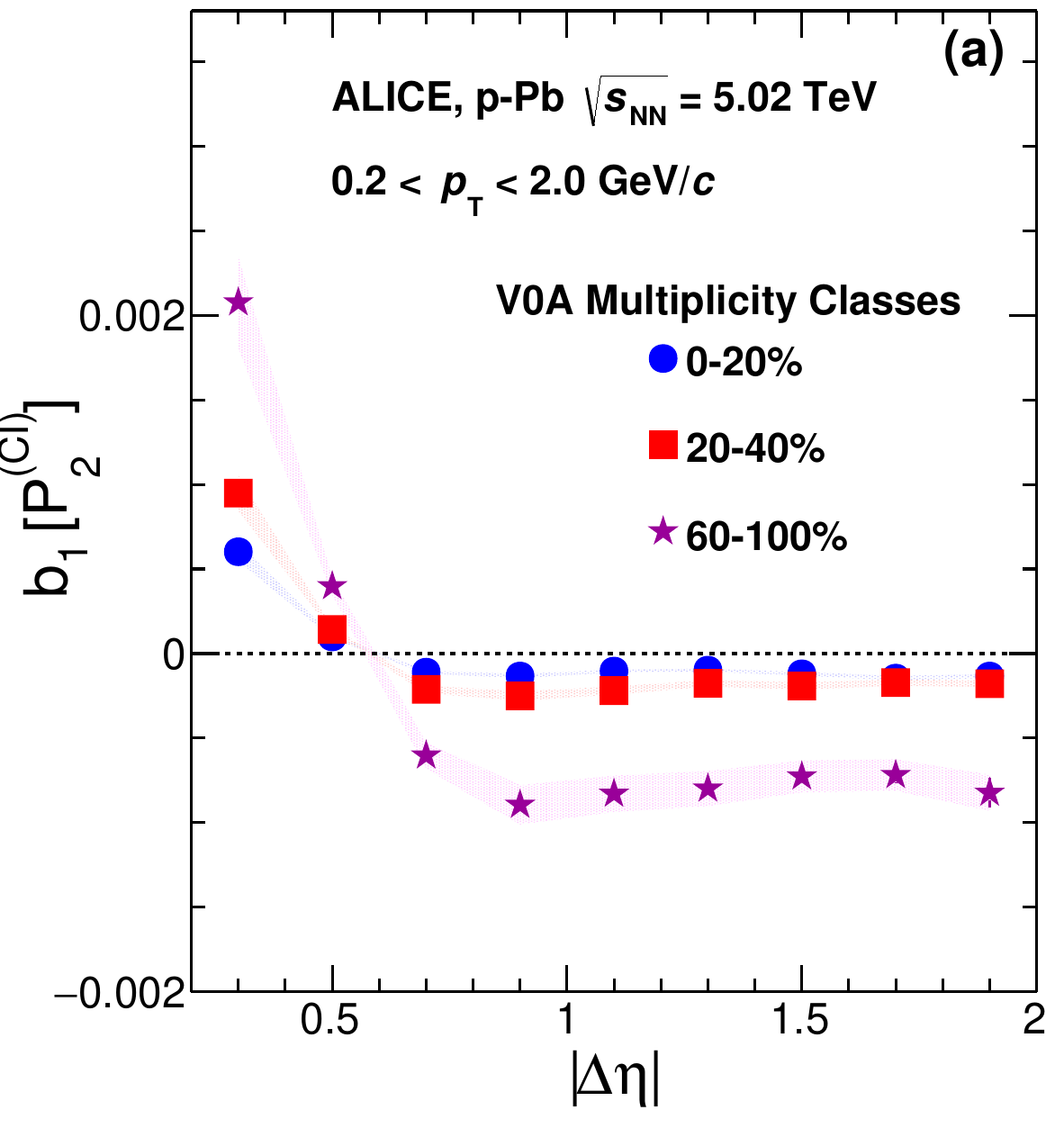}
\includegraphics[width=.45\linewidth]{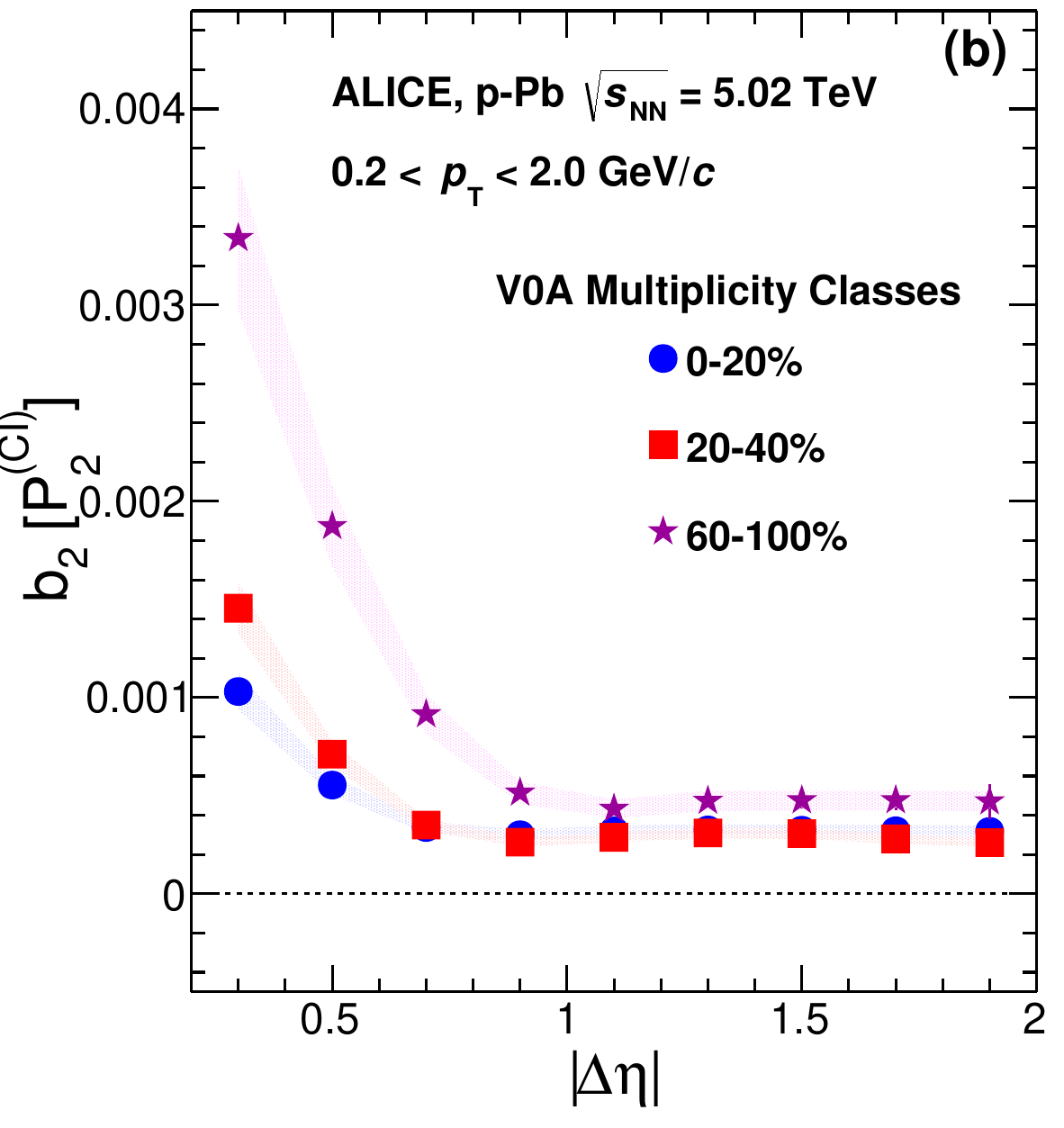}\\
\includegraphics[width=.45\linewidth]{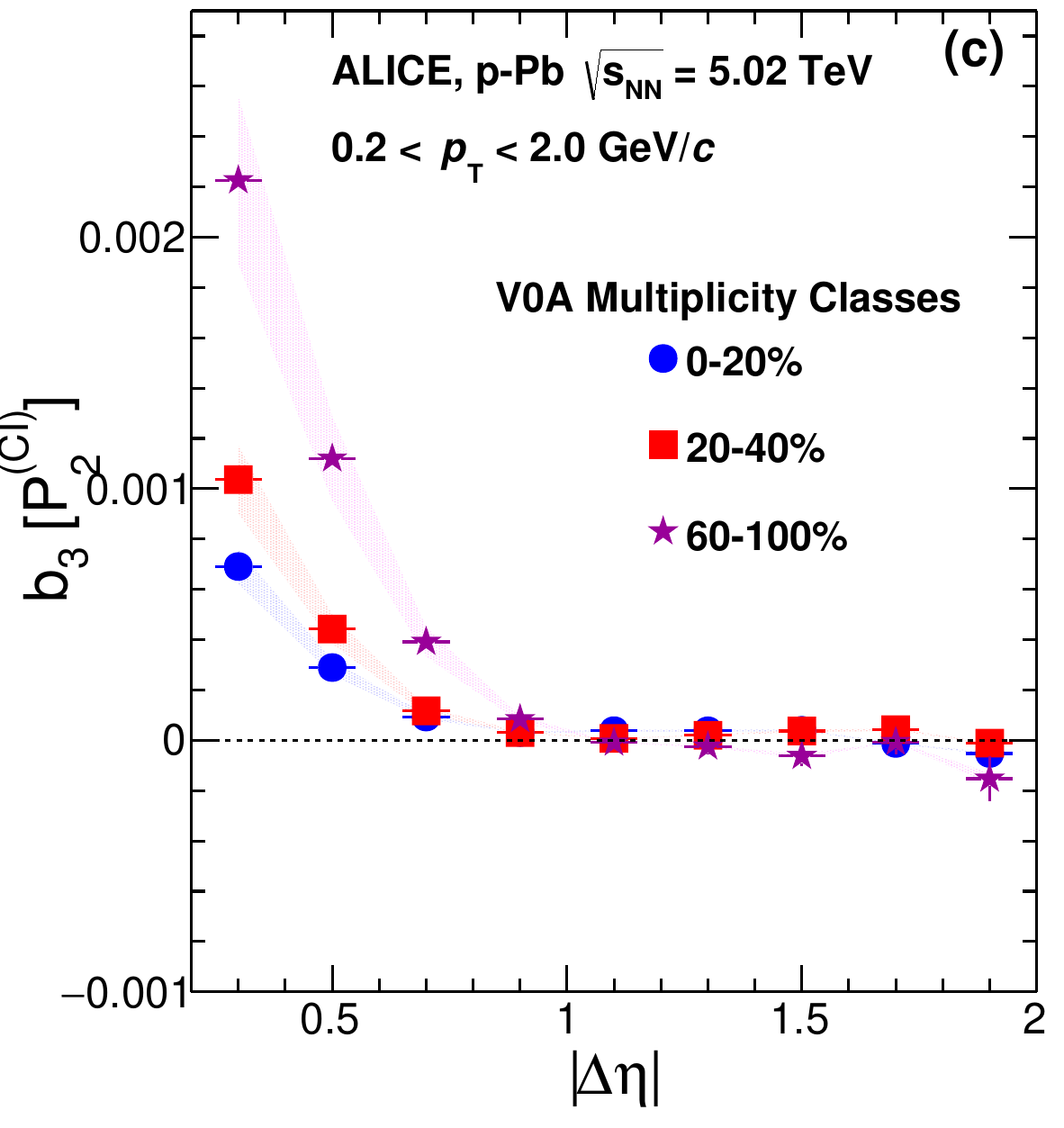}
\includegraphics[width=.45\linewidth]{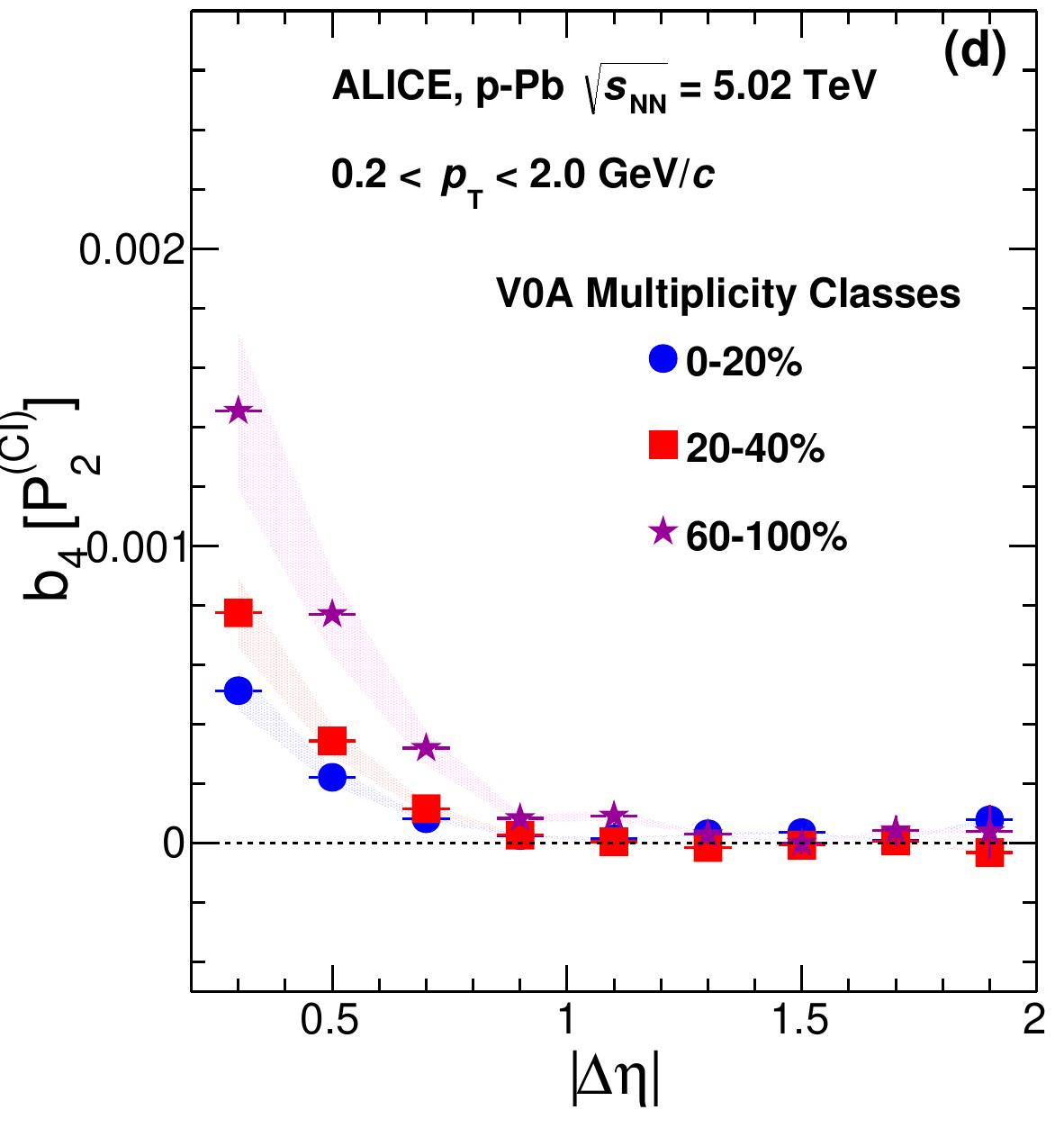}
\caption{Fourier coefficients, $b_{n}$, $n = 1,...,4$,  extracted from  $P_{2}^{\rm (CI)}$ correlation functions
measured  in    \pPb\  collisions at
$\snn = $ 5.02 TeV using three multiplicity classes. Vertical bars and shaded areas represent statistical and systematic  uncertainties, respectively.}
\label{Fig:bn_p2_ci_pPb}
\end{figure}

The $R_2$ and $P_2$ correlation functions shown in Fig.~\ref{Fig:Corr_2D_ChCI_pPb} exhibit non-trivial structures and dependences on $\Delta \varphi$. These may be due to a number of different particle production processes including resonance decays, coalescence of constituent quarks, string fragmentation, jets, and possibly several other mechanisms. In general, transverse anisotropies  associated with hydrodynamic flow and  differential attenuation of high $\pt $ particles by the anisotropic medium formed in   \pPb\  collisions are not readily expected  in small collision systems such as those produced in the minimum-bias or low multiplicity   \pPb\  collisions considered in this work. However, a number of recent works have reported evidence for collective motion  in high-multiplicty   \pPb\  collisions. It is thus of interest and valuable to characterize the azimuthal dependence of the correlation $R_2$ and $P_2$ in terms of Fourier decompositions  as a function of the relative pseudorapidity  $|\Delta\eta|$ of measured particles. Given non-flow effects are expected to dominate in minimum bias \pPb\ collisions, we report the coefficients $b_{n}$  calculated according to Eq.~(\ref{Eq:fourierDecomp})  rather than flow coefficients $v_n$. These are determined  based on projections of the $R_2$ and $P_2$ correlation functions onto $\Delta\varphi$ in several ranges of $|\Delta\eta|$.  The coefficients'  dependence on $|\Delta\eta|$ obtained from fits to the $R_{2}$ and
$P_{2}$ projections  are displayed in Figs.~\ref{Fig:bn_r2_ci_pPb}--\ref{Fig:bn_p2_ci_pPb} for  three different multiplicity classes.

\begin{figure}[!ht]
\centering
\includegraphics[width=.45\linewidth]{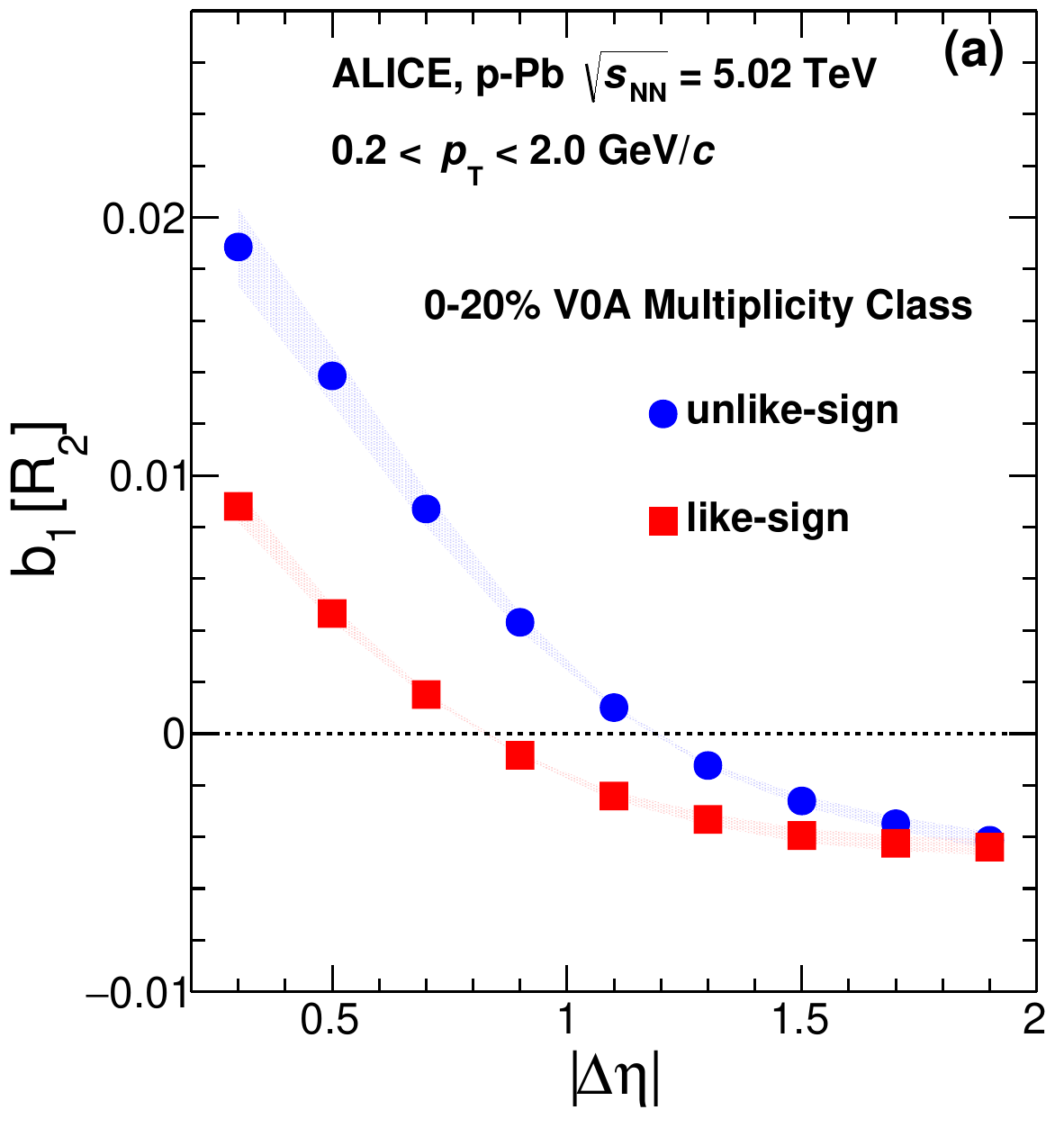}
\includegraphics[width=.45\linewidth]{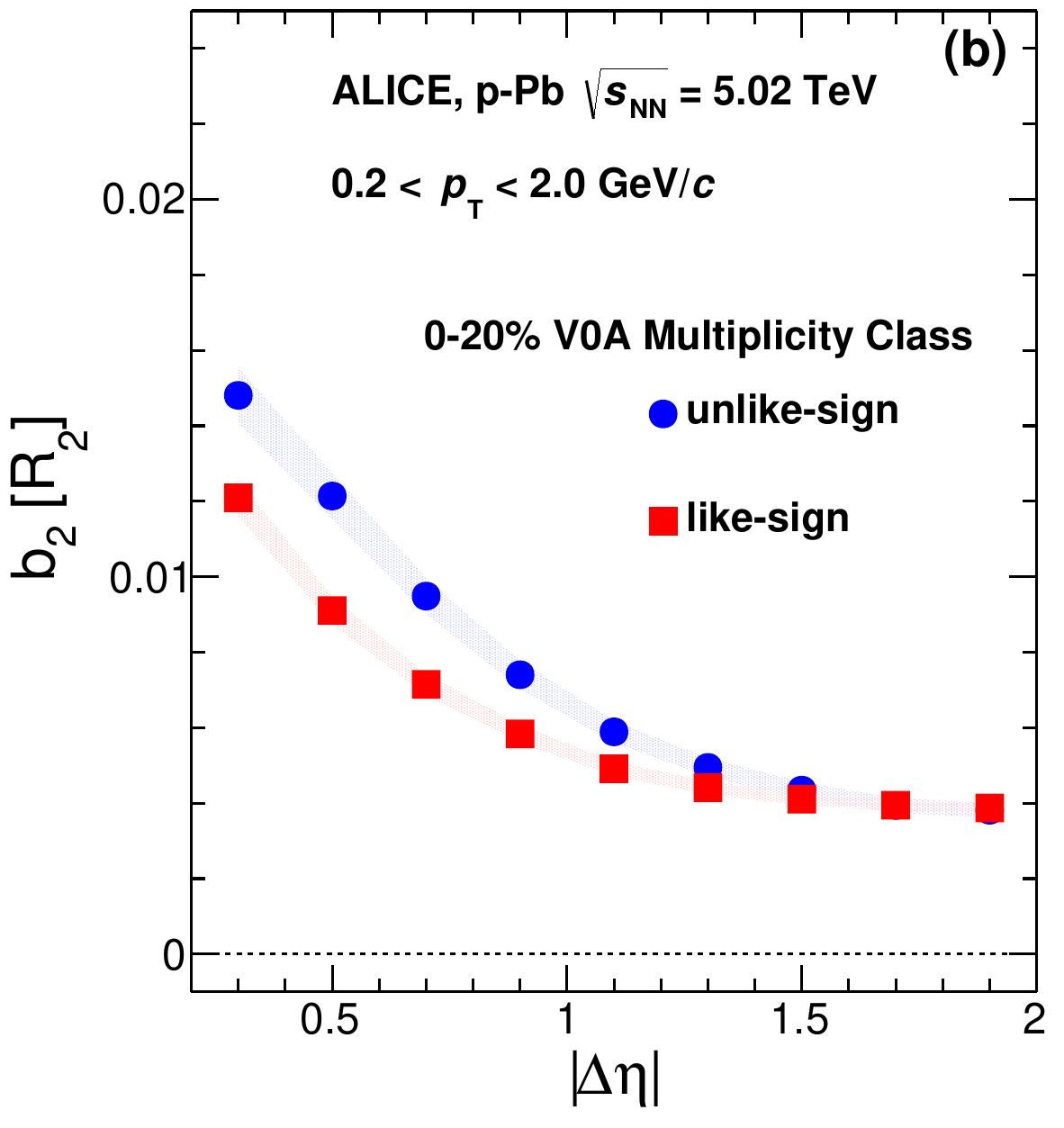}\\
\includegraphics[width=.45\linewidth]{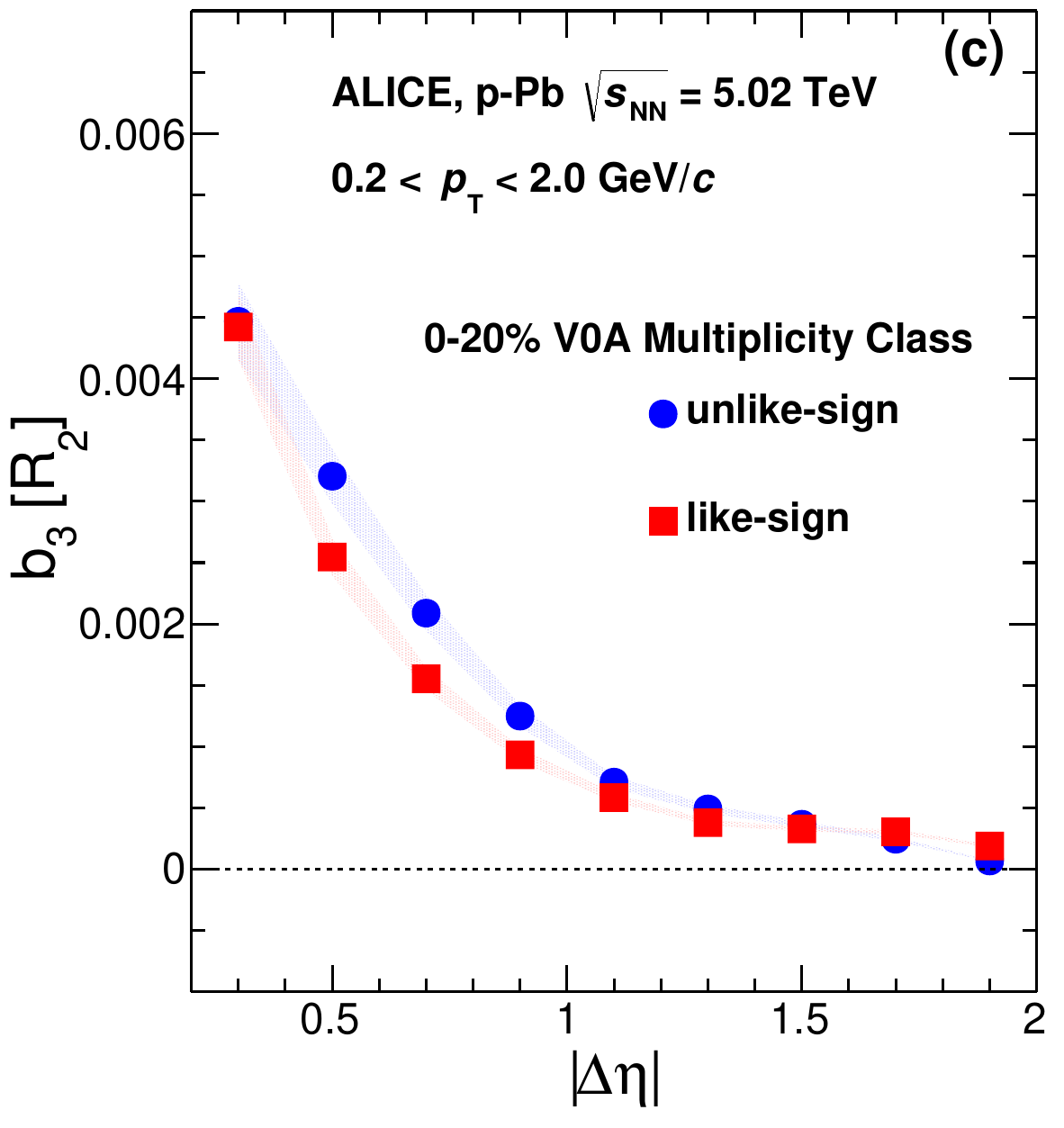}
\includegraphics[width=.45\linewidth]{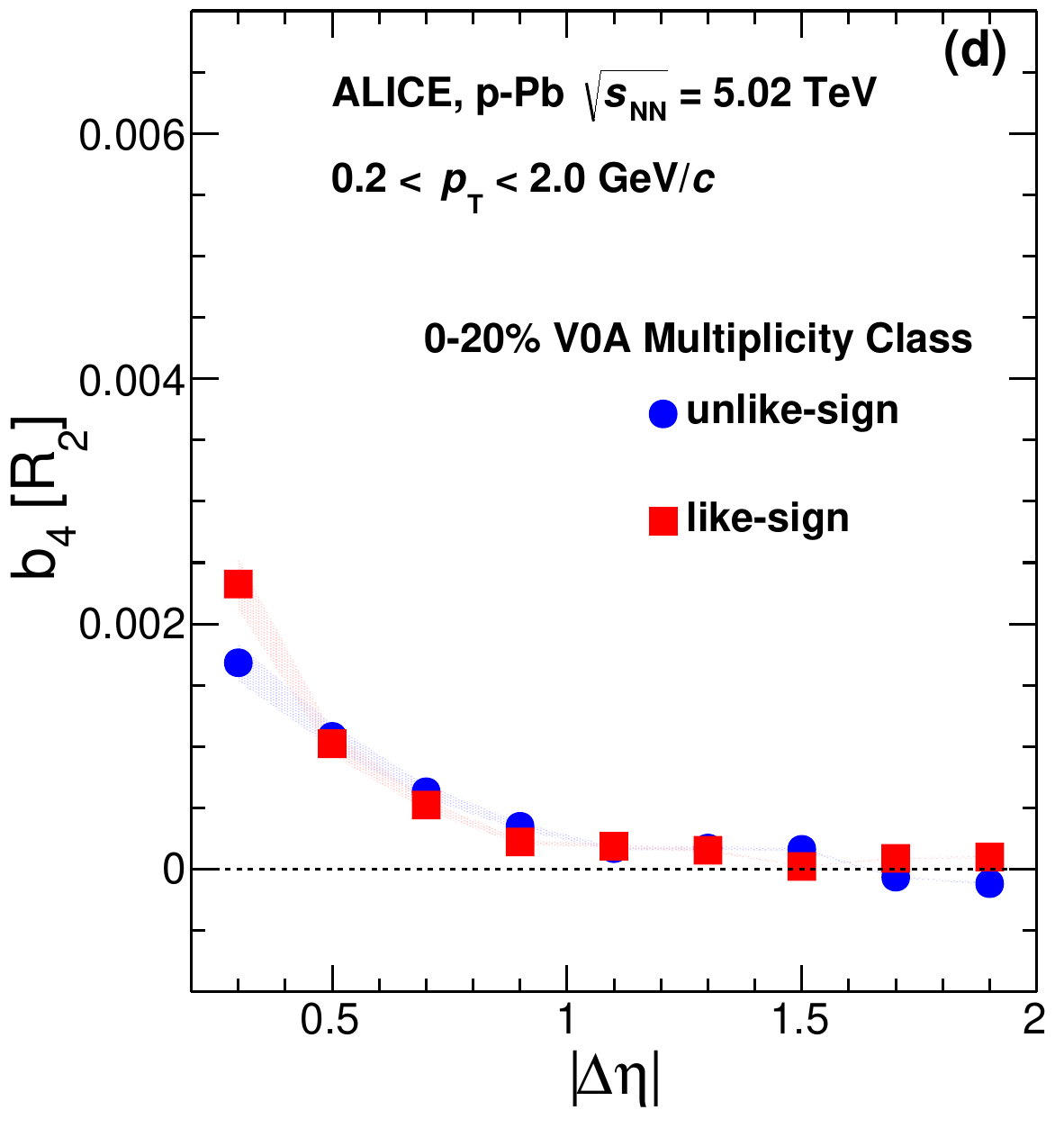}
\caption{Fourier coefficient, $b_{n}$, $n=1, \ldots, 4$, measured 
in $R_{2}$ in 0--20\% multiplicity class in \pPb\ collisions at
$\snn = $ 5.02 TeV. Vertical bars and shaded areas represent statistical and systematic  uncertainties, respectively.}
\label{Fig:bn_r2_usls_pPb}
\end{figure}

\begin{figure}[!ht]
\centering
\includegraphics[width=.45\linewidth]{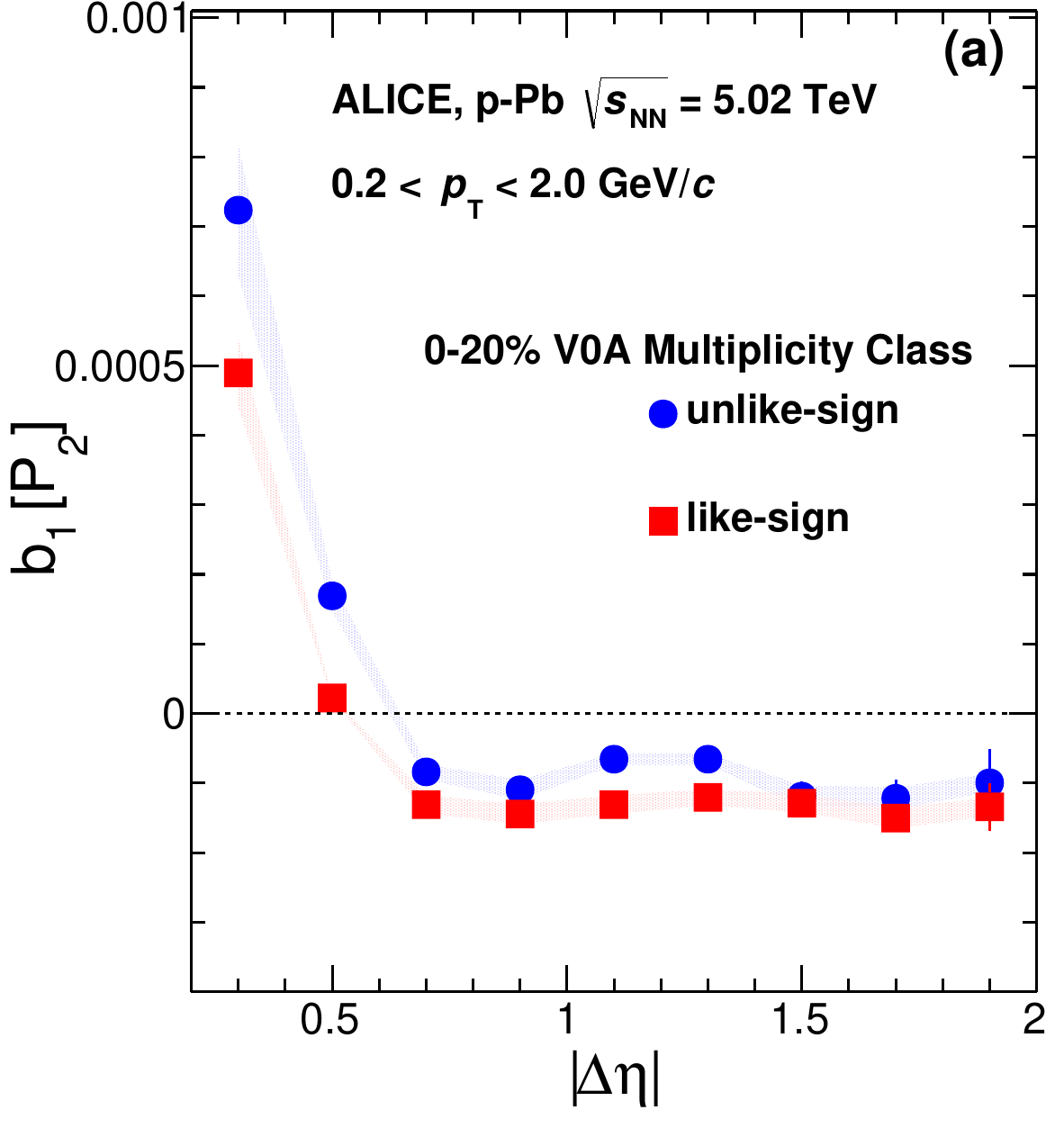}
\includegraphics[width=.45\linewidth]{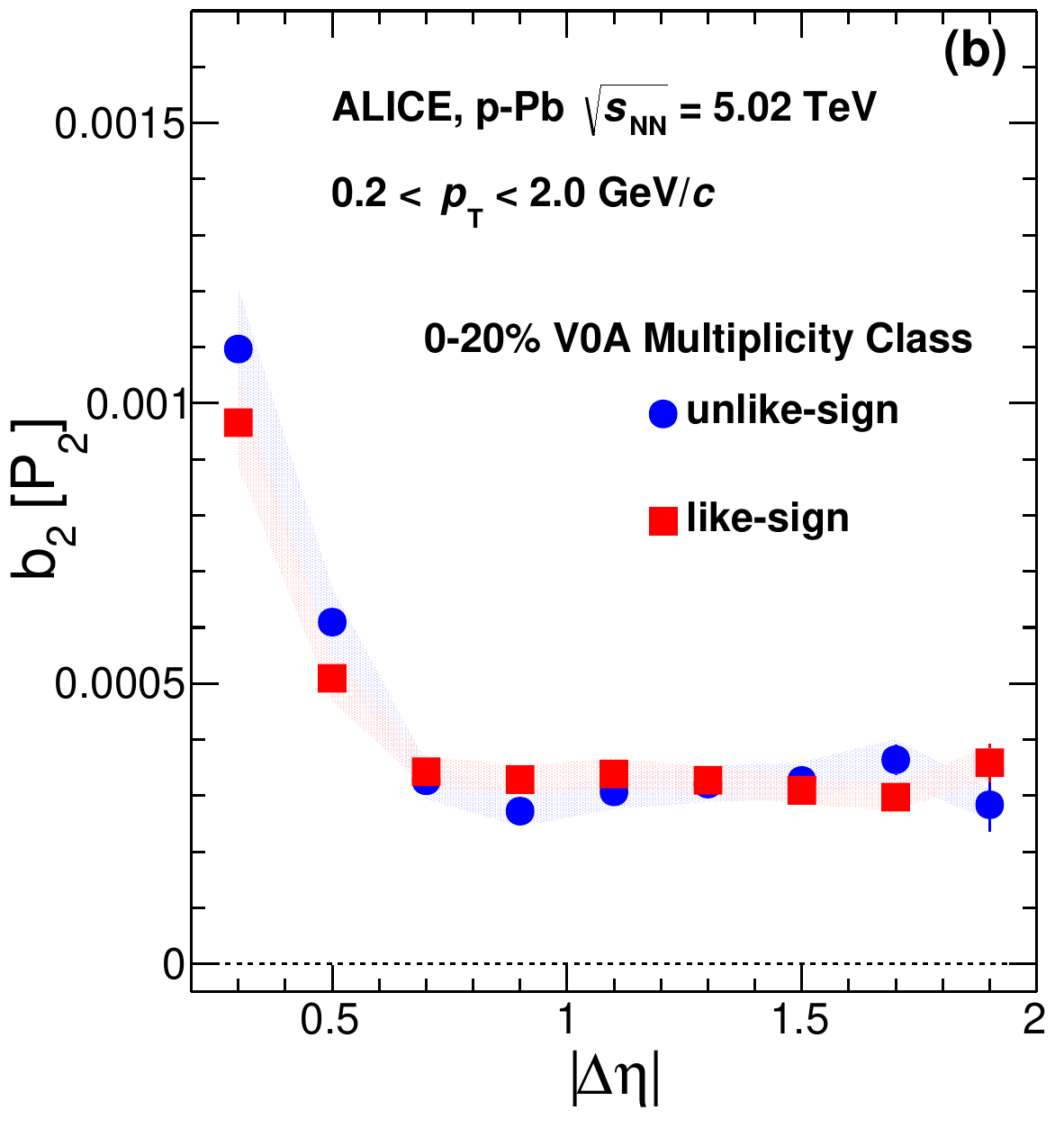}\\
\includegraphics[width=.45\linewidth]{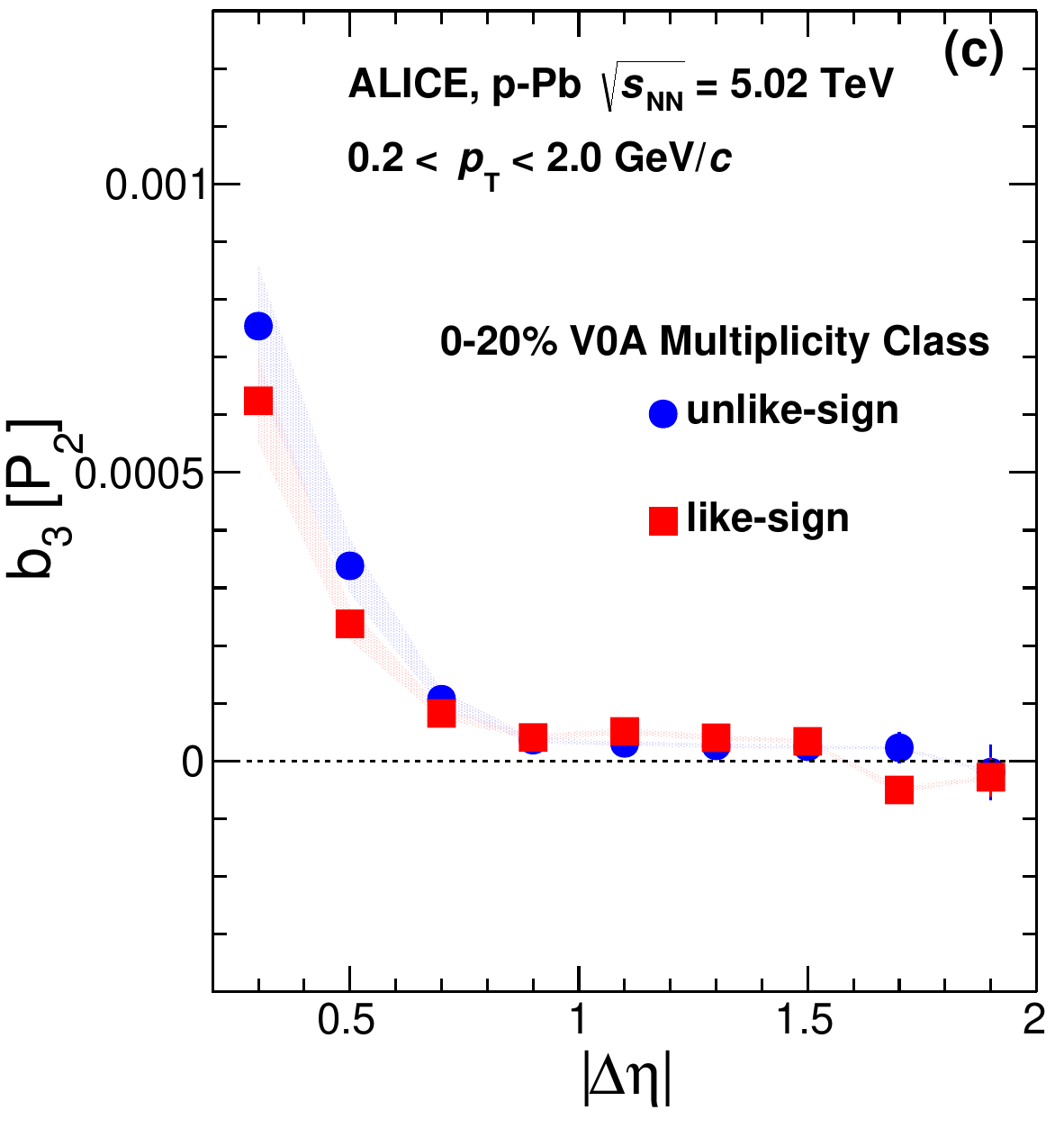}
\includegraphics[width=.45\linewidth]{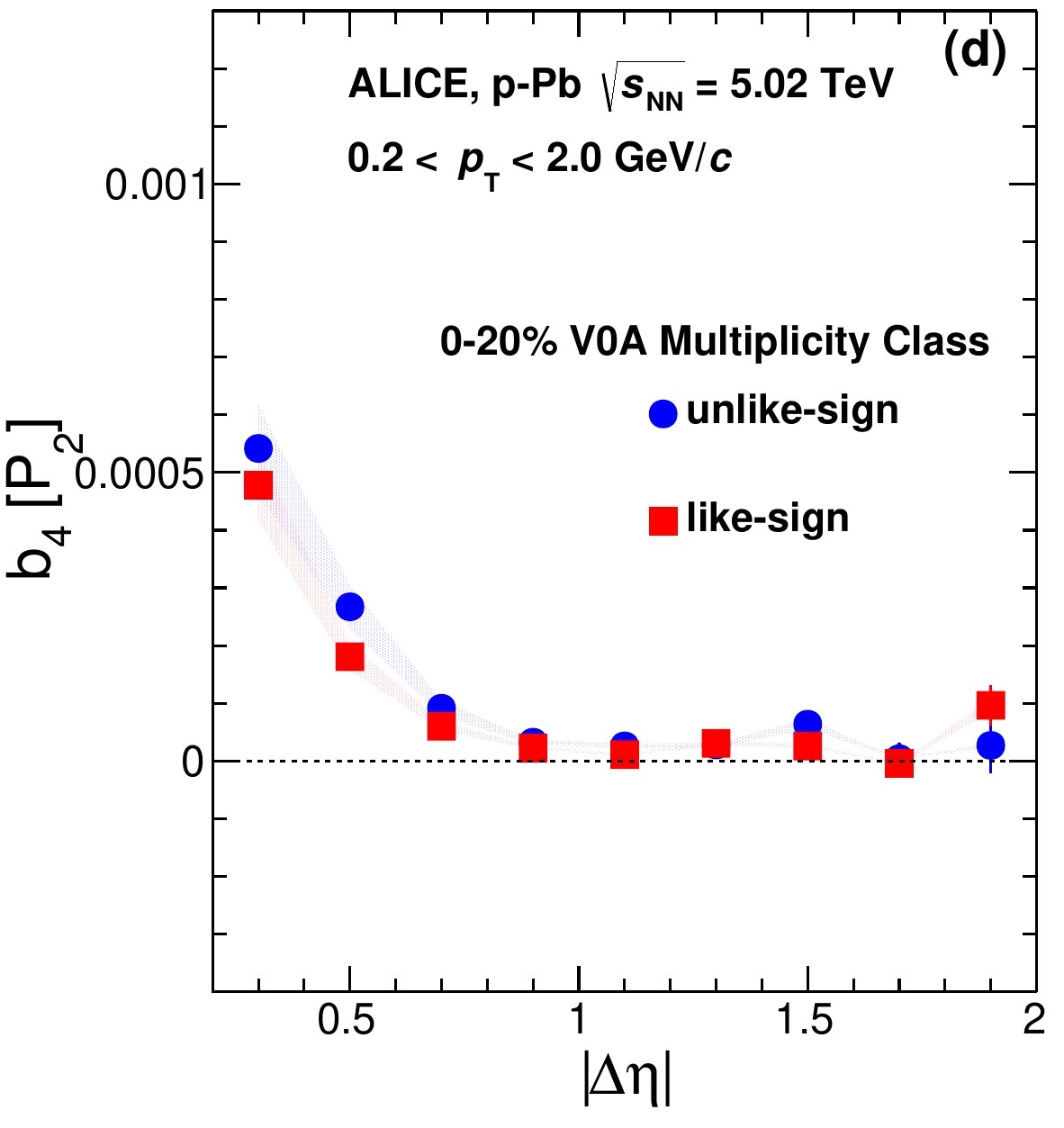}
\caption{Fourier coefficient, $b_{n}$, $n=1, \ldots, 4$, measured 
in $P_{2}$ in 0--20\% multiplicity class in \pPb\ collisions at
$\snn = $ 5.02 TeV. Vertical bars and shaded areas represent statistical and systematic  uncertainties, respectively.}
\label{Fig:bn_p2_usls_pPb}
\end{figure}

All in all, the coefficients $b_n$ obtained from fits to the $R_{2}$ and $P_{2}$
correlation functions measured at selected multiplicity classes
exhibit different dependences on $|\Delta\eta|$. The long range (i.e., in $|\Delta\eta|$)
of these correlation functions, in particular, is of interest to
understand the role of non-flow effects in measurements of flow.
Non-flow contributions  (e.g., those associated with resonance decays,  jets,  and momentum 
conservation) are expected to decrease with increasing  large $|\Delta\eta|$ gap. This can  be verified quantitatively 
based on the Fourier decompositions of $R_{2}$ and $P_{2}$
reported in Figs.~\ref{Fig:bn_r2_usls_pPb}--\ref{Fig:bn_p2_usls_pPb},  where one
observes that the coefficients, $b_{n}$
have  decreasing amplitudes for increasing $|\Delta\eta|$.
One notes, however, that the coefficient $b_2$ and coefficients of
higher order, $b_3$ and $b_4$, exhibit  qualitatively different dependences on $|\Delta\eta|$.
The higher order coefficients decrease rapidly,
within  $|\Delta\eta| < 1.5 (0.75)$ in $R_2^{\rm (CI)}$ ($P_2^{\rm (CI)}$) and become vanishingly small, within the statistical accuracy of this measurement, for larger
values of $|\Delta\eta|$, whereas $b_2$ coefficients' reduction with increasing $|\Delta\eta|$  saturates and reach a constant  value beyond $|\Delta\eta| \sim 1.5 (0.75)$.  
One   compares  the evolution of $b_n[R_2]$ and $b_n[P_2]$  coefficients with $|\Delta\eta|$ in more details.
The coefficients $b_{1}[R_2]$ measured in all three multiplicity classes, shown in  Fig.~\ref{Fig:bn_r2_ci_pPb} (a), exhibit a monotonic dependence on $|\Delta\eta|$, decreasing from positive values at $|\Delta\eta| =0.2$ to negative values at $|\Delta\eta| =1.9$, and crossing the axis (zero amplitude) at $|\Delta\eta| =1.0$. The positive values at $|\Delta\eta|\le0.9$ are determined by the presence of the strong near-side
peak, whereas negative values observed at large $|\Delta\eta|$ likely result from momentum conservation effects.  The coefficients $b_{1}[P_2]$, shown in  Fig.~\ref{Fig:bn_p2_ci_pPb} (a), exhibit similar monotonic trends as the $b_{1}[R_2]$ coefficients, with positive and negative values at short and large $|\Delta\eta|$ ranges, respectively, but their $|\Delta\eta|$ dependence crosses the axis and thus appear to vanish at approximately $|\Delta\eta|=0.6$ rather than the larger value $|\Delta\eta| =1.0$ observed in the case of the $R_2$ correlations. The lower crossing point value, $|\Delta\eta|=0.6$,  evidently results from
the much narrower near-side peaks observed in $P_2$ correlations relative to those found in the $R_2$ distributions.

One next considers graphs of $b_n$, $n\ge 2$, shown in Figs.~\ref{Fig:bn_r2_ci_pPb}--\ref{Fig:bn_p2_ci_pPb} (b-d),
extracted from $R_2^{\rm (CI)}$ and $P_2^{\rm (CI)}$ distributions. One finds that similarly to $b_1$ coefficients, the  $b_n[R_2]$ and $b_n[P_2]$ coefficients  all exhibit decreasing monotonic trends with increasing $|\Delta\eta|$. However, these coefficients
remain positive   in all three multiplicity classes and at all values of $|\Delta\eta|$, except for a few negative values of the $b_3$ and $b_4$ coefficients
observed at large $|\Delta\eta|$, which given their statistical accuracy are  consistent with positive values.
One notes, additionally, that the magnitude of the $b_n[R_2]$ coefficients decreases much slower with increasing $|\Delta\eta|$   than  the amplitude of the  $b_n[P_2]$ coefficients. Indeed, the $b_2[R_2]$
coefficients    appear to drop to a minimum value  at $|\Delta\eta| = 1.5-1.6$ while  $b_2[P_2]$ clearly reaches a
plateau  near $|\Delta\eta| = 0.6-0.7$. The third order coefficients  exhibit similar behavior, albeit, asymptotically reaching
much smaller values. The coefficient $b_3[P_2]$ clearly plateaus beyond $|\Delta\eta| = 0.6-0.7$ while
$b_3[R_2]$ is not clearly plateaued at $|\Delta\eta| = 1.8$. Similar trends are qualitatively  observed with the $b_4$ coefficients within  statistical uncertainties.

Overall, one finds that the $|\Delta\eta|$ dependence of the $b_n$ coefficients extracted in \pPb\ collisions for $R_2$ and $P_2$ correlation functions is rather similar to the evolution of the $v_n$ coefficients with $|\Delta\eta|$ observed in \PbPb\ collisions. Both sets of coefficients feature large values at small pair separations, decrease for increasing $|\Delta\eta|$, and   tend to plateau at approximately $|\Delta\eta| \sim 0.6 - 0.7$ in $P_2$ and  $|\Delta\eta| \sim 1.5$ in $R_2$. The non-flow component associated with the near-side peak is thus found to be suppressed in the case of $P_2$ for pair separations $0.7 <
|\Delta\eta| < 1.5$, implying that $\Delta \pt\Delta \pt$ averages to zero in that range. It is worth emphasizing, also, that $b_2$ remains constant and non-vanishing, in both $R_2$ and $P_2$ beyond $|\Delta\eta| \sim 1.5$ and  $|\Delta\eta| \sim 0.7$, respectively, thereby supporting the notion that collective behavior might be present in \pPb\ collisions~\cite{Aad:2012gla,Chatrchyan:2013nka,Aad:2013fja,Bozek:2010pb,Bozek:2012gr,Bozek:2013uha}. Unfortunately, the measurements presented in this work do not provide sufficient accuracy on  $b_3$ and $b_4$ to establish whether significant triangular and quadrangular  flow components are present in high multiplicity  \pPb\ collisions.

One further explores the long range behavior of the $R_2$ and $P_2$ correlation functions by
comparing the Fourier coefficients' $|\Delta\eta|$ dependence  of LS and US  correlations presented in Figs.~\ref{Fig:bn_r2_usls_pPb}--\ref{Fig:bn_p2_usls_pPb}, respectively. The presentation is limited to the
0--20\% multiplicity class but we  verified that correlations from lower multiplicity exhibit a similar behavior
as those shown.
One observes that the  coefficients obtained from US distributions, most particularly $b_1$ and $b_2$,
are significantly larger than those extracted from  LS distributions for rapidity difference smaller than $|\Delta\eta| \sim 1.5$, as evidently expected from the prominence of the near side US peaks observed in both $R_2$ and $P_2$ relative to the much smaller near-side structure encountered in LS distributions. One notes, however, that US and LS $R_2$ coefficients converge to essentially equal values at $|\Delta\eta| > 1.5$ and thus provide an indication that the correlation dynamics is charge agnostic at large relative pseudorapidities, a result also readily obvious from the $R_2^{\rm (CD)}$ presented in Fig. \ref{Fig:Corr_2D_ChCD_pPb}.
It is worth additionally noting that the differences between $b_3[R_2]$ and $b_4[R_2]$ of US and LS are rather
small for $|\Delta\eta|<1$ and essentially vanishing, within experimental uncertainties beyond $|\Delta\eta|\sim 1$.
The behavior and dependence of the $b_n[P_2]$ coefficients are qualitatively similar to those of $b_n[R_2]$ coefficients. One finds, however, that differences between US and LS coefficients are typically very small or vanishing for relative pseudorapidities as small as $|\Delta\eta| \sim 0.6$, again emphasizing the narrow peak observed in $P_2$ distributions relative to those measured in $R_2$ distributions.

\section{DISCUSSION}
\label{Sec:discussion}

\subsection{Charge insensitive non-flow contributions at large $|\Delta\eta|$}

Fourier decomposition analyses of   $R_2$ and $P_2$ correlation functions  measured in \PbPb\  collisions, shown in Figs.~\ref{Fig:vnP2_lsus_PbPb}--\ref{Fig:v2v3_r2VsP2_ci_7080}, reveal that beyond  $|\Delta\eta|\sim 0.9$, the coefficients $v_2$, $v_3$, and $v_4$ obtained  with LS and US pairs are identical within measurement uncertainties. This is confirmed also  by the inspection  of CD correlations, shown in Figs.~\ref{Fig:Corr_2D_ChCD_PbPb}--\ref{Fig:Corr_2D_ChCD_pPb}, which exhibit nearly vanishing amplitude in mid to central collisions beyond $|\Delta\eta|\sim 1.4$ and on the away-side, i.e., at $\Delta\varphi\sim \pi$. One can then consider a two-component model of these correlations consisting of a near-side component determined chiefly by  charge-dependent  particle production processes (such as resonance decays, $(+,-)$  pair creation in jets or via string hadronization, etc.) and a long range component essentially insensitive to particle charges. In  mid to central  \PbPb\  collisions, this long range component is  attributed to collective flow resulting in part from spatial anisotropy of the system and energy density and/or pressure gradients. However, the possibility  of a long-range non-flow contribution, i.e., non-collective in nature, cannot be eliminated. Indeed, long range and charge insensitive non-flow contributions may in part  arise from back-to-back jets, but they may also result from a superposition of long range particle correlations arising in simpler collision systems such as \pp\ and \pPb. The $R_2^{\rm (CD)}$ and $P_2^{\rm (CD)}$ distributions shown in Fig. \ref{Fig:Corr_2D_ChCD_pPb} reveal that two-particle correlations in   \pPb\  collisions also feature nearly vanishing correlation amplitude at large $|\Delta\eta|$ and on the away-side of these correlation functions. Recall from Sec. \ref{Sec:fd}, that  the Fourier decompositions of   $R_2$ and $P_2$ correlation functions of LS and US pairs feature essentially identical harmonic coefficients $b_n$, for $n$=2, 3, and 4, at large $|\Delta\eta|$.
Correlations in   \pPb\  collisions can then, at least approximately, be considered as a superposition of short range correlations leading to the production of the near-side peak observed in these correlations and a long range component insensitive to the charge of particles. It is  unclear whether this long range component reflects the production of a flowing medium in   \pPb\  collisions or whether it arises from non-collective particle production and transport. It is nonetheless of interest to consider how such a component would scale in \PbPb\ collisions if nucleon--nucleon (or parton--parton) interactions taking place in these collisions were completely independent of one another and in the absence of re-scattering of the particles these interactions produce. Indeed, assuming \PbPb\ collisions are such trivial superposition of   \pPb\  collisions, the long-range component of these   \pPb\  collisions can be considered, for practical intents, as a non-flow contribution to the correlation measured in \PbPb. One can then use a basic property of cumulants to determine an upper bound on non-flow effects in \PbPb\ arising from a superposition of   \pPb\   subprocesses. The normalized cumulants $R_2$ and $P_2$ scale inversely to the number of identical subprocesses. The non-flow contributions to the $v_n$ coefficients should then be of the order of $\sqrt{b_n}/\sqrt{m}$ where $m$ is the average number of wounded nucleons encountered at a given collision centrality in \PbPb\ collisions. Let us thus consider, as an example, a simple evaluation of an upper limit of contributions to elliptical flow measured in \PbPb\ collisions based on the long range values of $b_2$ in \pPb. In   \pPb\  collisions, one finds $b_2\sim 0.004$ at $|\Delta\eta|>1.5$.
Assuming that, on average,  a central \PbPb\ collision is equivalent to  approximately 200   \pPb\  collisions, the non-flow contribution to long range $v_2$ values is thus of the order of $\sqrt{0.004/200}=0.0045$. The measured $v_2$ for LS and US pairs in 0--5\% collision centralities amounts to $v_2=0.027$. Considering this ``non-flow" contributions adds  in 
quadrature with the flow term in \PbPb, one concludes non-flow contributions are of the order of   $\sim  1.5\%$ 
of the observed $v_2$ in this centrality. This conclusion is in qualitative agreement with
assessments of non-flow contributions obtained from other methods~\cite{Adam:2016nfo}.

\subsection{ Charge sensitive non-flow contributions at small $|\Delta\eta|$}

The two-component model invoked in the previous section to separate the near-side short-range correlation peaks and the long range correlations observed  in this work has been commonly used, in other works, to subtract the long-range component as a background, and to study the features of the near-side peak. However, the near-side peaks observed in $R_2$ and $P_2$ exhibit rather different properties and one may then wonder whether a two-component model is actually appropriate. Indeed, one finds that the near-side peaks observed in $P_2$ correlations, in both  \PbPb\ and   \pPb\  collisions, cover a different $|\Delta\eta|$ range than the peaks observed
in $R_2$ distributions. Accordingly, one finds  that the LS and US $v_n$ and $b_n$ coefficients measured in \PbPb\ and   \pPb\  collisions, respectively, reach a plateau at much smaller $|\Delta\eta|$ in $P_2$ distributions than in $R_2$ distributions. This is rather remarkable given that both observables are proportional, effectively, to integrals over the 0.2 to 2.0  \gevc\   momentum range of the two-particle density $\rho_2({\bf p}_1,{\bf p}_2)$ albeit with different coefficients (unity for $R_2$ and $\Delta \pt \Delta \pt $ for $P_2$). One would thus expect the two correlation observables to feature similar near-side structures and dependence on $|\Delta\eta|$. The observed difference between the shapes, not just the strengths, must then arise from $P_2$'s dependence on $\Delta \pt \Delta \pt $. In fact, given  this coefficient is not positive definite, correlated pairs may  yield either  positive or negative contributions to $P_2$. The narrower peak observed in $P_2$ implies that pairs in the range $0.5 < |\Delta \eta| < 0.9$, where $P_2$ is suppressed relative to $R_2$, receives, on average, vanishing contributions from the  $\Delta \pt \Delta \pt $
coefficient, while the range  $|\Delta \eta| < 0.5$ is positive definite on average. Conceivably, the near-side might itself consists of two components, one ``regular" component with non-vanishing $\la\Delta \pt \Delta \pt \ra$ present in both $R_2$ and $P_2$, and one component with  vanishing $\la\Delta \pt \Delta \pt \ra$ contributing only to $R_2$. However, it is difficult to identify  particle production processes that might feature such properties. It is possible, on the other hand, that certain processes might  feature vanishing $\la\Delta \pt \Delta \pt \ra$ over a limited range of phase space. Consider, for instance, the decay of resonances such as the $\rho^0$-meson into a pair of $\pi^+\pi^-$ mesons. In-flight decays of  $\rho^0$-mesons produce  kinematically focused $\pi^+\pi^-$ pairs, which are detected at small relative angles ($\Delta\phi$ and $\Delta\eta$) in the laboratory frame. Correlated pions  from such decays could feature  positive or negative values of  $\Delta \pt \Delta \pt $ depending on the orientation of the decay relative to the direction of their parent $\rho^0$-meson. Likewise, particles composing  jets might also contribute differentially, with $|\Delta\eta|$, to $P_2$. The core
of jets (particles emitted at small angles relative to  the jet axis) typically involve large momenta, i.e., particles with momenta well in excess of the inclusive average $\la \pt \ra$. They would thus make a strong positive contribution to $P_2$. Particles emitted at large angles, relative to the jet axis, typically feature lower momenta. They might then contribute equally negative and positive terms to $\Delta \pt \Delta \pt $ and thus yield
a vanishing average. Particles of the jet outer edges  would evidently have positive contributions to $R_2$ and thus produce a near-side peak characteristic of the width of jets but their vanishing $\Delta \pt \Delta \pt $ average  might effectively produce
a narrower peak in $P_2$ relative to that observed in $R_2$.

One can speculate further about the role of jets in near-side correlations based on the $R_2^{\rm (CD)}$ and  $P_2^{\rm (CD)}$ distributions shown in Figs.~\ref{Fig:Corr_2D_ChCD_PbPb}--\ref{Fig:Corr_2D_ChCD_pPb}. In \PbPb\ collisions, the observed longitudinal narrowing of $R_2^{\rm (CD)}$ distributions with increasing collision centrality may be interpreted as evidence, in part, for strong radial flow and two stage particle emission. Indeed, correlated particles emitted from a radially boosted source are kinematically focused, i.e., emitted at smaller relative rapidity. Similarly, late stage particle emission, after the system has cooled down, may also produce particles with  smaller relative rapidity. The $R_2^{\rm (CD)}$ correlation function is thus expected  to narrow considerably under the combination of the two effects. Careful modeling of the correlation functions shall be required, however, to interpret the observed narrowing of  $R_2^{\rm (CD)}$ and disentangle the relative contributions of radial flow and  late stage emission.

Additionally, in light of the narrower width of $P_2^{\rm (CI)}$ distributions relative to those of $R_2^{\rm (CI)}$ and the role of jet-like contributions in these correlation functions, as discussed above, one should also examine the role of jet-like contributions to $R_2^{\rm (CD)}$ and $P_2^{\rm (CD)}$  distributions. It is in fact interesting that the longitudinal width of $P_2^{\rm (CD)}$ remains essentially independent of collision centrality, thereby hinting that it might be insensitive to effects associated with radial flow and  two-stage particle production. A dominance of jet-like contributions to this correlation could then be used to study the impact of the medium on jets. That would likely require, however, a much larger dataset to reduce statistical uncertainties and enable more precise corrections for instrumental effects, which currently limit the precision of the measurement reported in this work.

\section{SUMMARY AND CONCLUSION}
\label{Sec:summary}

Measurements of two-particle differential number-correlation functions $R_2$ and transverse-momentum correlation functions $P_2$ obtained in  \PbPb\ collisions at $\snn = $ 2.76 TeV and in \pPb\ collisions at $\snn = $ 5.02 TeV were presented. Measurements were reported as a function of collision  centrality and multiplicity for these two collision systems, respectively, for charged particles in the range $|\eta|<$ 1.0 and $0.2 < \pt < 2.0$  \gevc. Measurements of correlation functions  for like-sign (LS) and unlike-sign (US) particle pairs  were first carried out separately and combined  to obtain charge-independent (CI) and charge-dependent (CD) correlation  functions.  The $R_2$ and $P_2$ correlators exhibit similar features, most notably a relatively strong  near-side peak centered at $|\Delta\eta| \sim \Delta\varphi \sim 0$, and a weaker away-side ridge (at $\Delta\varphi = \pi$) with a width larger than the $\eta$ acceptance (2 units)  in low-multiplicity event classes. Both correlation observables also exhibit  strong harmonic modulations in mid-central to central \PbPb\  collisions. However, there are also interesting and revealing differences. One finds, both in \PbPb\ and   \pPb\  collisions, that the near-side peak of $P_2$ is much narrower in $|\Delta\eta|$ and $\Delta\varphi$  than  observed with $R_2$. One also observes, in the 5\% most central \PbPb\ collisions,  that the away-side of $P_2$ features a dip structure at $\Delta\varphi \sim \pi$,
and side band peaks at $\Delta\varphi \sim \pi\pm\pi/3$ extending across $|\Delta\eta|<2$.  Such a modulated structure is not present in the 5\% most central \PbPb\ collisions measured in this work
for $R_2$ but  was  observed for number correlations, similar to $R_2$, only in very  central collisions (0--2\%) thereby indicating that $P_2$ is somewhat more sensitive to the presence of a third-harmonic (triangular) flow component.

The  width of the near-side peak of the $R_2$ and $P_2$ charge-independent and  charge-dependent correlation functions were studied in order to better understand the relative contributions of non-flow and flow effects to particle correlations. In \PbPb,  the
longitudinal width, $\la \Delta\eta^k\ra^{1/k}$, of both $R_2^{\rm (CI)}$ and $P_2^{\rm (CI)}$ exhibits sizable growth for increasing collision centrality. However, no significant
dependence of the CI correlation widths was observed in \pPb. In contrast, one finds that the width of $R_2^{\rm (CD)}$ correlation functions
significantly narrow with increasing collision centrality in \PbPb, or produced particle multiplicity in \pPb, while only a modest decrease of the width
of the near-side $P_2^{\rm (CD)}$ peak could be ascertained within the current analysis. One furthermore observes that the $\Delta\varphi$ width of the near-side peak of 
$R_2^{\rm (CD)}$  exhibits a significant decrease with increasing produced-particle multiplicity  in  \PbPb. The 
decrease is more modest in   \pPb\  collisions for $R_2^{\rm (CD)}$, while the observed
azimuthal width of the near-side peak of $P_2^{\rm (CD)}$ is consistent with a modest decrease with increasing multiplicity.

The narrowing of
the near-side of $R_2^{\rm (CD)}$ is consistent with the narrowing of the balance function already reported and can be interpreted, in part,
as an effect of radial flow and two-stage hadronization. However, finite diffusion effects, which  broaden the correlation functions,
are also expected in long-lived collision systems. The observed broadening of  $R_2^{\rm (CI)}$ and $P_2^{\rm (CI)}$, with increasing collision
centrality in \PbPb\ collisions,  might in part result from such diffusive effects, but other processes
influencing the strength of long-range longitudinal correlations must be considered.
The dependence of the $\Delta\varphi$ width of the near-side peak of $R_2^{\rm (CD)}$ and $P_2^{\rm (CD)}$ were studied vs. increasing pair
separation in $\Delta\eta$. They exhibit a non-monotonic dependence on the pair separation, which might in part be caused by diffusion effects, although the role
of differential radial flow may not be excluded without specific models of these effects.
In fact, one anticipates that the observed centrality and pair-separation dependence  of the width of the near side peaks of $R_2^{\rm (CD)}$ and $P_2^{\rm (CD)}$
shall provide important constraints in the formulation of models of the collision dynamics, which might help to better constrain the
contributions of radial flow, diffusion, and two-stage emission in \PbPb\ collisions, most particularly.

The need to better understand the roles of non-flow and flow also prompted the analysis in terms of  $|\Delta\eta|$ pair separation ($\eta$ gap) dependent
Fourier decompositions of the $\Delta\varphi$ behavior of the $R_2$ and $P_2$ correlation functions.
Significant differences in the dependence of the harmonic and flow coefficients between the correlator $R_2$ and $P_2$ were found,
owing  to the fact, most likely, that the measured $P_2$ correlation functions feature a much narrower near-side peak than their corresponding
$R_2$ counterparts. Indeed, one observes that the $v_n$ coefficients measured in $P_2$ correlations reach a plateau at much smaller
pair separation than those observed in $R_2$ correlations. These differences indicate that the $R_2$ and $P_2$ correlation functions  exhibit distinct sensitivities
to flow and non-flow effects
and could then be exploited, in theoretical models, to obtain  better insight into  particle production and transport dynamics  in
heavy-ion collisions.  Long-range non-flow effects may also exist, however, and  the magnitude of the $b_2$  coefficients observed at large pair separation
in   \pPb\  collisions was used to obtain an upper limit of ~1.5\% for  non-flow contributions to $v_2$ in the 5\% most central \PbPb\ collisions.

\newenvironment{acknowledgement}{\relax}{\relax}
\begin{acknowledgement}
\section*{Acknowledgements}

The ALICE Collaboration would like to thank all its engineers and technicians for their invaluable contributions to the construction of the experiment and the CERN accelerator teams for the outstanding performance of the LHC complex.
The ALICE Collaboration gratefully acknowledges the resources and support provided by all Grid centres and the Worldwide LHC Computing Grid (WLCG) collaboration.
The ALICE Collaboration acknowledges the following funding agencies for their support in building and running the ALICE detector:
A. I. Alikhanyan National Science Laboratory (Yerevan Physics Institute) Foundation (ANSL), State Committee of Science and World Federation of Scientists (WFS), Armenia;
Austrian Academy of Sciences and Nationalstiftung f\"{u}r Forschung, Technologie und Entwicklung, Austria;
Ministry of Communications and High Technologies, National Nuclear Research Center, Azerbaijan;
Conselho Nacional de Desenvolvimento Cient\'{\i}fico e Tecnol\'{o}gico (CNPq), Universidade Federal do Rio Grande do Sul (UFRGS), Financiadora de Estudos e Projetos (Finep) and Funda\c{c}\~{a}o de Amparo \`{a} Pesquisa do Estado de S\~{a}o Paulo (FAPESP), Brazil;
Ministry of Science \& Technology of China (MSTC), National Natural Science Foundation of China (NSFC) and Ministry of Education of China (MOEC) , China;
Ministry of Science and Education, Croatia;
Ministry of Education, Youth and Sports of the Czech Republic, Czech Republic;
The Danish Council for Independent Research | Natural Sciences, the Carlsberg Foundation and Danish National Research Foundation (DNRF), Denmark;
Helsinki Institute of Physics (HIP), Finland;
Commissariat \`{a} l'Energie Atomique (CEA) and Institut National de Physique Nucl\'{e}aire et de Physique des Particules (IN2P3) and Centre National de la Recherche Scientifique (CNRS), France;
Bundesministerium f\"{u}r Bildung, Wissenschaft, Forschung und Technologie (BMBF) and GSI Helmholtzzentrum f\"{u}r Schwerionenforschung GmbH, Germany;
General Secretariat for Research and Technology, Ministry of Education, Research and Religions, Greece;
National Research, Development and Innovation Office, Hungary;
Department of Atomic Energy Government of India (DAE), Department of Science and Technology, Government of India (DST), University Grants Commission, Government of India (UGC) and Council of Scientific and Industrial Research (CSIR), India;
Indonesian Institute of Science, Indonesia;
Centro Fermi - Museo Storico della Fisica e Centro Studi e Ricerche Enrico Fermi and Istituto Nazionale di Fisica Nucleare (INFN), Italy;
Institute for Innovative Science and Technology , Nagasaki Institute of Applied Science (IIST), Japan Society for the Promotion of Science (JSPS) KAKENHI and Japanese Ministry of Education, Culture, Sports, Science and Technology (MEXT), Japan;
Consejo Nacional de Ciencia (CONACYT) y Tecnolog\'{i}a, through Fondo de Cooperaci\'{o}n Internacional en Ciencia y Tecnolog\'{i}a (FONCICYT) and Direcci\'{o}n General de Asuntos del Personal Academico (DGAPA), Mexico;
Nederlandse Organisatie voor Wetenschappelijk Onderzoek (NWO), Netherlands;
The Research Council of Norway, Norway;
Commission on Science and Technology for Sustainable Development in the South (COMSATS), Pakistan;
Pontificia Universidad Cat\'{o}lica del Per\'{u}, Peru;
Ministry of Science and Higher Education and National Science Centre, Poland;
Korea Institute of Science and Technology Information and National Research Foundation of Korea (NRF), Republic of Korea;
Ministry of Education and Scientific Research, Institute of Atomic Physics and Romanian National Agency for Science, Technology and Innovation, Romania;
Joint Institute for Nuclear Research (JINR), Ministry of Education and Science of the Russian Federation and National Research Centre Kurchatov Institute, Russia;
Ministry of Education, Science, Research and Sport of the Slovak Republic, Slovakia;
National Research Foundation of South Africa, South Africa;
Centro de Aplicaciones Tecnol\'{o}gicas y Desarrollo Nuclear (CEADEN), Cubaenerg\'{\i}a, Cuba and Centro de Investigaciones Energ\'{e}ticas, Medioambientales y Tecnol\'{o}gicas (CIEMAT), Spain;
Swedish Research Council (VR) and Knut \& Alice Wallenberg Foundation (KAW), Sweden;
European Organization for Nuclear Research, Switzerland;
National Science and Technology Development Agency (NSDTA), Suranaree University of Technology (SUT) and Office of the Higher Education Commission under NRU project of Thailand, Thailand;
Turkish Atomic Energy Agency (TAEK), Turkey;
National Academy of  Sciences of Ukraine, Ukraine;
Science and Technology Facilities Council (STFC), United Kingdom;
National Science Foundation of the United States of America (NSF) and United States Department of Energy, Office of Nuclear Physics (DOE NP), United States of America.    
\end{acknowledgement}

\newpage
\clearpage
\bibliography{dptdpt_prc_V16}
\bibliographystyle{utphys}   

\newpage
\appendix
\section{THE ALICE COLLABORATION}
\label{app:collab}

\begingroup
\small
\begin{flushleft}
S.~Acharya\Irefn{org139}\And 
F.T.-.~Acosta\Irefn{org20}\And 
D.~Adamov\'{a}\Irefn{org93}\And 
J.~Adolfsson\Irefn{org80}\And 
M.M.~Aggarwal\Irefn{org98}\And 
G.~Aglieri Rinella\Irefn{org34}\And 
M.~Agnello\Irefn{org31}\And 
N.~Agrawal\Irefn{org48}\And 
Z.~Ahammed\Irefn{org139}\And 
S.U.~Ahn\Irefn{org76}\And 
S.~Aiola\Irefn{org144}
% ...
\And
\renewcommand\labelenumi{\textsuperscript{\theenumi}~}

\section*{Affiliation notes}
\renewcommand\theenumi{\roman{enumi}}
\begin{Authlist}
\item \Adef{org*}Deceased
\item \Adef{orgI}Dipartimento DET del Politecnico di Torino, Turin, Italy
\item \Adef{orgII}M.V. Lomonosov Moscow State University, D.V. Skobeltsyn Institute of Nuclear, Physics, Moscow, Russia
\item \Adef{orgIII}Department of Applied Physics, Aligarh Muslim University, Aligarh, India
\item \Adef{orgIV}Institute of Theoretical Physics, University of Wroclaw, Poland
\end{Authlist}

\section*{Collaboration Institutes}
\renewcommand\theenumi{\arabic{enumi}~}
\begin{Authlist}
\item \Idef{org1}A.I. Alikhanyan National Science Laboratory (Yerevan Physics Institute) Foundation, Yerevan, Armenia
\item \Idef{org2}Bogolyubov Institute for Theoretical Physics, National Academy of Sciences of Ukraine, Kiev, Ukraine
\item \Idef{org3}Bose Institute, Department of Physics  and Centre for Astroparticle Physics and Space Science (CAPSS), Kolkata, India
\item \Idef{org4}Budker Institute for Nuclear Physics, Novosibirsk, Russia
\item \Idef{org5}California Polytechnic State University, San Luis Obispo, California, United States
\item \Idef{org6}Central China Normal University, Wuhan, China
\item \Idef{org7}Centre de Calcul de l'IN2P3, Villeurbanne, Lyon, France
\item \Idef{org8}Centro de Aplicaciones Tecnol\'{o}gicas y Desarrollo Nuclear (CEADEN), Havana, Cuba
\item \Idef{org9}Centro de Investigaci\'{o}n y de Estudios Avanzados (CINVESTAV), Mexico City and M\'{e}rida, Mexico
\item \Idef{org10}Centro Fermi - Museo Storico della Fisica e Centro Studi e Ricerche ``Enrico Fermi', Rome, Italy
\item \Idef{org11}Chicago State University, Chicago, Illinois, United States
\item \Idef{org12}China Institute of Atomic Energy, Beijing, China
\item \Idef{org13}Chonbuk National University, Jeonju, Republic of Korea
\item \Idef{org14}Comenius University Bratislava, Faculty of Mathematics, Physics and Informatics, Bratislava, Slovakia
\item \Idef{org15}COMSATS Institute of Information Technology (CIIT), Islamabad, Pakistan
\item \Idef{org16}Creighton University, Omaha, Nebraska, United States
\item \Idef{org17}Department of Physics, Aligarh Muslim University, Aligarh, India
\item \Idef{org18}Department of Physics, Pusan National University, Pusan, Republic of Korea
\item \Idef{org19}Department of Physics, Sejong University, Seoul, Republic of Korea
\item \Idef{org20}Department of Physics, University of California, Berkeley, California, United States
\item \Idef{org21}Department of Physics, University of Oslo, Oslo, Norway
\item \Idef{org22}Department of Physics and Technology, University of Bergen, Bergen, Norway
\item \Idef{org23}Dipartimento di Fisica dell'Universit\`{a} 'La Sapienza' and Sezione INFN, Rome, Italy
\item \Idef{org24}Dipartimento di Fisica dell'Universit\`{a} and Sezione INFN, Cagliari, Italy
\item \Idef{org25}Dipartimento di Fisica dell'Universit\`{a} and Sezione INFN, Trieste, Italy
\item \Idef{org26}Dipartimento di Fisica dell'Universit\`{a} and Sezione INFN, Turin, Italy
\item \Idef{org27}Dipartimento di Fisica e Astronomia dell'Universit\`{a} and Sezione INFN, Bologna, Italy
\item \Idef{org28}Dipartimento di Fisica e Astronomia dell'Universit\`{a} and Sezione INFN, Catania, Italy
\item \Idef{org29}Dipartimento di Fisica e Astronomia dell'Universit\`{a} and Sezione INFN, Padova, Italy
\item \Idef{org30}Dipartimento di Fisica `E.R.~Caianiello' dell'Universit\`{a} and Gruppo Collegato INFN, Salerno, Italy
\item \Idef{org31}Dipartimento DISAT del Politecnico and Sezione INFN, Turin, Italy
\item \Idef{org32}Dipartimento di Scienze e Innovazione Tecnologica dell'Universit\`{a} del Piemonte Orientale and INFN Sezione di Torino, Alessandria, Italy
\item \Idef{org33}Dipartimento Interateneo di Fisica `M.~Merlin' and Sezione INFN, Bari, Italy
\item \Idef{org34}European Organization for Nuclear Research (CERN), Geneva, Switzerland
\item \Idef{org35}Faculty of Electrical Engineering, Mechanical Engineering and Naval Architecture, University of Split, Split, Croatia
\item \Idef{org36}Faculty of Engineering and Science, Western Norway University of Applied Sciences, Bergen, Norway
\item \Idef{org37}Faculty of Nuclear Sciences and Physical Engineering, Czech Technical University in Prague, Prague, Czech Republic
\item \Idef{org38}Faculty of Science, P.J.~\v{S}af\'{a}rik University, Ko\v{s}ice, Slovakia
\item \Idef{org39}Frankfurt Institute for Advanced Studies, Johann Wolfgang Goethe-Universit\"{a}t Frankfurt, Frankfurt, Germany
\item \Idef{org40}Gangneung-Wonju National University, Gangneung, Republic of Korea
\item \Idef{org41}Gauhati University, Department of Physics, Guwahati, India
\item \Idef{org42}Helmholtz-Institut f\"{u}r Strahlen- und Kernphysik, Rheinische Friedrich-Wilhelms-Universit\"{a}t Bonn, Bonn, Germany
\item \Idef{org43}Helsinki Institute of Physics (HIP), Helsinki, Finland
\item \Idef{org44}High Energy Physics Group,  Universidad Aut\'{o}noma de Puebla, Puebla, Mexico
\item \Idef{org45}Hiroshima University, Hiroshima, Japan
\item \Idef{org46}Hochschule Worms, Zentrum  f\"{u}r Technologietransfer und Telekommunikation (ZTT), Worms, Germany
\item \Idef{org47}Horia Hulubei National Institute of Physics and Nuclear Engineering, Bucharest, Romania
\item \Idef{org48}Indian Institute of Technology Bombay (IIT), Mumbai, India
\item \Idef{org49}Indian Institute of Technology Indore, Indore, India
\item \Idef{org50}Indonesian Institute of Sciences, Jakarta, Indonesia
\item \Idef{org51}INFN, Laboratori Nazionali di Frascati, Frascati, Italy
\item \Idef{org52}INFN, Sezione di Bari, Bari, Italy
\item \Idef{org53}INFN, Sezione di Bologna, Bologna, Italy
\item \Idef{org54}INFN, Sezione di Cagliari, Cagliari, Italy
\item \Idef{org55}INFN, Sezione di Catania, Catania, Italy
\item \Idef{org56}INFN, Sezione di Padova, Padova, Italy
\item \Idef{org57}INFN, Sezione di Roma, Rome, Italy
\item \Idef{org58}INFN, Sezione di Torino, Turin, Italy
\item \Idef{org59}INFN, Sezione di Trieste, Trieste, Italy
\item \Idef{org60}Inha University, Incheon, Republic of Korea
\item \Idef{org61}Institut de Physique Nucl\'{e}aire d'Orsay (IPNO), Institut National de Physique Nucl\'{e}aire et de Physique des Particules (IN2P3/CNRS), Universit\'{e} de Paris-Sud, Universit\'{e} Paris-Saclay, Orsay, France
\item \Idef{org62}Institute for Nuclear Research, Academy of Sciences, Moscow, Russia
\item \Idef{org63}Institute for Subatomic Physics, Utrecht University/Nikhef, Utrecht, Netherlands
\item \Idef{org64}Institute for Theoretical and Experimental Physics, Moscow, Russia
\item \Idef{org65}Institute of Experimental Physics, Slovak Academy of Sciences, Ko\v{s}ice, Slovakia
\item \Idef{org66}Institute of Physics, Homi Bhabha National Institute, Bhubaneswar, India
\item \Idef{org67}Institute of Physics of the Czech Academy of Sciences, Prague, Czech Republic
\item \Idef{org68}Institute of Space Science (ISS), Bucharest, Romania
\item \Idef{org69}Institut f\"{u}r Kernphysik, Johann Wolfgang Goethe-Universit\"{a}t Frankfurt, Frankfurt, Germany
\item \Idef{org70}Instituto de Ciencias Nucleares, Universidad Nacional Aut\'{o}noma de M\'{e}xico, Mexico City, Mexico
\item \Idef{org71}Instituto de F\'{i}sica, Universidade Federal do Rio Grande do Sul (UFRGS), Porto Alegre, Brazil
\item \Idef{org72}Instituto de F\'{\i}sica, Universidad Nacional Aut\'{o}noma de M\'{e}xico, Mexico City, Mexico
\item \Idef{org73}iThemba LABS, National Research Foundation, Somerset West, South Africa
\item \Idef{org74}Johann-Wolfgang-Goethe Universit\"{a}t Frankfurt Institut f\"{u}r Informatik, Fachbereich Informatik und Mathematik, Frankfurt, Germany
\item \Idef{org75}Joint Institute for Nuclear Research (JINR), Dubna, Russia
\item \Idef{org76}Korea Institute of Science and Technology Information, Daejeon, Republic of Korea
\item \Idef{org77}KTO Karatay University, Konya, Turkey
\item \Idef{org78}Laboratoire de Physique Subatomique et de Cosmologie, Universit\'{e} Grenoble-Alpes, CNRS-IN2P3, Grenoble, France
\item \Idef{org79}Lawrence Berkeley National Laboratory, Berkeley, California, United States
\item \Idef{org80}Lund University Department of Physics, Division of Particle Physics, Lund, Sweden
\item \Idef{org81}Nagasaki Institute of Applied Science, Nagasaki, Japan
\item \Idef{org82}Nara Women{'}s University (NWU), Nara, Japan
\item \Idef{org83}National and Kapodistrian University of Athens, School of Science, Department of Physics , Athens, Greece
\item \Idef{org84}National Centre for Nuclear Research, Warsaw, Poland
\item \Idef{org85}National Institute of Science Education and Research, Homi Bhabha National Institute, Jatni, India
\item \Idef{org86}National Nuclear Research Center, Baku, Azerbaijan
\item \Idef{org87}National Research Centre Kurchatov Institute, Moscow, Russia
\item \Idef{org88}Niels Bohr Institute, University of Copenhagen, Copenhagen, Denmark
\item \Idef{org89}Nikhef, National institute for subatomic physics, Amsterdam, Netherlands
\item \Idef{org90}NRC Kurchatov Institute IHEP, Protvino, Russia
\item \Idef{org91}NRNU Moscow Engineering Physics Institute, Moscow, Russia
\item \Idef{org92}Nuclear Physics Group, STFC Daresbury Laboratory, Daresbury, United Kingdom
\item \Idef{org93}Nuclear Physics Institute of the Czech Academy of Sciences, \v{R}e\v{z} u Prahy, Czech Republic
\item \Idef{org94}Oak Ridge National Laboratory, Oak Ridge, Tennessee, United States
\item \Idef{org95}Ohio State University, Columbus, Ohio, United States
\item \Idef{org96}Petersburg Nuclear Physics Institute, Gatchina, Russia
\item \Idef{org97}Physics department, Faculty of science, University of Zagreb, Zagreb, Croatia
\item \Idef{org98}Physics Department, Panjab University, Chandigarh, India
\item \Idef{org99}Physics Department, University of Jammu, Jammu, India
\item \Idef{org100}Physics Department, University of Rajasthan, Jaipur, India
\item \Idef{org101}Physikalisches Institut, Eberhard-Karls-Universit\"{a}t T\"{u}bingen, T\"{u}bingen, Germany
\item \Idef{org102}Physikalisches Institut, Ruprecht-Karls-Universit\"{a}t Heidelberg, Heidelberg, Germany
\item \Idef{org103}Physik Department, Technische Universit\"{a}t M\"{u}nchen, Munich, Germany
\item \Idef{org104}Research Division and ExtreMe Matter Institute EMMI, GSI Helmholtzzentrum f\"ur Schwerionenforschung GmbH, Darmstadt, Germany
\item \Idef{org105}Rudjer Bo\v{s}kovi\'{c} Institute, Zagreb, Croatia
\item \Idef{org106}Russian Federal Nuclear Center (VNIIEF), Sarov, Russia
\item \Idef{org107}Saha Institute of Nuclear Physics, Homi Bhabha National Institute, Kolkata, India
\item \Idef{org108}School of Physics and Astronomy, University of Birmingham, Birmingham, United Kingdom
\item \Idef{org109}Secci\'{o}n F\'{\i}sica, Departamento de Ciencias, Pontificia Universidad Cat\'{o}lica del Per\'{u}, Lima, Peru
\item \Idef{org110}Shanghai Institute of Applied Physics, Shanghai, China
\item \Idef{org111}St. Petersburg State University, St. Petersburg, Russia
\item \Idef{org112}Stefan Meyer Institut f\"{u}r Subatomare Physik (SMI), Vienna, Austria
\item \Idef{org113}SUBATECH, IMT Atlantique, Universit\'{e} de Nantes, CNRS-IN2P3, Nantes, France
\item \Idef{org114}Suranaree University of Technology, Nakhon Ratchasima, Thailand
\item \Idef{org115}Technical University of Ko\v{s}ice, Ko\v{s}ice, Slovakia
\item \Idef{org116}Technische Universit\"{a}t M\"{u}nchen, Excellence Cluster 'Universe', Munich, Germany
\item \Idef{org117}The Henryk Niewodniczanski Institute of Nuclear Physics, Polish Academy of Sciences, Cracow, Poland
\item \Idef{org118}The University of Texas at Austin, Austin, Texas, United States
\item \Idef{org119}Universidad Aut\'{o}noma de Sinaloa, Culiac\'{a}n, Mexico
\item \Idef{org120}Universidade de S\~{a}o Paulo (USP), S\~{a}o Paulo, Brazil
\item \Idef{org121}Universidade Estadual de Campinas (UNICAMP), Campinas, Brazil
\item \Idef{org122}Universidade Federal do ABC, Santo Andre, Brazil
\item \Idef{org123}University College of Southeast Norway, Tonsberg, Norway
\item \Idef{org124}University of Cape Town, Cape Town, South Africa
\item \Idef{org125}University of Houston, Houston, Texas, United States
\item \Idef{org126}University of Jyv\"{a}skyl\"{a}, Jyv\"{a}skyl\"{a}, Finland
\item \Idef{org127}University of Liverpool, Liverpool, United Kingdom
\item \Idef{org128}University of Tennessee, Knoxville, Tennessee, United States
\item \Idef{org129}University of the Witwatersrand, Johannesburg, South Africa
\item \Idef{org130}University of Tokyo, Tokyo, Japan
\item \Idef{org131}University of Tsukuba, Tsukuba, Japan
\item \Idef{org132}Universit\'{e} Clermont Auvergne, CNRS/IN2P3, LPC, Clermont-Ferrand, France
\item \Idef{org133}Universit\'{e} de Lyon, Universit\'{e} Lyon 1, CNRS/IN2P3, IPN-Lyon, Villeurbanne, Lyon, France
\item \Idef{org134}Universit\'{e} de Strasbourg, CNRS, IPHC UMR 7178, F-67000 Strasbourg, France, Strasbourg, France
\item \Idef{org135} Universit\'{e} Paris-Saclay Centre d¿\'Etudes de Saclay (CEA), IRFU, Department de Physique Nucl\'{e}aire (DPhN), Saclay, France
\item \Idef{org136}Universit\`{a} degli Studi di Foggia, Foggia, Italy
\item \Idef{org137}Universit\`{a} degli Studi di Pavia, Pavia, Italy
\item \Idef{org138}Universit\`{a} di Brescia, Brescia, Italy
\item \Idef{org139}Variable Energy Cyclotron Centre, Homi Bhabha National Institute, Kolkata, India
\item \Idef{org140}Warsaw University of Technology, Warsaw, Poland
\item \Idef{org141}Wayne State University, Detroit, Michigan, United States
\item \Idef{org142}Westf\"{a}lische Wilhelms-Universit\"{a}t M\"{u}nster, Institut f\"{u}r Kernphysik, M\"{u}nster, Germany
\item \Idef{org143}Wigner Research Centre for Physics, Hungarian Academy of Sciences, Budapest, Hungary
\item \Idef{org144}Yale University, New Haven, Connecticut, United States
\item \Idef{org145}Yonsei University, Seoul, Republic of Korea
\end{Authlist}
\endgroup
\end{document}